\documentclass[twocolumn]{aastex631}

\hypersetup{linkcolor=blue,citecolor=blue,filecolor=cyan,urlcolor=magenta}

\usepackage{tabularx}
\usepackage{url}
\usepackage{graphicx}
\usepackage[T1]{fontenc}
\usepackage{ae,aecompl}
\usepackage{booktabs}
\usepackage{natbib}
\usepackage{enumitem}
\usepackage{mathtools}
\usepackage{amssymb}
\usepackage{amsmath}
\usepackage{ulem}
\usepackage{epstopdf}
\usepackage{color}
\usepackage{xcolor}
\usepackage{needspace}
\usepackage{soul}

\usepackage{blindtext}
\usepackage{longtable}
\usepackage{tabu}
\usepackage{appendix}
\usepackage{rotating}

\usepackage{lipsum} 

\def\unit #1{\,{\rm #1}}

\newcommand\kms{\rm \,\unit{km\,s^{-1}}}

\newcommand\cmsqi{\rm \,\unit{cm^{-2}}}

\newcommand\kev{\rm \,\unit{keV}}

\newcommand\xiunit{\rm \,erg\,cm\,s^{-1}}

\newcommand\ledd{L_{\rm Edd}}
\newcommand\lambdaedd{\lambda_{\rm Edd}}

\newcommand\lbol{L_{\rm  bol}}

\newcommand\nh{\rm N_{H}}

\newcommand\ks{\, \rm ks}

\newcommand\ev{\unit{\, eV}}
\usepackage{amsmath} 

\newcommand\swift{{\it Swift}}
\newcommand\xmm{{\it XMM-Newton}}

\newcommand\nustar{{\it NuSTAR}}

\newcommand\zwicky{{\it ZTF}}

\newcommand{\onees}{{\sc {1ES 1927+654}}}

\def\aap{A\&A} \def\apjl{ApJ}  
 \def\apjs{ApJS}

\shorttitle{1ES 1927+654, multi-wavelength study }

\journalinfo{Accepted for publication in ApJ}
\begin{document}


\title{The emergence of X-ray emission lines during relativistic radio-jet formation in the changing-look active galactic nucleus 1ES~1927+654}


\author[0000-0002-9163-8653]{Dev R. Sadaula} 
\affiliation{Astrophysics Science Division, NASA Goddard Space Flight Center, Greenbelt, MD 20771, USA.}
\affiliation{Center for Space Science and Technology, University of Maryland Baltimore County, 1000 Hilltop Circle, Baltimore, MD 21250, USA.}
\affiliation{Center for Research and Exploration in Space Science and Technology, NASA/GSFC, Greenbelt, Maryland 20771, USA}

\author[0000-0003-2714-0487]{Sibasish Laha} 
\affiliation{Astrophysics Science Division, NASA Goddard Space Flight Center, Greenbelt, MD 20771, USA.}
\affiliation{Center for Space Science and Technology, University of Maryland Baltimore County, 1000 Hilltop Circle, Baltimore, MD 21250, USA.}
\affiliation{Center for Research and Exploration in Space Science and Technology, NASA/GSFC, Greenbelt, Maryland 20771, USA}

\author[0000-0002-7676-9962]{Eileen T. Meyer}
\affiliation{Department of Physics, University of Maryland Baltimore County, 1000 Hilltop Circle Baltimore, MD 21250, USA}

\author[0000-0003-4727-2209]{Onic I. Shuvo}
\affiliation{Department of Physics, University of Maryland Baltimore County, 1000 Hilltop Circle Baltimore, MD 21250, USA}

\author[0000-0001-6523-6522]{Main Pal}
\affiliation{Department of Physics, Sri Venkateswara College, University of Delhi, Benito Juarez Road, Dhaula Kuan,  New Delhi -- 110021, India}

\author[0000-0003-4790-2653]{Ritesh Ghosh}
\affiliation{MKHS, Murshidabad, West Bengal, India 742401, India }

\author[0000-0002-1094-3147]{Matteo Guainazzi}
\affiliation{European Space Agency (ESA), European Space Research and Technology Centre (ESTEC), Keplerlaan 1, 2201 AZ Noordwijk, The Netherlands}

\author[0000-0001-9879-7780]{Fabio Pacucci}
\affiliation{Center for Astrophysics $\vert$ Harvard \& Smithsonian, Cambridge, MA 02138, USA} 
\affiliation{Black Hole Initiative, Harvard University, Cambridge, MA 02138, USA}

\author[0000-0002-4622-4240]{Stefano Bianchi}
\affiliation{Dipartimento di Matematica e Fisica, Universit\`a degli Studi Roma Tre, Via della Vasca Navale 84, 00146, Roma, Italy}

\author[0009-0006-4968-7108]{Luigi Gallo}
\affiliation{Department of Astronomy and Physics, Saint Mary's University, 923 Robie Street, Halifax, B3H 3C3, Canada}

\author[0000-0001-9475-5292]{Rostom Mbarek}
\altaffiliation{Neil Gehrels Fellow}
\affiliation{Joint Space-Science Institute, University of Maryland, College Park, MD, USA}
\affiliation{Department of Astronomy, University of Maryland, College Park, MD, USA}

\author[0000-0001-9725-5509]{Amelia M. Hankla}
\altaffiliation{Neil Gehrels Fellow}
\affiliation{Joint Space-Science Institute, University of Maryland, College Park, MD 20742, USA}
\affiliation{Department of Astronomy, University of Maryland, College Park, MD, USA}

\author[0000-0002-1239-2721]{Fabio La Franca}
\affiliation{Dipartimento di Matematica e Fisica, Universit\`a degli Studi Roma Tre, Via della Vasca Navale 84, 00146, Roma, Italy}

\author[0000-0000-0000-0000]{Tahir Yaqoob}
\affiliation{Department of Physics, University of Maryland Baltimore County, 1000 Hilltop Circle Baltimore, MD 21250, USA}

\author[0000-0003-4127-0739]{Megan Masterson}
\affiliation{MIT Kavli Institute for Astrophysics and Space Research, Massachusetts Institute of Technology, Cambridge, MA 02139, USA}

\author[0000-0003-0172-0854]{Erin Kara}
\affiliation{MIT Kavli Institute for Astrophysics and Space Research, Massachusetts Institute of Technology, Cambridge, MA 02139, USA}

\author[0000-0002-4992-4664]{Missagh Mehdipour}
\affiliation{Department of Astronomy, University of Michigan, 1085 South University Avenue, Ann Arbor, MI, 48109, USA}
\affiliation{Space Telescope Science Institute, 3700 San Martin Dr, Baltimore, MD 21218, USA}

\author[0000-0001-5231-2645]{Claudio Ricci}
\affiliation{Instituto de Estudios Astrof\'isicos, Facultad de Ingenier\'ia y Ciencias, Universidad Diego Portales, Av. Ej\'ercito Libertador 441, Santiago, Chile}
\affiliation{Kavli Institute for Astronomy and Astrophysics, Peking University, Beijing 100871, China}

\author[0000-0003-3828-2448]{Javier A Garcia}
\affiliation{Astrophysics Science Division, NASA Goddard Space Flight Center, Greenbelt, MD 20771, USA.}

\author[0000-0002-5779-6906]{Timothy R. Kallman}
\affiliation{Astrophysics Science Division, NASA Goddard Space Flight Center, Greenbelt, MD 20771, USA.}

\author[0000-0002-1118-8470]{Ralf Ballhausen}
\affiliation{Astrophysics Science Division, NASA Goddard Space Flight Center, Greenbelt, MD 20771, USA.}
\affiliation{Department of Astronomy, University of Maryland, College Park, MD, USA}

\author[0000-0003-0936-8488]{Mitchell C.~Begelman}
\affiliation{JILA, University of Colorado and National Institute of Standards and Technology, 440 UCB, Boulder, CO 80309-0440, USA.}

\author[0000-0001-7801-0362]{Alexander Philippov}
\affiliation{Department of Physics, University of Maryland, College Park, MD, USA}

\author[0000-0002-8377-9667]{Suvendu Rakshit}
\affiliation{Aryabhatta Research Institute of Observational Sciences (ARIES), Manora Peak, Nainital, 263002 India}

\author[0000-0003-0543-3617]{Francesca Panessa}
\affiliation{INAF -- Istituto di Astrofisica e Planetologia Spaziali, Via del Fosso del Cavaliere 100, Roma, 00133, Italy}

\author[0000-0001-9735-4873]{Ehud Behar}
\affiliation{Department of Physics, Technion, Haifa 32000, Israel}

\author[0000-0003-1673-970X]{S. Bradley Cenko}
\affiliation{Astrophysics Science Division, NASA Goddard Space Flight Center, Greenbelt, MD 20771, USA.}
\affiliation{Joint Space-Science Institute, University of Maryland, College Park, MD 20742, USA}

\author[0000-0001-5742-5980]{Federica Ricci}
\affiliation{Dipartimento di Matematica e Fisica, Universit\`a degli Studi Roma Tre, Via della Vasca Navale 84, 00146, Roma, Italy}
\affiliation{INAF-Osservatorio Astronomico di Roma, via Frascati 33, 00040 Monteporzio Catone, Italy}

\author[0009-0006-7483-0463]{Ilaria Villani}
\affiliation{Dipartimento di Matematica e Fisica, Universit\`a degli Studi Roma Tre, Via della Vasca Navale 84, 00146, Roma, Italy}
\affiliation{INAF-Osservatorio Astronomico di Roma, via Frascati 33, 00040 Monteporzio Catone, Italy}

\author[0000-0002-0982-0561]{James N. Reeves}
\affiliation{Department of Physics, Institute for Astrophysics and Computational Sciences, The Catholic University of America, Washington, DC, USA.}
\affiliation{INAF, Osservatorio Astronomico di Brera, Merate, Lecco, Italy.}



\begin{abstract}

We present results from a comprehensive multi-wavelength monitoring campaign of the changing-look active galactic nucleus 1ES~1927+654 during the onset and evolution of a relativistic radio jet $\sim$(May 2022--August 2025), using observations from \xmm{} \swift{}, TNG, ZTF, VLA, and VLBA. The soft X-ray emission lines at $\sim 0.56\kev$ and $\sim 1\kev$ have appeared with variable strength and width during the formation of the nascent jet. We note that the $\sim 1$~keV feature has been persisting since the post-2017 flare phase. We also report the detection of a broad ($\sim 800\ev$) FeK emission feature at $(6-7)\kev$ in the $\sim 70\ks$ stacked EPIC-pn spectra, marking the first such detection, which historically was lacking in this source. The joint spectral fitting of \xmm{} EPIC-pn and RGS data reveals the presence of ionized absorbers in 2022 ($\log{\xi/\xiunit}\sim 1.5\pm 0.3$, $\nh \sim 2.5\pm 0.9\times 10^{20}\cmsqi$), but weaker than that detected during the high accretion state in 2018 (Eddington ratio, $\lambdaedd>1$). The absorption features further weakened in 2023-2025 and were marginally detectable ($\nh \le 10^{20}\cmsqi$). The entire scenario is suggestive of a real-time transition of the accretion flow (from $\lambdaedd>1$ to $\lambdaedd\sim 0.3$) during which the winds become weaker and the jet starts to form and evolve. Furthermore, both the soft X-ray $(0.3-2)\kev$ and 5~GHz radio fluxes, which increased by factors of $\sim 10$ and $\sim 60$, respectively, since 2022, have recently plateaued at elevated levels, indicating a stabilized accretion disk, corona, and jet configuration.

\end{abstract}


\section{Introduction} \label{sec:intro}
Active Galactic Nuclei (AGN) are among the most energetic objects in the universe, characterized by the rapid accretion of matter onto supermassive black holes (SMBHs) \citep{dressler1989, rees1984}. Some of these systems not only emit huge luminosity across the wide electromagnetic band but also exhibit powerful outflows in the form of jets and winds \citep{Kellermann1989,cre03,fabian2012,king2015}. The jets are collimated streams of relativistic high-energy particles in a magnetically confined structure extending beyond the host galaxy. Whereas the winds are dispersed conical outflow \citep{Nicastro1999,Krongold2007} and presumably originated from the accretion disk \citep{pro04} or inner region of the torus \citep{Krolik2001}. Both the jets and the winds are  important mechanisms by which the SMBH provides feedback to its surroundings, thereby leading to galaxy-SMBH coevolution across cosmic time
\citep{Ferrarese2000,elv00,cre03, fabian2012,laha2021natAs}. The ionized winds from an AGN are mostly detected as blue-shifted ionized absorption in the optical-UV-X-ray spectra. Studies show that $\ge $50\% of AGNs show signatures of ionized absorption \citep{laha2021natAs}. Typically, the column density and ionization parameter of ionized outflows range from $(10^{20} -  10^{22}) \cmsqi$  and log$(\xi/ \xiunit)$= 0.5 - 2.0 \citep{laha2014WAX, tombesi2010,king2015,sadaula2023,sadaula2024,Sadaula2025}. These outflows are commonly known as warm absorbers~(WAs). The jets, on the other hand, manifest themselves through strong radio signatures \citep{Kellermann1989,Sikora2007, Hardcastle2019}. The radio emission in jets comes from synchrotron processes of relativistic electrons accelerated in the magnetic fields \citep{Blandford1979,Begelman1984,Harris2006}.

Changing-Look AGNs (CL AGNs) \citep{lama2015, ricci2022} are a subclass of active galaxies that exhibit dramatic transitions ($\sim 5-100$ times rise in optical flux in a matter of months to years) in their spectral properties over months to years. Even though the origin of the CL-AGNs is still not clearly understood, these changes are believed to be driven by intrinsic variations in the accretion process, such as rapid fluctuations in the accretion rate, disk instabilities, or changes in the structure of the X-ray corona \citep{noda2018, dext2019}. These systems provide a unique opportunity to study AGN activities in human timescales, such as the nature of accretion, outflow, obscuration, etc., which otherwise happen at longer time scales $\sim (10^{5}-10^{6})$ years \citep{Marconi2004,2015MNRAS.451.2517S}.
 
The AGN 1ES 1927+654, which was detected and cataloged in the Einstein Slew Survey \citep{Elvis1992} and studied thoroughly in X-rays \citep{Boller2003, Gallo2013}. This source is unique and enigmatic and has been identified as a changing-look AGN since Dec 2017 \citep{trak19,Ricci2020,Ricci2021,laha2022,masterson2022,Tess_hinkle_1es1927}. This source went through a UV/optical outburst in December 2017, marking it as a changing-look AGN \citep{trak19}, which was previously known as a true type 2 AGN \citep{Boller2003,Gallo2013}.  The destruction and reconstruction of the X-ray corona have been reported in this source and for the first time in an AGN \citep{Ricci2020,masterson2022,laha2022}. This source then came back to its normal state in all wavelength bands \citep{laha2022} after 1200 days of the initial burst. However, starting in May 2022, the source has exhibited strange multi-band behavior, including (a) a strong rise in the soft X-ray excess by a factor of $\sim 10$, (b) an exponential increase in the core radio flux (in Feb 2023 by a factor of $\sim 60$ in just $\sim 4$ months) followed by a spatially resolved radio jet in Feb 2024; (c) a consistent X-ray quasi-periodic oscillation (QPO) with an increasing frequency ($\sim 0.9 -2.2$) mHz \citep{meyer2025, laha2025, masterson2025, Ricci2021}, and (d) a moderate accretion rate of $\lambdaedd\sim 0.3$ \citep{ghosh2023} where $\lambdaedd=\lbol/\ledd$, $\ledd$ is the Eddington luminosity and $\lbol$ is the bolometric luminosity. These behaviors prompted high-cadence observations in radio/optical/UV/X-rays using several state-of-the-art telescopes.
 
In this paper, we report the results obtained from multi-wavelength observations from \xmm{}, \swift{}, Telescope Nationale Galileo (TNG), Zwicky Transient Facility (ZTF), Very Large Array (VLA), and Very Long Baseline Array (VLBA) with a focus on the X-ray spectral study of \onees{} using XMM-Newton's Reflection Grating Spectrometer (RGS) and European Photon Imaging Camera (EPIC-pn) during the time period May 2022 to May 2025. The emission lines at $\sim 0.56 \kev$ and $\sim 1 \kev$ in the soft X-ray were identified, and FeK emission features in $(6-7)\kev$ have been detected. We also investigated ionized outflow and found it to be present in the 2022 observation and weakened in the subsequent years. These studies coincide with the time period when the radio jet had formed (Feb. 2023), the QPO was discovered (May 2022), and a high soft X-ray state (May 2022) occurred. The soft X-ray $(0.3-2)\kev$ and 5 GHz radio fluxes are plateaued, indicating they are now at saturation level, and QPO is still persisting  and plateaued at $\sim 2.5$ mHZ \citep{Masterson2026}.

 
The paper is organized as follows. Section~\ref{sec:observations} describes the observations and data reduction procedures. The results from the multi-wavelength light curves and spectral analyses are presented in Section~\ref{section:results}. Section~\ref{section:discussion} discusses the physical implications of our findings, and Section~\ref{sec:conclusion} summarizes the main conclusions of this work.

\section{Observations and data reprocessing}
\label{sec:observations}
We have used observational data sets obtained from archives as well as Director's Discretionary Time (DDT) and Guest Investigator (GI) programs from multiple facilities, such as \xmm{}, \swift{} and \zwicky{}, \textit{TNG}, \textit{VLA} and \textit{VLBA}, in this work. Here we discuss each of them in detail.

\subsection{XMM-Newton}
\begin{table}[h!]
\centering
\caption{XMM-Newton observations of 1ES 1927+654 that are included in this work.}
\begin{tabular}{cccc}
\hline
\textbf{Obsid} & \textbf{Date} & \textbf{Exposure} & \textbf{Short ID} \\
&\textbf{(yyyy-mm-dd)}&\textbf{(s)}&\\
\hline
0671860201 & 2011-05-20 & 19748 & x0\\\hline
0902590201 & 2022-07-26 & 16602 & x1\\
0902590301 & 2022-07-28 & 12009 & x2\\
0902590401 & 2022-07-30 & 12006 & x3\\
0902590501 & 2022-08-01 & 14094 & x4\\
0915390701 & 2023-02-21 & 20249 & x5\\
0931791401 & 2023-08-07 & 21049 & x6\\
0932392001 & 2024-03-04 & 15220 & x7\\
0932392101 & 2024-03-12 & 20606 & x8\\
0953010401 & 2024-07-19 & 15595 & x9\\
0953010501 & 2024-07-27 & 17035 & x10\\
0953010901 & 2024-10-13 & 8582 & x11\\
0953010601 & 2024-10-21 & 16710 & x12\\
0953010801 & 2025-01-19 & 13742 & x13\\
0953010701 & 2025-01-25 & 15241 & x14\\
0970190101 & 2025-04-30 & 16201 & x15\\
0970190301 & 2025-05-02 & 13324 & x16\\
0970190201 & 2025-05-05 & 12552 & x17\\
\hline
\end{tabular}
\label{Table:XMM_obs}
\end{table}

We have included $17$  \xmm{} \citep{2001A&A...365L...1J} observations in the years 2022, 2023, 2024, and 2025, a period that encompasses the radio jet formation and the X-ray QPO phase in this source. The 2011 \xmm{} observation of the source was included  for comparison because it is only the pre-flare X-ray observation (prior to the changing-look event in Dec 2017). Table \ref{Table:XMM_obs} shows the details (observation ID, date of observation, total exposure time, and a short ID) of the observations. The observations on Oct-2024 and May-2025 were obtained through DDT (PI: M. Masterson). A follow-up timing study on this source with \xmm{} observations is presented in detail by Masterson et al. \citep{Masterson2026}. 

We used the \xmm{} Science Analysis System (SAS v19.0.0) to process the Observation Data Files (ODFs) from all observations. The {\tt EVSELECT} task was used to select the single and double events for the PN detector ({\tt PATTERN<4}). We created light curves from the event files for each observation to account for the high background flaring using a rate cutoff of $<0.4 \rm \, count\, s^{-1}$. The data above the cutoff is most likely due to the background flare for this source \citep{laha2014WAX}. We found significant particle background flares in the Aug-2023 observations, but fewer in other epochs. We found that the Feb 2023, Aug 2023, and March 2024 observations show some pileup in the soft X-rays (using the SAS task {\tt epatplot}). For these observations, we selected an annular region for source photons with inner and outer radii of $8$ arcsec and $30$ arcsec, respectively, centered on the source. In all cases, the background photons were extracted from appropriate regions away from the source. The response matrices were generated using the SAS tasks {\tt arfgen} and {\tt rmfgen}. The spectra were grouped using the command {\tt ftgrppha} with a minimum of 20 counts in each energy bin for the PN spectrum.

The RGS data were reprocessed using the SAS task {\it rgsproc}, with the source’s optical coordinates used as a reference to calculate the position-dependent wavelength scale. The particle background was filtered using light curves extracted from CCD9, which has the highest sensitivity to proton events and the lowest effective area for X-ray photons. Only data below the background count rate of 1 cts/s were considered to obtain the good time interval. The spectra were grouped using the command {\tt ftgrppha} with a minimum of 1 count in each energy bin for the RGS spectrum.

\subsection{\swift{}}
\label{subsec:swift}

\onees{} has been observed by \emph{the Neil Gehrels Swift Observatory} X-ray Telescope \citep[XRT,][]{burrows2005} at a high cadence from May 2022 to Aug 2025 under DDT and \emph{Swift}-GI programs (PI: S.Laha). The details of the \swift{} observations are listed in Table  \ref{tab:swift_obs_tab1} and Table  \ref{tab:swift_obs_tab2}. The details of the prior \swift{} observations before April 2024 are given in previous papers \citep{laha2022,ghosh2023,laha2025}. We follow the same abbreviations for \swift{} observation IDs as used in \cite{laha2025}.  We have also obtained the UV flux density using \swift{} UVOT~\citep{2005SSRv..120...95R} observations happening simultaneously with \swift{} XRT. We refer to \cite{ghosh2023,laha2025} for a full description of UVOT data reprocessing and analysis. The UV flux density quoted in this work (Figure \ref{fig:swift_lc_zoom}, Figure \ref{fig:swift_lc_main}, Table \ref{tab:swift_fit}) has been corrected for the absorption by the dust and gas of the Milky Way galaxy.

\begin{table}
\centering
\caption{Swift Observations of 1ES~1927+654, included in this work (Apr. 2024 to Apr. 5, 2025).}
\begin{tabular}{cccc}
\hline
\textbf{Obsid} & \textbf{Date} & \textbf{Exposure} & \textbf{Short ID} \\
&\textbf{(yyyy-mm-dd)}&\textbf{(s)}&\\
\hline
00016519003 & 2024-04-18 & 953 & S145 \\
00016519004 & 2024-04-25 & 874 & S146 \\
00097300002 & 2024-04-25 & 669 & S146A \\
00097300003 & 2024-04-30 & 1543 & S147 \\
00016519005 & 2024-05-02 & 916 & S148 \\
00097300004 & 2024-05-09 & 976 & S149 \\
00097300005 & 2024-05-24 & 661 & S150 \\
00097300006 & 2024-05-30 & 801 & S151 \\
00089836001 & 2024-06-10 & 2255 & S152 \\
00097300009 & 2024-06-20 & 886 & S153 \\
00097300010 & 2024-06-27 & 1020 & S154 \\
00097300011 & 2024-07-04 & 1068 & S155 \\
00097300012 & 2024-07-14 & 919 & S156 \\
00097300013 & 2024-07-18 & 941 & S157 \\
00097300014 & 2024-07-25 & 943 & S158 \\
00097300015 & 2024-08-01 & 841 & S159 \\
00097300016 & 2024-08-08 & 1388 & S160 \\
00097300017 & 2024-08-15 & 719 & S161 \\
00097300018 & 2024-08-22 & 958 & S162 \\
00097300020 & 2024-09-10 & 1008 & S163 \\
00097300021 & 2024-09-12 & 916 & S164 \\
00089836002 & 2024-09-20 & 1433 & S165 \\
00097300023 & 2024-09-26 & 938 & S166 \\
00097300024 & 2024-10-03 & 1041 & S167 \\
00097300025 & 2024-10-10 & 876 & S168 \\
00097300026 & 2024-10-24 & 651 & S169 \\
00097300027 & 2024-10-31 & 886 & S170 \\
00097300028 & 2024-11-07 & 909 & S171 \\
00097300029 & 2024-11-14 & 819 & S172 \\
00097300030 & 2024-11-21 & 941 & S173 \\
00097300031 & 2024-11-28 & 769 & S174 \\
00097300032 & 2024-12-12 & 921 & S175 \\
00097300033 & 2024-12-19 & 923 & S176 \\
00097300034 & 2024-12-26 & 1073 & S177 \\
00097300036 & 2025-01-09 & 709 & S178 \\
00097300037 & 2025-01-16 & 948 & S179 \\
00097300038 & 2025-02-06 & 933 & S180 \\
00097300039 & 2025-02-13 & 931 & S181 \\
00097300040 & 2025-02-20 & 1001 & S182 \\
00097300041 & 2025-02-27 & 826 & S183 \\
00097300042 & 2025-03-06 & 951 & S184 \\
00097300043 & 2025-03-22 & 919 & S185 \\
00097802001 & 2025-04-01 & 936 & S186 \\
00010682145 & 2025-04-02 & 1046 & S187 \\
00010682146 & 2025-04-03 & 881 & S188 \\
00010682147 & 2025-04-04 & 889 & S189 \\
00010682148 & 2025-04-05 & 769 & S190 \\
\hline
\end{tabular}
\label{tab:swift_obs_tab1}
\end{table}

\begin{table}
\centering
\caption{Swift observations of \onees{}, included in this work (Apr. 7, 2025 to Aug. 2025).}
\begin{tabular}{cccc}
\hline
\textbf{Obsid} & \textbf{Date} & \textbf{Exposure} & \textbf{Short ID} \\
&\textbf{(yyyy-mm-dd)}&\textbf{(s)}&\\
\hline
00010682150 & 2025-04-07 & 746 & S191 \\
00010682151 & 2025-04-08 & 904 & S192 \\
00097802002 & 2025-04-08 & 699 & S192A \\
00010682152 & 2025-04-09 & 866 & S193 \\
00010682154 & 2025-04-11 & 988 & S194 \\
00010682155 & 2025-04-12 & 966 & S195 \\
00097802003 & 2025-04-15 & 1021 & S196 \\
00097802004 & 2025-04-22 & 931 & S197 \\
00097802006 & 2025-05-13 & 1813 & S198 \\
00097802008 & 2025-05-24 & 354 & S199 \\
00097802009 & 2025-05-27 & 899 & S200 \\
00097802010 & 2025-05-29 & 539 & S201 \\
00097802011 & 2025-06-03 & 686 & S202 \\
00097802012 & 2025-06-10 & 1128 & S203 \\
00097802013 & 2025-06-26 & 574 & S204 \\
00097802014 & 2025-07-01 & 1056 & S205 \\
00097802015 & 2025-07-08 & 933 & S206 \\
00097802016 & 2025-07-15 & 1021 & S207 \\
00097802017 & 2025-07-22 & 1226 & S208 \\
00097802018 & 2025-07-29 & 889 & S209 \\
00097802019 & 2025-08-03 & 581 & S210 \\
00097802020 & 2025-08-14 & 806 & S211 \\
00097802021 & 2025-08-19 & 923 & S212 \\
00097802022 & 2025-08-26 & 804 & S213 \\
\hline
\end{tabular}
\textbf{Note}: Refer to \citet{laha2022}, \citet{ghosh2023} and \citet{laha2025} for all the previous Swift observations.
\label{tab:swift_obs_tab2}
\end{table}

\subsection{Zwicky Transient Facility}
We obtained the optical r-band photometric data from the Zwicky Transient Facility (ZTF) for the period starting Jan 2018 to August 2025 through the public release website DR 21\footnote{\url{https://www.ztf.caltech.edu/ztf-public-releases.html}}  \citep{graham_ztf_2019,bellm_ztf_2019}. We have used the same set of data in our earlier work, \citealt{laha2022,laha2025} and we refer to those works for a full detail of the observations. We note that the host galaxy contribution is not subtracted from the r-band flux. The data presented here are on a nearly daily cadence. The light curves are given in Figures \ref{fig:swift_lc_zoom} and \ref{fig:swift_lc_main} below.

\subsection{Telescopio Nazionale Galileo}
We conducted two spectroscopic monitoring campaigns of the AGN \object{1ES 1927+654} (PI: La Franca, Ricci, Villani) using the Telescopio Nazionale Galileo (TNG) with the DOLORES instrument.

The first campaign was carried out between May and November 2023 and consisted of seven observing epochs, obtained during gray-dark time and typically separated by $\gtrsim 2$ weeks, in order to probe spectral and flux variability on short timescales.  
The second campaign began in October 2025 and is part of an ongoing monitoring program that continued until March 2026, providing additional spectroscopic observations beyond the epochs presented here.

Observations were generally obtained under good atmospheric conditions, with airmass $< 1.5$ and seeing in the range ($1.1\!-\!1.4$) arcsec.

All spectra were acquired with DOLORES using the LR-B grism (spectral resolution $R \simeq 600$) and a $1.2$ arcsec slit. For each epoch, three individual exposures of 360~s were obtained, resulting in a total integration time of 1080~s. All observations were performed at the parallactic angle.

\begin{figure*}
    \centering
    \includegraphics[width=0.99\textwidth]{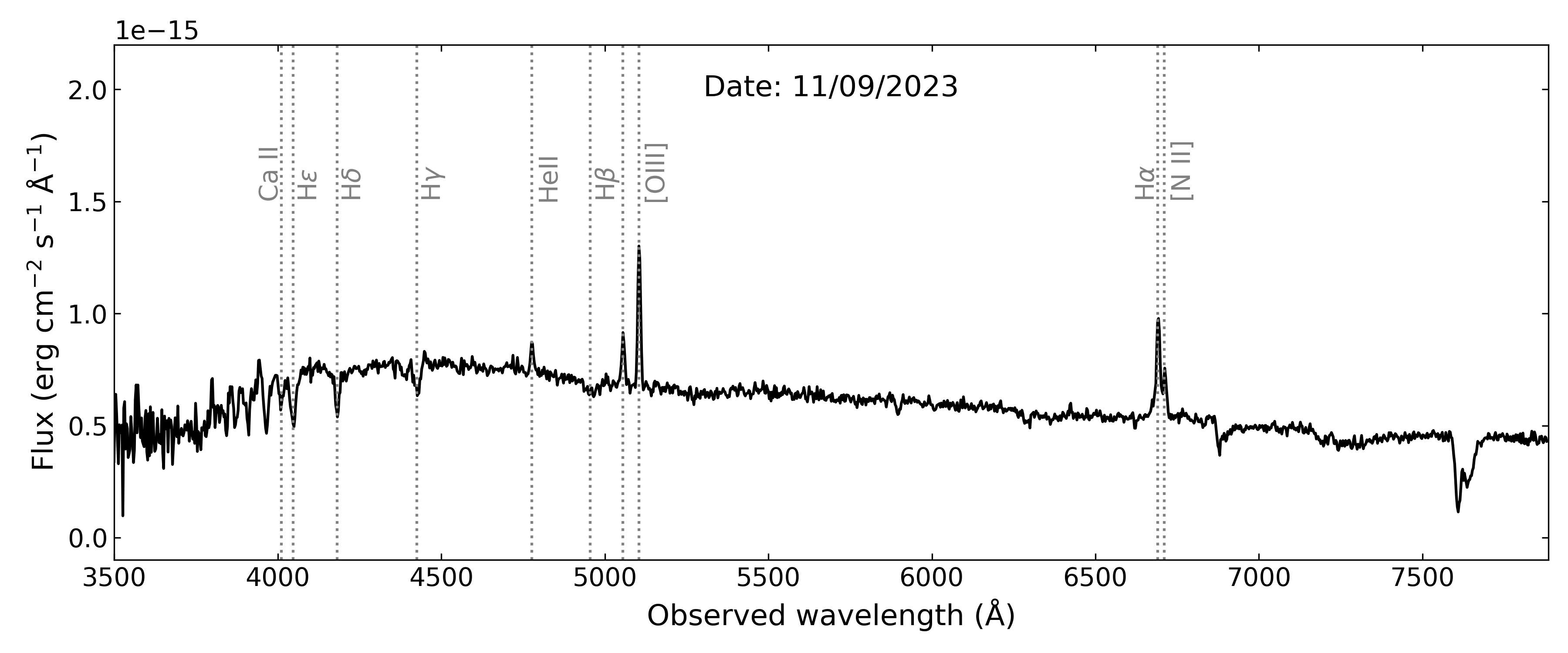}

            \caption{Optical spectrum of \onees{} obtained with TNG/DOLORES on September 11, 2023, during our observational campaign following the soft X-ray rise and radio outburst. The broad-line region (BLR) component is clearly absent. The observed spectrum is shown in black. Gray dotted lines indicate the positions of typical AGN emission lines and host galaxy absorption lines. See Table \ref{tab:TNG_spec_2023} for details.} 
    \label{fig:TNG_optical_spectra}
\end{figure*}

Standard calibration frames (bias, flat fields, and arc-lamp exposures) were obtained during the same nights as the science observations. Data reduction was carried out using standard procedures within \texttt{IRAF}, including bias subtraction, flat-field correction, sky subtraction, and cosmic-ray removal. Wavelength calibration was performed using Ar, Ne+Hg+Kr, and He lamps for each observing run. Flux calibration was obtained using the spectrophotometric standard star LDS~749B, observed on nights adjacent to the target observations.
The spectrum from 2023 is presented in Fig. \ref{fig:TNG_optical_spectra}.

\subsection{VLBA and VLA}
This source has been continuously monitored by the Very Long Baseline Array (VLBA) and the Very Large Array (VLA) since Feb 2023. The details of the source evolution in radio are presented in our previous works \citealt{meyer2025,laha2025}, hence we refer the readers to these for data reduction and analysis steps. In this work we present the 5 GHz radio light curve obtained from the VLBA data for the entire period encompassing the pre-flare to August 2025.

\begin{figure*}
    \centering
    \includegraphics[width=0.85\linewidth]{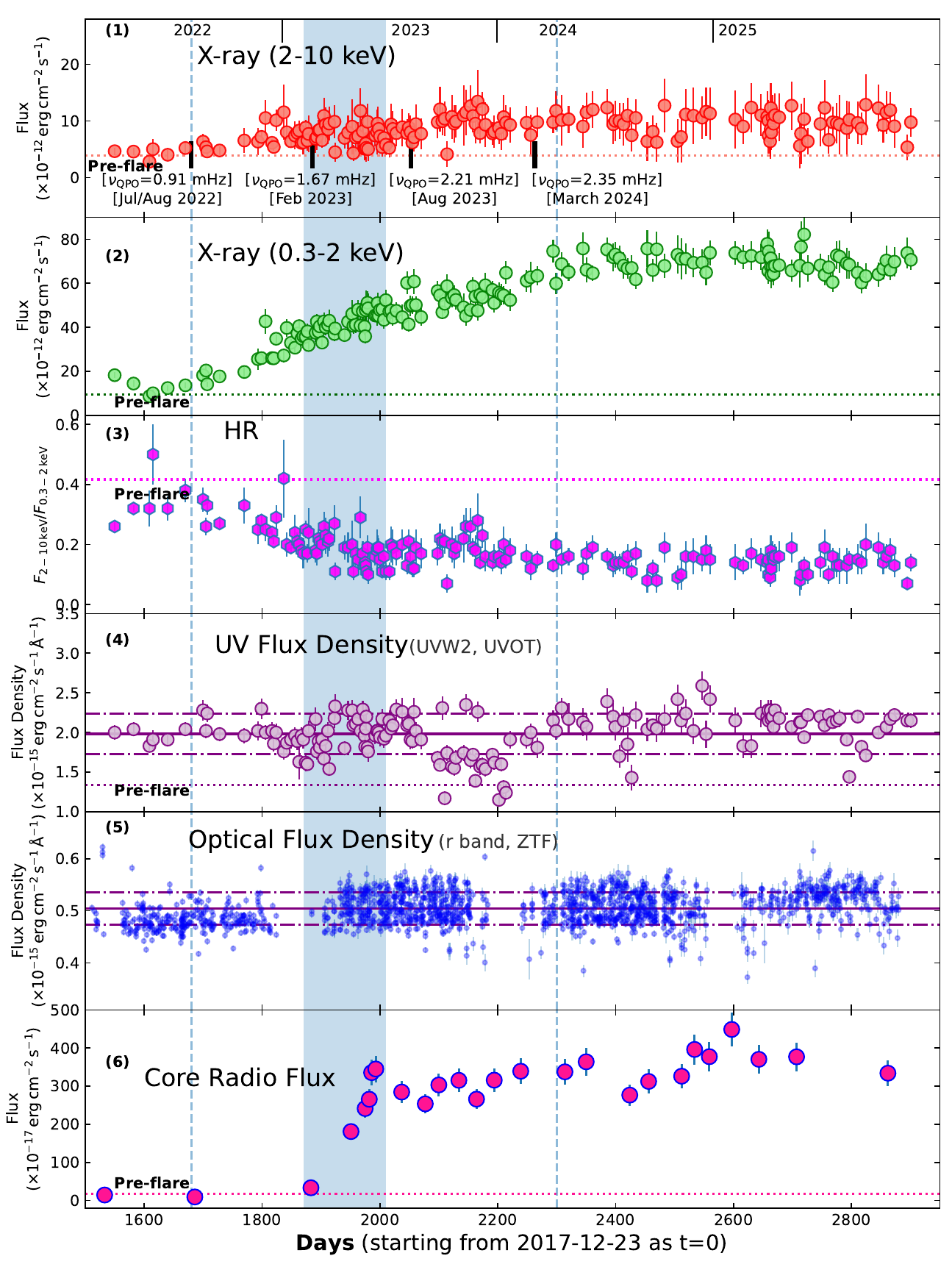}
    \caption{The light curves for 1ES 1927+654 in different wavelength bands from May 2022 to August 2025. The reference date is 2017-12-23, when the UV/optical burst was initially reported by \citet{trak19}. The first and second panels are  $(2-10) \kev$ and $(0.3-2) \kev$ Swift XRT data light curves, respectively. The third panel is the hardness ratio $(F_{(2-10)\kev}/F_{(0.3-2)\kev})$, the fourth panel is the Swift UVOT UVW2 band UV light curve, the fifth is the optical r-band (ZTF) flux density,  and the sixth panel is the {\it VLBA} core C-band ($\sim 5 GHz$) radio flux \cite{meyer2024}. The vertical bold dark lines in the first panels are the date and frequency of QPO detection and $(2-10) \kev$ as reported by \citet{masterson2025}. The first vertical dashed blue line corresponds to the date (May 2022) when the soft X-ray started to rise, while the second vertical blue dashed line corresponds to the date (April 2024) when the soft X-ray started to plateau. For the details, refer to our previous paper \citep{laha2025}.}
    \label{fig:swift_lc_zoom}
\end{figure*}

\section{Analysis and Results}
\label{section:results}
In this section, we present the details of data analysis of multi-wavelength observations, focusing on the X-ray spectra and the key results.

\subsection{The \swift{} and VLBA light curves}

We have included multi-wavelength light curves in X-ray, UV, optical, and radio as shown in Figures \ref{fig:swift_lc_zoom} and \ref{fig:swift_lc_main}. Figure \ref{fig:swift_lc_zoom} shows the evolution of the source from April 2022 to August 2025, whereas Figure \ref{fig:swift_lc_main} extends from May 2018 to August 2025, encompassing the entire time period after a violent event occurred in December 2017. We also note that the light curves  before April 2024 are already presented in previous works \citep{laha2022, ghosh2023, laha2025, meyer2025}. 

The first panel, panel (1), in Figure \ref{fig:swift_lc_zoom} shows the $2-10 \kev$ X-ray (hard X-ray) light curve demonstrating a slow increase (< 2 times) from May 2022 until August 2025. The second panel (2) shows $0.3-2 \kev$ the X-ray (soft X-ray) light curve, which has increased steadily to nearly $\sim 10$ times that of its pre-flare value and plateaued after April 2024.

The third panel of figure \ref{fig:swift_lc_zoom} shows the ratio of hard to soft X-ray, the hardness ratio ($HR=(F_{2-10\kev}/F_{0.3-2\kev})$). This shows that the source is in a very soft state since April 2024 compared to the time before soft X-rays started to rise in 2022. The obvious reasons are the large increase in the soft X-ray flux and the steeper power law index, $\Gamma$. 

The fourth panel shows the UV flux density, which shows a variation within the $30 \%$ of its mean value. The fifth panel shows the r-band optical flux, and the last panel shows the core radio flux in c-band ($\sim 5 \ $GHz). We note that the radio flux has nearly plateaued after April 2024 to a value $\sim 60$ times that of the pre-flare value. For details on this figure, refer to \citealt{laha2022,laha2025}.

\subsection{The Optical spectra}

The optical spectrum of \onees{} in 2023 is shown in Fig. \ref{fig:TNG_optical_spectra} (we refer to \citet{laha2022} for an analysis of earlier epochs). The details of the prominent emission lines are presented in Table \ref{tab:TNG_spec_2023}. The spectra exhibit comparable intensity levels and spectral slopes in the red wavelength range ($5500$--$7000$~\AA) compared to those in 2021 \citep{laha2022}. This shows that the optical properties did not exhibit any significant change just before the rise in the soft X-ray emission in 2022, nor in the period following the soft X-ray increase. We have also added the optical spectra for different epochs of 2024 and 2025 in Appendix \ref{sec:optical spectra evolution}. The optical properties remained nearly the same throughout the period from 2022 to 2025.

\begin{table}[ht!]
\centering
\caption{Spectroscopic line measurements of 1ES~1927+654 in 2023.}
\begin{tabular}{l@{\hskip 6pt}c@{\hskip 4pt}c@{\hskip 4pt}c@{\hskip 4pt}c}
\hline
Line & Center & FWHM & Flux & S/N \\
     & {\scriptsize(\AA)} & {\scriptsize (km s$^{-1}$)} & {\scriptsize ($10^{-17}$ erg cm$^{-2}$ s$^{-1}$)} &  \\
\hline
HeII 4687  & 4686.7$\pm$1.0 & 600$\pm$17 & 162.7$\pm$4.7 & 9 \\
H$\beta$ & 4862.4$\pm$1.0 & 600$\pm$17 & 22.5$\pm$0.7  & 1 \\
OIII 4959& 4960.0$\pm$1.0 & 600$\pm$17 & 294.9$\pm$8.5 & 14 \\
OIII 5007& 5007.9$\pm$1.0 & 600$\pm$17 & 893$\pm$25    & 42 \\
H$\alpha$   & 6564.8$\pm$1.0 & 581$\pm$20 & 723$\pm$24    & 27 \\
NII 6549 & 6550.0$\pm$1.0 & 581$\pm$20 & 106.6$\pm$3.6 & 4 \\
NII 6585 & 6585.4$\pm$1.0 & 581$\pm$20 & 49$\pm$12     & 12 \\
\hline
\end{tabular}
\label{tab:TNG_spec_2023}
\end{table}

\subsection{X-ray Spectral analysis}\label{sec:spec_analysis}

In this section, we aim to explore two primary aspects: (1) the presence or absence of an ionized outflow and (2) the detection and characterization of possible emission lines in the X-ray spectra (both high-resolution RGS and broadband EPIC-pn). All spectral analyses were performed through simultaneous fitting of the RGS1, RGS2, and EPIC-pn spectra (where possible) with all model parameters tied during the fitting process except the scaling factor, as shown in Figure \ref{fig:simult_fit_jul26}. For spectral fitting, the energy ranges used are $(0.4-2)\kev$ for RGS and $(0.3-10)\kev$ for PN spectra.

For emission line analysis, we follow two different ways:
\begin{enumerate}
\item {Individual Spectral Analysis}: We first analyze each of the individual \xmm{} observations listed in Table \ref{Table:XMM_obs} between July 2022 and May 2025 separately.

\item Stacked Spectral Analysis: To improve the signal-to-noise ratio (SNR), we stack the spectra from multiple observations within each year.  This results in combined exposures of approximately 55 ks for 2022, 41 ks for 2023, 91 ks for 2024, and 70 ks for 2025. The stacking of the observations on a yearly basis was one of the choices to minimize the time gaps between observations. All the observations in 2022 were taken within a week time period. There were only two observations in 2023, which were taken within half a year apart. The observations in 2024 and 2025 are separated by a few months.
 
The main reason for stacking the RGS and EPIC-pn spectra is to allow us to search for possible weak emission features that may not be detected in individual observations. We understand stacking may introduce additional spurious features, so we deal with these with some caution. 

\end{enumerate}

\subsubsection{Warm Absorber}

The warm ionized gas absorption in the observed spectra is estimated using the table model WARMABS \footnote{available from the webpage\\ https://heasarc.gsfc.nasa.gov/docs/software/xstar/xstar.html}. The model spectra were calculated using standard XSTAR \citep{kall01} and source average spectral energy density (SED). The details of SED are given in Appendix \ref{sec:creating_ionized_absorber_model}. The turbulent velocity was chosen to be $1000$ km s$^{-1}$. We also performed a simulation to show that the WA properties with RGS spectra can be constrained for a WA as small as $5\times 10^{19}\cmsqi$ using our model, as shown in Figure \ref{fig:sim_spec_rgs_wa}.

To validate our table model, we performed a spectral fit of the June 2018 \xmm{} observation. This epoch was previously analyzed by \citet{Ricci2021}, who identified an ionized outflow with a column density of $N_{\text{H}} \approx 2.2 \times 10^{20}\cmsqi$, an ionization parameter of $\log(\xi) \approx 2.9 \xiunit$, and an observed redshift of $z \approx -0.22$.

Adopting their multi-component model, \texttt{const$\times$tbabs$\times$ztbabs$\times$warmabs(zgauss + zgauss + bbody + pow)}, we conducted a joint fit of the EPIC-PN and RGS spectra. PN spectra in the joint fit are included in order to constrain the broadband continuum, whereas RGS is to constrain the soft X-ray narrow features. Our results are broadly consistent with the literature, yielding $N_{\text{H}} \approx 6 \times 10^{20}\,\text{cm}^{-2}$, $\log(\xi) \approx 2.2/\,\text{erg}\,\text{cm}\,\text{s}^{-1}$, and $z \approx -0.30$. The marginal discrepancies in the best-fit parameters of ionized outflow are likely attributable to the inclusion of high-resolution RGS data, which provides greater sensitivity to the ionized absorption features compared to the EPIC-MOS spectra.

We investigated the presence of ionized absorption features in the soft X-ray band ($0.4-2)\kev$ by adding the {\tt warmabs} model we created to our baseline continuum tbabs$\times$ztbabs(bbody + pow), which we name the {\textbf `baseline model'}. Here {\tt tbabs} models the galactic absorption $(\mathrm{N_H}$ frozen to $6.42 \times 10^{20}$~cm$^{-2}$, \citealt{kalberla2005}), {\tt ztbabs} describes the host-galaxy absorption, {\tt bbody} describes the soft X-ray excess emission, and {\tt pow} describes the coronal emission. This source has consistently shown the signature of neutral gas absorption from, seemingly, the host galaxy, even though the origin of this absorption is not clear\citep{Boller2003,Gallo2013,ricci2017,laha2022} and we have used {\tt ztbabs} to account this absorption.

\begin{figure*}
    \centering
    \includegraphics[width=0.75\linewidth]{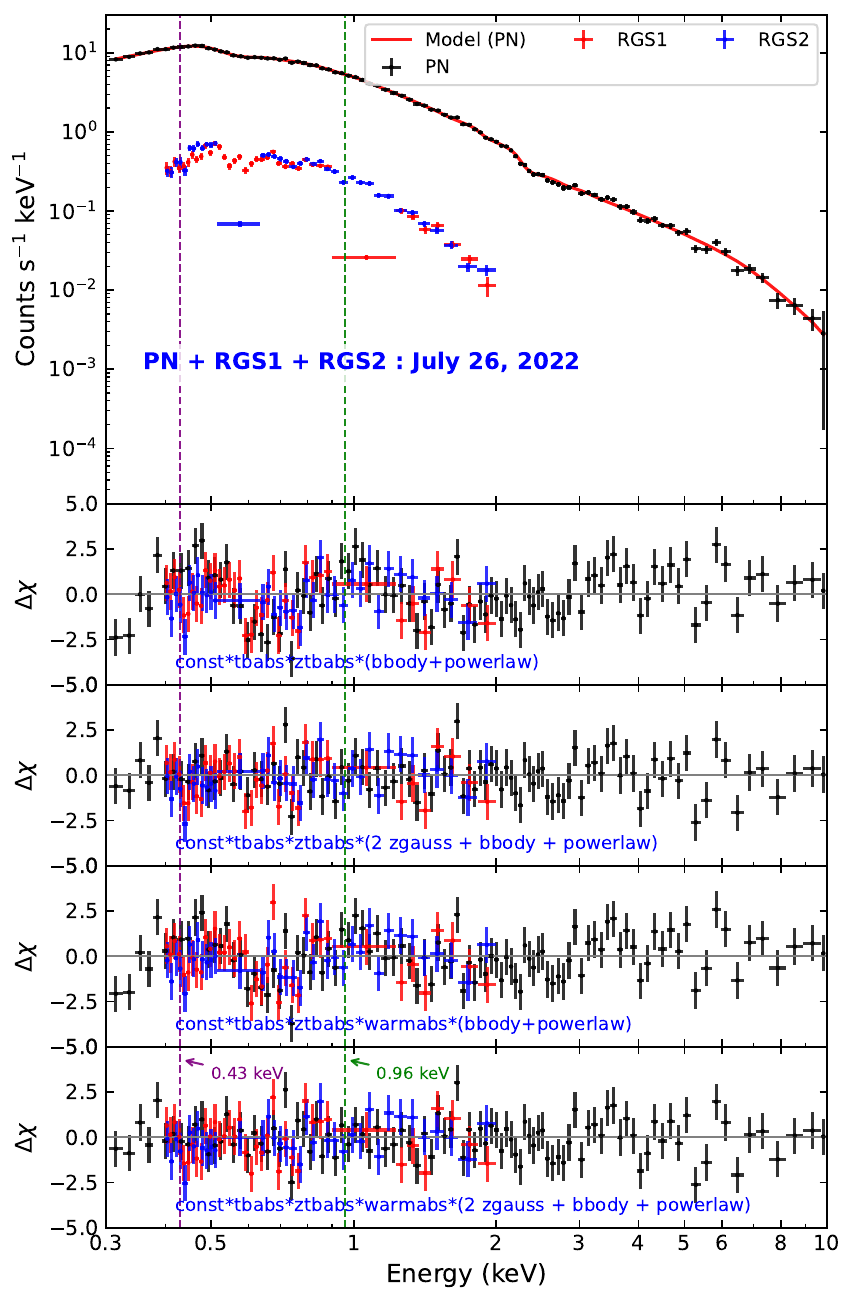}
    \caption{Simultaneous spectral fitting of the XMM-Newton EPIC-pn (black), RGS1 (red), and RGS2 (blue) data for the July 26, 2022 observation. The top panel shows the observed spectrum along with the best-fit model, while the lower four panels display the residuals ($\Delta\chi$) for different spectral models applied sequentially. From top to bottom, the models are: (1) the baseline continuum model, $\mathrm{const}\times\mathrm{tbabs}\times\mathrm{ztbabs}\times(\mathrm{bbody}+\mathrm{powerlaw})$; (2) the baseline model with two Gaussian emission lines at $\sim$0.45 keV and $\sim$1 keV; (3) the baseline model including an ionized absorber modeled with $\mathrm{warmabs}$; and (4) the full model combining both the emission lines and the ionized absorber. All the spectra are heavily rebinned for visualization. The residuals highlight the presence of soft X-ray emission features and demonstrate that the inclusion of Gaussian emission lines significantly improves the fit. The features below $2\kev$ can also be modeled by including a warm absorber but are statistically less significant than emission lines. 
    \label{fig:simult_fit_jul26}}
\end{figure*}

Figure~\ref{fig:simult_fit_jul26} illustrates the step-by-step spectral modeling procedure for the July 26, 2022, observation to test whether the soft X-ray residuals are better described by emission features, ionized absorption, or a combination of both. We fitted the EPIC-pn, RGS1, and RGS2 spectra simultaneously, tying all physical model parameters between the instruments and allowing only the cross-normalization constants to vary. We first applied the baseline continuum model, \texttt{const$\times$tbabs$\times$ztbabs(bbody + pow)}, which gives $C/{\rm dof}=4594/4582$ and leaves residuals below $\sim2$ keV as shown in the second panel of Figure \ref{fig:simult_fit_jul26}. We then added two Gaussian emission lines to the baseline model which improves the fit to $C/{\rm dof}=4495/4576$, corresponding to $\Delta C\sim 99$ for six additional free parameters as shown in the third panel of Figure \ref{fig:simult_fit_jul26}. The best-fit line energies are $E\simeq0.43$ keV and $E\simeq0.96$ keV, showing that the dominant residuals can be efficiently modeled as emission features. 

As an alternative, we replaced the Gaussian lines with a warm absorber component and fitted the model \texttt{const$\times$tbabs$\times$ztbabs$\times$warmabs(bbody + powerlaw)}. This gives $C/{\rm dof}=4561/4579$, or $\Delta C\sim 33$ for three additional free parameters as shown in the fourth panel of Figure \ref{fig:simult_fit_jul26}. Finally, we fitted the full model, including both the emission lines and the warm absorber, \texttt{const$\times$tbabs$\times$ztbabs$\times$warmabs(zgauss+ zgauss + bbody + powerlaw)}, obtaining $C/{\rm dof}=4489/4573$ as shown in the fifth panel of Figure \ref{fig:simult_fit_jul26}. We found only a modest improvement of $\Delta C\sim 6$ relative to the continuum-plus-two-Gaussian model, despite the addition of the warm absorber. It demonstrates that the soft X-ray features in the July 26, 2022 spectrum are statistically better described by emission lines than by a warm absorber alone and that once the emission lines are included, the data do not require a significant warm absorber component. Hence, we adopt the warm absorber {\tt warmabs} on the top of the baseline model to analyze the ionized outflow in the rest of the observations.

\begin{figure*}
    \centering
    \includegraphics[width=0.49\linewidth]{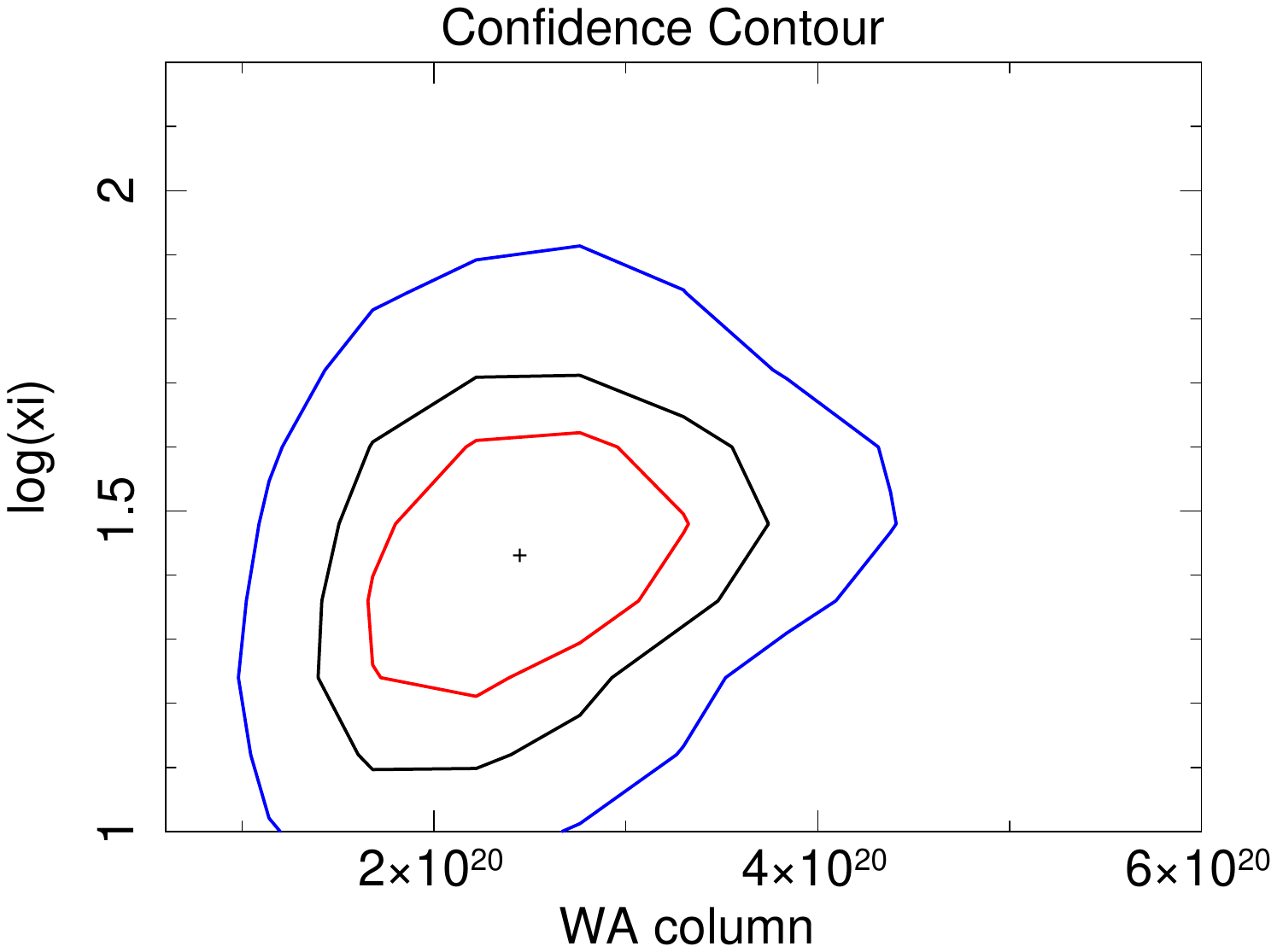}
    \includegraphics[width=0.49\linewidth]{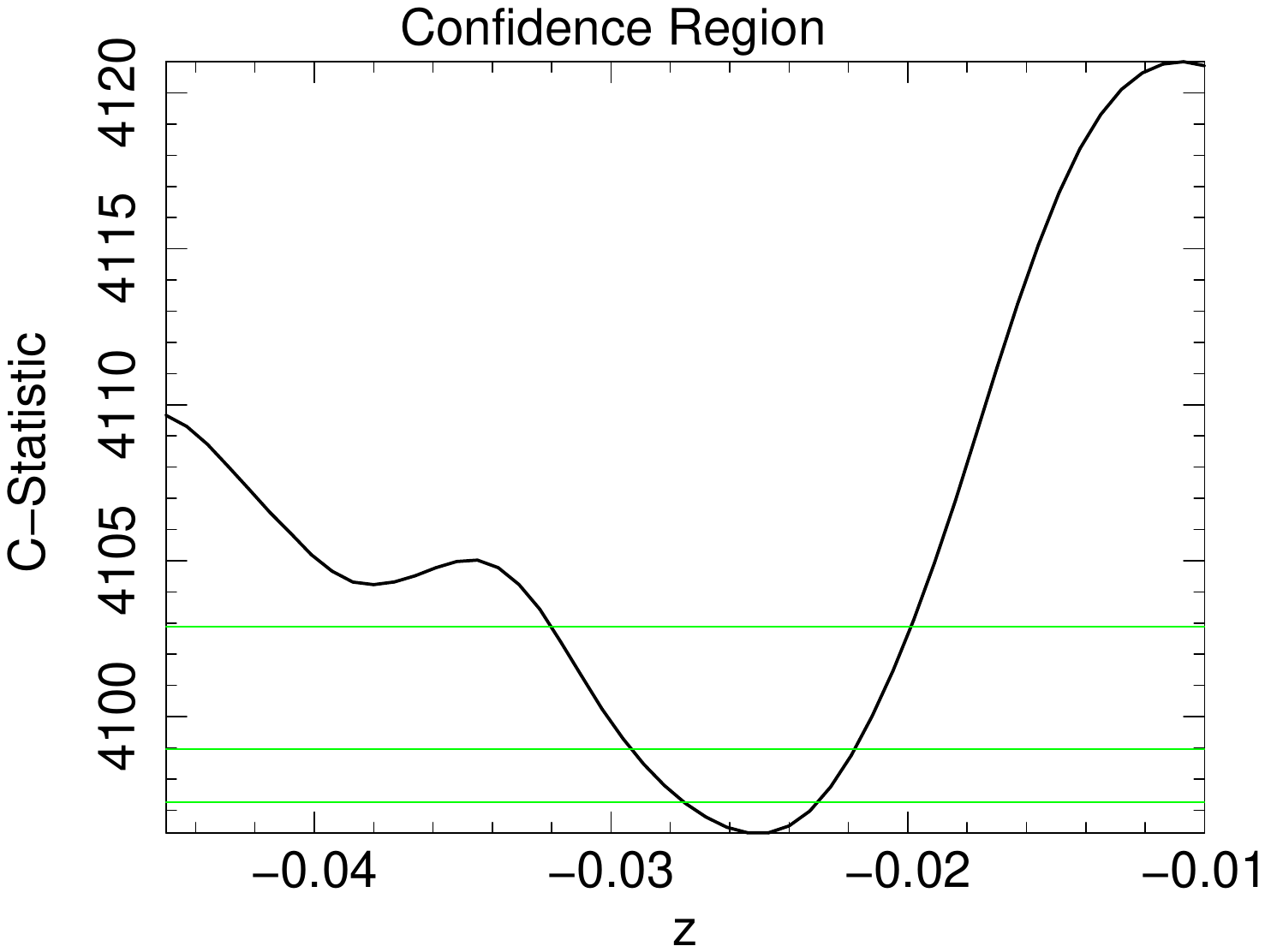}
    \caption{\textbf{Left: }Confidence contours for the ionized absorber (\texttt{warmabs}) parameters: ionization parameter ($\log \xi$) vs. equivalent hydrogen column density (WA column) for the July 26, 2022, observation. The contours represent the 67\% (red), 90\% (black), and 99\% (blue) confidence levels calculated using the C-statistic. The crosshair indicates the best-fit values obtained from the simultaneous RGS+PN fit. The plot illustrates the constraints on the column density and log$\xi$ and the degree of degeneracy between them.} \textbf{Right:} The 1-D distribution of the C-statistic as a function of the absorber redshift $z$ for the July 26, 2022, observation.  The green lines represent the 67\%, 90\%, and 99\% confidence levels from the bottom. The well-defined minimum at $z \approx -0.025$ provides robust evidence for a blueshifted outflow, corresponding to a velocity of $v \approx 7500$ km s$^{-1}$.
    \label{fig:contour_Jul26}
\end{figure*}

To further explore the constraints and potential degeneracies between these parameters, we generated confidence contours for the column density versus the ionization parameter for the model \texttt{const$\times$tbabs$\times$ztbabs$\times$warmabs(bbody + pow)}. Figure \ref{fig:contour_Jul26} displays these contours for the July 26, 2022, observation at the 67\% (red), 90\% (black), and 99\% (blue) confidence levels. The contours indicate a degree of parameter degeneracy, where a range of $N_H$ and $\log \xi$ values can provide comparable fits to the data. Even though the features below $2\kev$ can be modeled with emission lines, this contour shows the presence of warm ionized gas in the source with at least 90\% confidence for this particular observation.

The right panel of Figure \ref{fig:contour_Jul26} presents the 1-D confidence profile for the absorber redshift $z$. The parabolic shape of the C-statistic curve confirms that the blueshift is well-constrained, with a clear minimum at $z \approx -0.025$. This indicates that the outflow velocity is robustly detected, further supporting the presence of a blueshifted warm absorber component in the July 26, 2022, data.

The warm absorber components consistently exhibit significant blueshifts across the 2022--2025 epochs. As detailed in Table \ref{tab:wa_fits}, the measured blueshifts range from $z \approx 0.013$ to $z \approx 0.051$. For the high-signal July 26, 2022 observation, the detected blueshift of $z = 0.025 \pm 0.004$ corresponds to an outflow velocity of $v \approx 7500$ km s$^{-1}$. In the July 19, 2024 observation, we find a maximum blueshift of $z = 0.051 \pm 0.003$, indicating velocities reaching $\sim 15,300$ km s$^{-1}$, which is at the border of the traditional regime of Ultra-Fast Outflows (UFOs) despite the lower ionization state. These types of outflows are called an "entrained UFO" \citep{Serafinelli2019}. In four observations (2022-07-28, 2024-03-12, 2025-01-19, 2025-04-30), only the upper limit of the z ( $\sim 0.006=1800 \kms)$ was obtained from the fitting. 

The 2022 observations show the highest WA column $N_H \approx (1.26-2.65) \times 10^{20}$ cm$^{-2}$ coinciding with the initial rise of the soft X-ray flux \citep{ghosh2023, laha2025} and nascent jet formation \citep{meyer2025}. However, a weaker WA with a smaller statistical improvement is present in the subsequent observations from 2023 to 2025. We also note that a multiplicative constant ({\tt const}) was included to account for cross-calibration uncertainties, fixed at unity for EPIC-pn, and allowed to vary for RGS1 and RGS2, yielding values of $\sim0.90$ and $\sim0.93$.

\vspace{1cm}
\subsubsection{Ultra Fast Outflow (UFO)}
\label{subsubsec:ufo}
UFOs are characterized by blue-shifted absorption lines indicative of high ionization, high velocity, and high column density, powerful AGN-driven winds \citep{tombesi2010,laha2021natAs}. We fitted a UFO with the initial fitting parameters, $\log(\xi) = 3.5$ and $N_{\mathrm{H}} = 10^{23} \ \mathrm{cm}^{-2}$. The final XSPEC model was tbabs$\times$ztbabs$\times$UFO(bbody + pow). The speed of the UFO was taken to be 0.20$c$. The ionization parameter $\log(\xi)$ was bounded from $(2-4.5) /\xiunit$, whereas the column density $N_{\mathrm{H}}$ is bounded between $10^{22}$ and $10^{24} \ \mathrm{cm}^{-2}$. These values are typical of UFOs. The details of the individual fitting with a UFO are shown in Table \ref{tab:ufo_fits} in the appendix. The statistics did not improve; rather, they worsened in many observations, suggesting the absence of UFOs.

\begin{table*}[h]
\centering
\caption{Best-fit parameters, obtained from simultaneous fitting of the RGS1, RGS2, and PN spectra
for individual observations for the model \texttt{const$\times$tbabs$\times$ztbabs$\times$warmabs(bbody + pow)}.}
\label{tab:wa_fits}
\begin{tabular}{cccccccccc}
\hline
\textbf{Date} & \textbf{ztbabs} & \textbf{${N_H}^W$} & \textbf{$\log(\xi)$} & \textbf{$z$} & \textbf{kT} & \textbf{kT Norm} & \textbf{$\Gamma$} 
& \textbf{$\Gamma$ Norm} \tablenotemark{a}& \textbf{$\Delta C / \Delta \mathrm{dof}$} \\
 & $( 10^{20}$ cm$^{-2})$ & $( 10^{20}$ cm$^{-2})$ & & & (keV) & $(\times 10^{-4})$ & & $(\times 10^{-2})$ & \\
\hline
2022-07-26 & $2.65 \pm 0.61$ & $2.47 \pm 0.93$ & $1.43 \pm 0.24$ & $0.025 \pm 0.004$ & $0.13 \pm 0.01$ & $0.77 \pm 0.11$ & $2.74 \pm 0.04$ & $0.51 \pm 0.02$ & 33/3 \\
2022-07-28 & $2.10 \pm 0.62$ & $2.60 \pm 1.00$ & $1.13 \pm 0.25$ & $<0.006$ & $0.14 \pm 0.01$ & $0.83 \pm 0.13$ & $2.72 \pm 0.04$ & $0.52 \pm 0.02$ & 32/3 \\
2022-07-30 & $2.17 \pm 0.63$ & $3.00 \pm 2.25$ & $2.26 \pm 0.49$ & $0.013 \pm 0.006$ & $0.14 \pm 0.01$ & $0.66 \pm 0.09$ & $2.73 \pm 0.04$ & $0.54 \pm 0.02$ & 11/3 \\
2022-08-01 & $1.26 \pm 0.80$ & $2.37 \pm 1.00$ & $1.26 \pm 0.31$ & $0.048 \pm 0.012$ & $0.14 \pm 0.01$ & $0.88 \pm 0.09$ & $2.58 \pm 0.06$ & $0.48 \pm 0.02$ & 30/3 \\

\hline
2023-02-21 & $4.18 \pm 0.50$ & $0.95 \pm 0.41$ & $1.55 \pm 0.36$ & $0.050 \pm 0.004$ & $0.15 \pm 0.01$ & $3.13 \pm 0.16$ & $3.07 \pm 0.04$ & $1.07 \pm 0.04$ & 17/3 \\
2023-08-07 & $5.03 \pm 0.46$ & $0.61 \pm 0.43$ & $1.78 \pm 0.57$ & $0.049 \pm 0.003$ & $0.15 \pm 0.01$ & $5.00 \pm 0.18$ & $3.16 \pm 0.04$ & $1.32 \pm 0.05$ & 10/3 \\
2024-03-04 & $4.73 \pm 0.53$ & $0.30 \pm 0.29$ & $1.23 \pm 0.66$ & $0.048 \pm 0.019$ & $0.15 \pm 0.01$ & $5.88 \pm 0.20$ & $3.16 \pm 0.04$ & $1.37 \pm 0.06$ & 5/3 \\
2024-03-12 & $4.73 \pm 0.47$ & $0.59 \pm 0.57$ & $1.91 \pm 0.45$ & $< 0.006$ & $0.15 \pm 0.01$ & $5.72 \pm 0.17$ & $3.18 \pm 0.04$ & $1.35 \pm 0.05$ & 8/3 \\
2024-07-19 & $5.01 \pm 0.48$ & $0.62 \pm 0.36$ & $1.46 \pm 0.38$ & $0.051 \pm 0.003$ & $0.15 \pm 0.01$ & $6.18 \pm 0.22$ & $3.18 \pm 0.04$ & $1.43 \pm 0.05$ & 12/3 \\
2024-07-27 & $5.17 \pm 0.48$ & $1.26 \pm 0.84$ & $2.33 \pm 0.35$ & $0.015 \pm 0.007$ & $0.15 \pm 0.01$ & $6.04 \pm 0.19$ & $3.20 \pm 0.04$ & $1.42 \pm 0.05$ & 10/3 \\
2024-10-13 & $4.38 \pm 0.67$ & $1.50 \pm 0.88$ & $2.33 \pm 0.30$ & $0.015 \pm 0.003$ & $0.15 \pm 0.01$ & $6.03 \pm 0.25$ & $3.13 \pm 0.06$ & $1.34 \pm 0.07$ & 11/3 \\
2024-10-21 & $4.60 \pm 0.50$ & $0.66 \pm 0.83$ & $2.22 \pm 0.61$ & $0.017 \pm 0.008$ & $0.15 \pm 0.01$ & $6.03 \pm 0.20$ & $3.12 \pm 0.04$ & $1.43 \pm 0.06$ & 4/3 \\
2025-01-19 & $4.51 \pm 0.56$ & $1.30 \pm 0.95$ & $2.33 \pm 0.43$ & $< 0.006$ & $0.16 \pm 0.01$ & $6.12 \pm 0.20$ & $3.15 \pm 0.05$ & $1.39 \pm 0.06$ & 9/3 \\
2025-01-25 & $5.08 \pm 0.52$ & $0.55 \pm 0.59$ & $1.80 \pm 0.47$ & $0.040 \pm 0.004$ & $0.15 \pm 0.01$ & $5.96 \pm 0.20$ & $3.20 \pm 0.04$ & $1.36 \pm 0.06$ & 8/3 \\
2025-04-30 & $4.69 \pm 0.52$ & $< 0.64$ & $1.47 \pm 0.96$ & $< 0.006$ & $0.15 \pm 0.01$ & $5.97 \pm 0.20$ & $3.21 \pm 0.04$ & $1.36 \pm 0.06$ & 3/3 \\
2025-05-02 & $3.88 \pm 0.56$ & $1.74 \pm 0.80$ & $2.33 \pm 0.20$ & $0.019 \pm 0.003$ & $0.16 \pm 0.01$ & $6.20 \pm 0.20$ & $3.10 \pm 0.05$ & $1.26 \pm 0.06$ & 17/3 \\
2025-05-05 & $4.71 \pm 0.58$ & $0.77 \pm 0.58$ & $2.16 \pm 0.62$ & $0.016 \pm 0.005$ & $0.15 \pm 0.01$ & $6.12 \pm 0.22$ & $3.18 \pm 0.05$ & $1.43 \pm 0.07$ & 5/3 \\
\hline
\end{tabular}

\textbf{Note}: The turbulent speeds for the warm absorber model were taken to be $1000 \kms$. The positive $\Delta C$ value at the last column shows the improvement in the fit statistics after adding the WA component to a baseline model. The listed $z$ values are shown as positive magnitudes for convenience, but they correspond to \emph{blueshifted} warm-absorber velocities with a negative sign in front of them. For small shifts, the outflow speed is approximately $v \approx zc$. For example, $z=0.025$ corresponds to $v \approx 0.025\,c \approx 7500$ km s$^{-1}$. We have separated 2022 and other years by a horizontal line, as strong WA is detected only in 2022.
\tablenotetext{a}{Power-law normalization in units of photons keV$^{-1}$ cm$^{-2}$ s$^{-1}$ at 1 keV.}

\end{table*}

\subsubsection{Soft X-ray emission lines in individual observation}
\label{subsubsection:emission_line_indiv_obs}
During the previous violent changing-look phase of the source \onees{}, studies have detected a transient broad emission line at $\sim 1\kev$ \citep{Ricci2021,masterson2022}, which was interpreted as a reflection of relativistic winds during a super-Eddington period. Here, we explore the presence of emission lines in the soft X-rays during the period May 2022-August 2025.

After fitting the baseline model to the RGS+PN individual observations in 2023, 2024, and 2025, we found positive residuals in the energy ranges $\sim 0.5\kev$, $\sim 1\kev$ (in both PN and RGS spectra) and $\sim 7 \kev$ (in PN spectra). However, for 2022 observations, the soft X-ray residuals were found at $\sim 0.47 \kev $ 1 $\sim 1 \kev$ as shown in Figure \ref{fig:simult_fit_jul26}. The individual RGS1 and RGS2 spectra and their best fits are presented in Figures \ref{fig:rgs_set1}, \ref{fig:rgs_set2}, \ref{fig:rgs_set3}, \ref{fig:rgs_set4}, \ref{fig:rgs_set5}, \ref{fig:rgs_set6}, \ref{fig:rgs_set7}, \ref{fig:rgs_set8} and \ref{fig:rgs_set9} in the Appendix. Also, the individual PN spectra and their best fit are presented in Figures \ref{fig:pn_set1}, \ref{fig:pn_set2}, \ref{fig:pn_set3}, \ref{fig:pn_set4}, \ref{fig:pn_set5}, \ref{fig:pn_set6}, \ref{fig:pn_set7}, \ref{fig:pn_set8}, and \ref{fig:pn_set9} in the Appendix.

We used Gaussian profiles to model these emission features in the individual observation fits (see Tables \ref{Table:indiv_fit_2022_2023}, \ref{Table:indiv_fit_2024} and \ref{Table:indiv_fit_2025}). Note that we have quoted the statistical improvement in fit statistics ($\Delta C$) on the addition of these Gaussian model components in each case to demonstrate how significant these emission lines are. The improvement in the addition of the Gaussian $\sim 0.56\kev$ was significant in most of the observations with $\Delta C$> 20. Whereas the $\sim 1 \kev$ Gaussian improves the fit statistics significantly in 2024 and 2025 observations only. The $0.56\kev$ emission feature, presumably the signature of OVII resonance, inter-combination, and forbidden transitions, \citep{laha2011Mrk704} appears to be the line width ranging from $\sigma\sim (5-60) \ev$ as shown in Tables \ref{Table:indiv_fit_2022_2023}, \ref{Table:indiv_fit_2024} and \ref{Table:indiv_fit_2025}). The $\sim 1\kev$ feature, on the other hand, is broad $\sim (80-120) \ev$. The parameter values quoted in Tables \ref{Table:indiv_fit_2022_2023}, \ref{Table:indiv_fit_2024} and \ref{Table:indiv_fit_2025} are the best fit values. The uncertainty on line E and line normalization, however, was estimated by freezing the line width $\sigma$ at its best fit values in all the cases because the lines were weak and all the parameters could not be constrained simultaneously.

\subsubsection{Soft X-ray emission lines in stacked spectra}\label{subsubsec:emission_line_stacked}

\begin{figure*}
    \centering
    \includegraphics[width=0.49\linewidth]{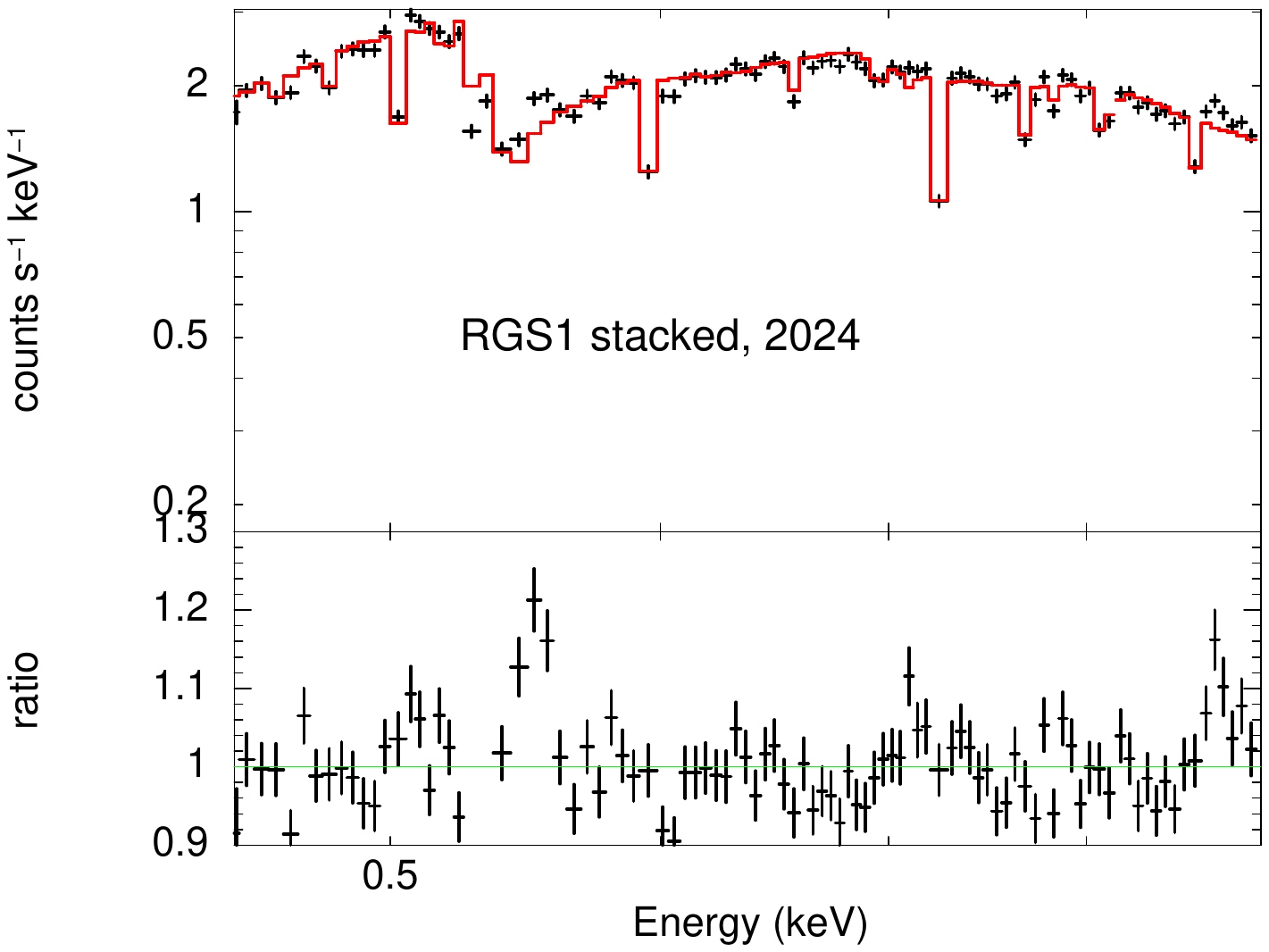}
    \includegraphics[width=0.49\linewidth]{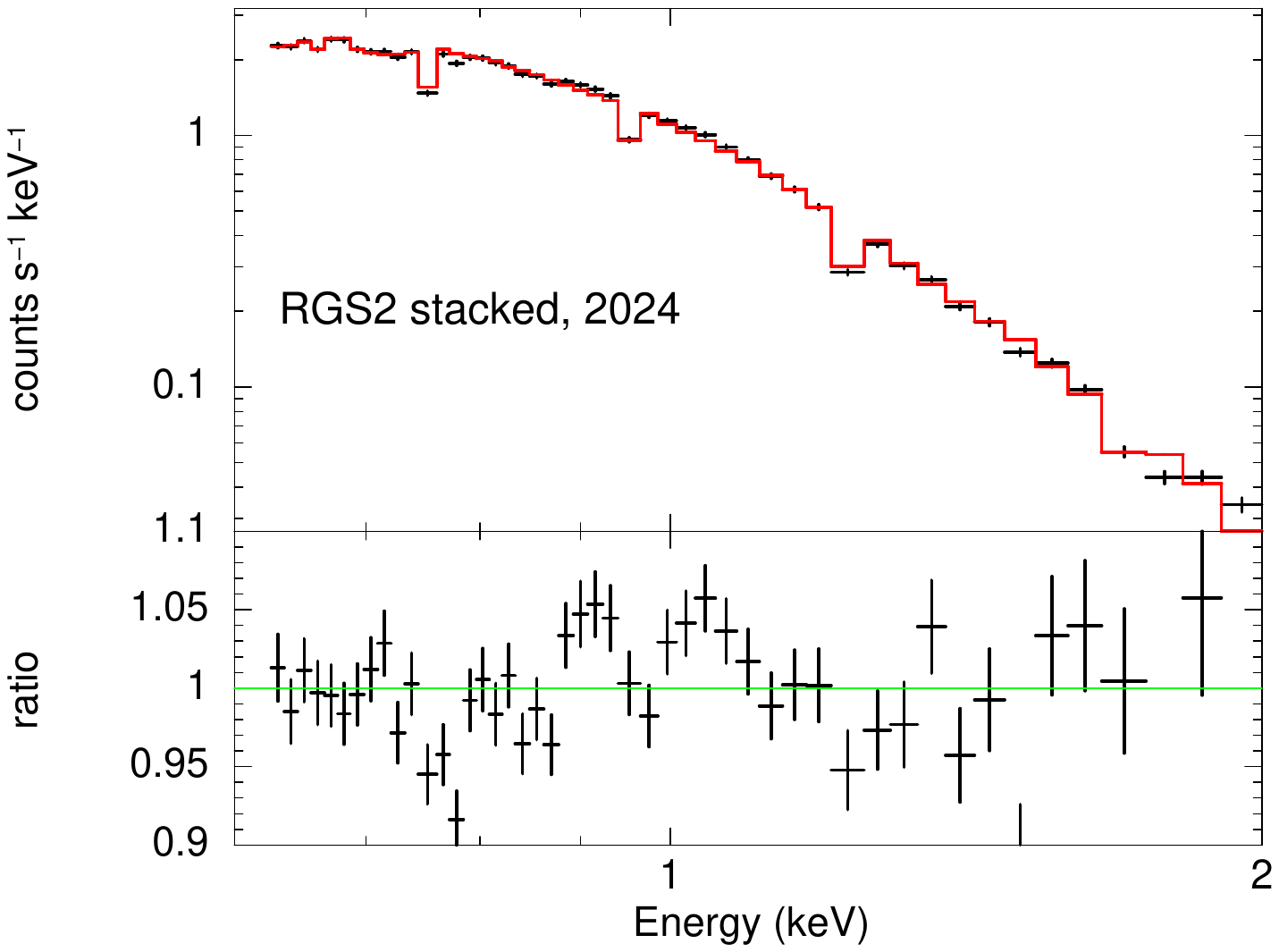}
    \caption{Figure showing the RGS1 and RGS2 data and fit for one observation taken of Stack, 2024, to the emission feature at $\sim 0.56\kev$ and at $\sim 1\kev$ . The continuum parameters (black body temperature and power law index) were obtained with PN spectrum, and fixed those values while fitting RGS only. The spectra were re-binned for visualization.  }
    \label{fig:rgs1_rgs2_only}
\end{figure*}

\begin{figure*}
    \centering
    \vbox{
    \hbox{
    \includegraphics[width=0.5\linewidth]{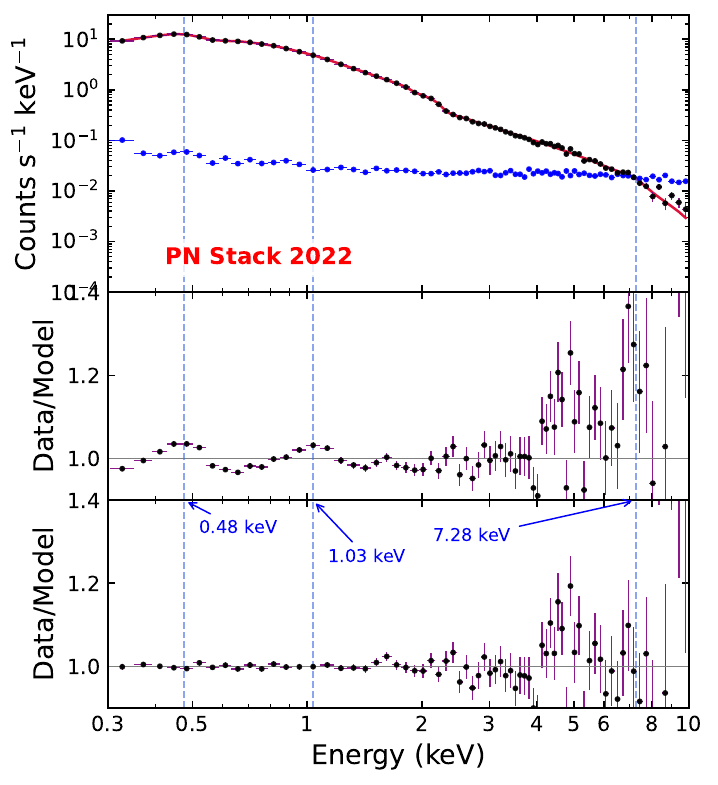}
    \includegraphics[width=0.5\linewidth]{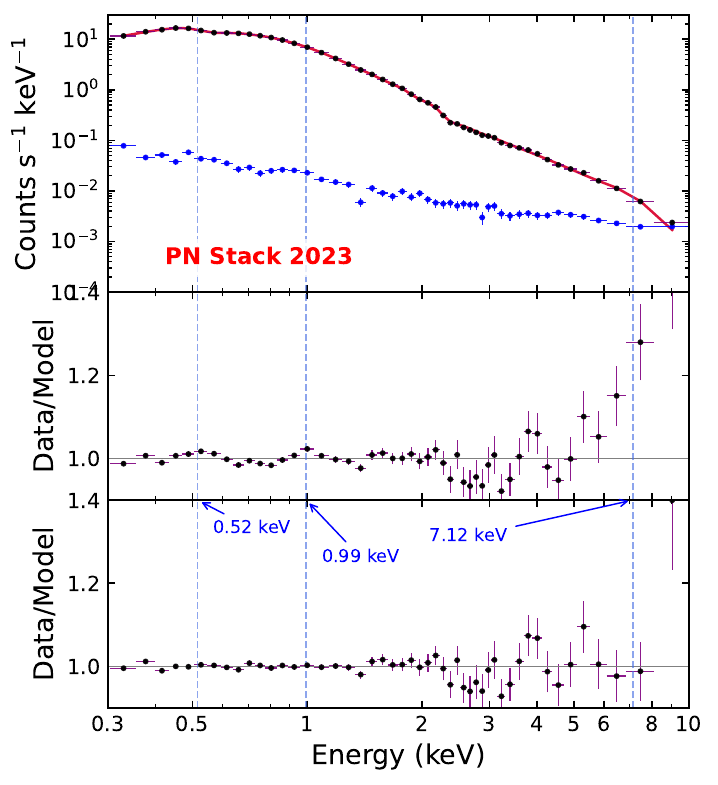}
    }

    \centering
    \hbox{
    \includegraphics[width=0.5\linewidth]{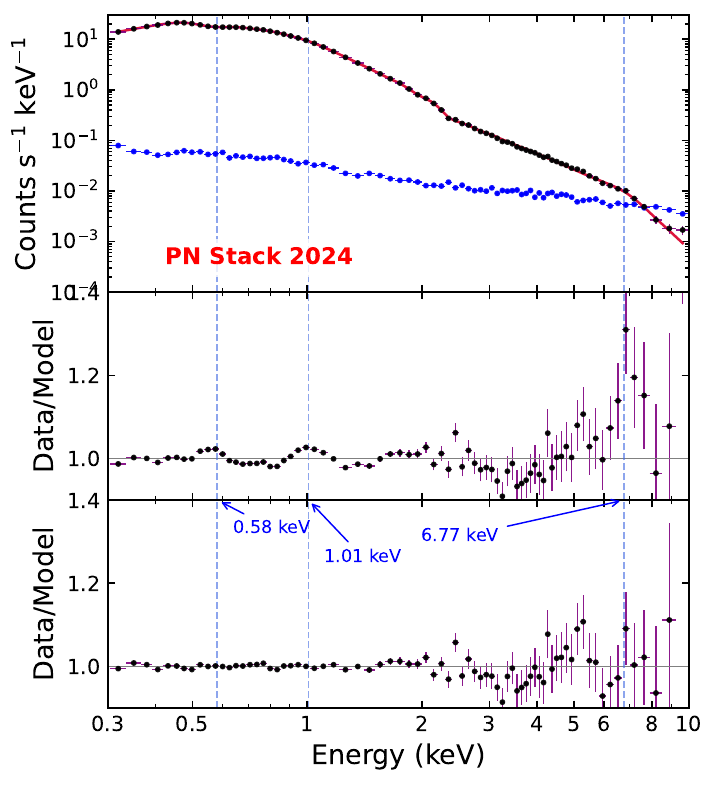}
    \includegraphics[width=0.5\linewidth]{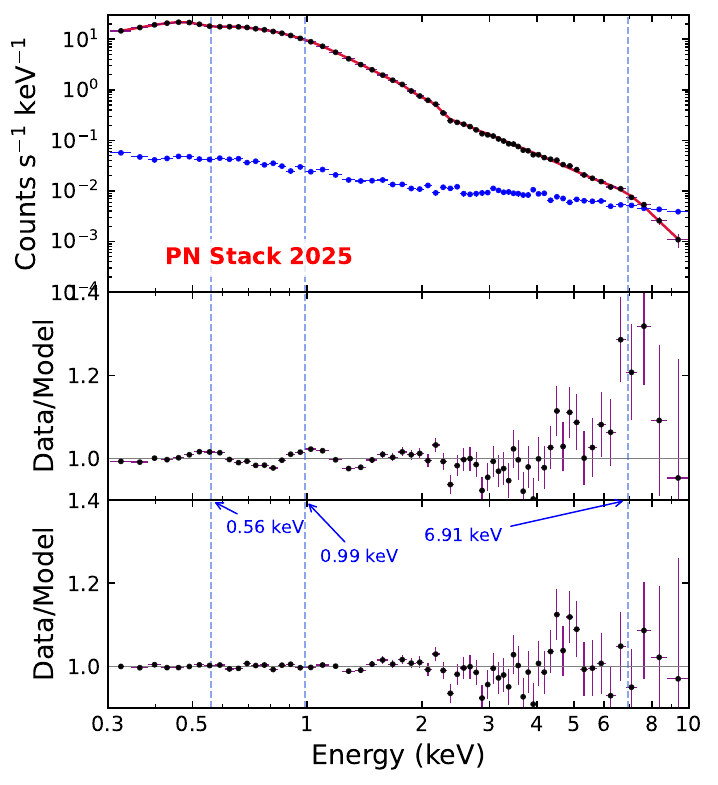}
    }
    }
    \caption{ The best-fit model, with residuals before and after fitting the emission lines for the stacked EPIC-pn spectra in the years 2022, 2023, 2024 and 2025 (labeled as such). The top panel for every figure shows the best fit model and the PN spectrum. The blue data points on the top panel of every figure denote the X-ray background spectra. The middle panel shows the fit residuals when we removed the three emission lines at $\sim 0.56\kev$, $0.99\kev$ and $7\kev$ from the fit (denoted by three vertical dashed lines),to demonstrate their presence. The bottom panel in every figure shows the residuals after we incorporated the three emission lines in our fit (hence the best fit). The FeK emission line was nearly absent in 2022 but is present very strongly in 2023, and the intensity dropped in 2024 and stayed the same in 2025. We note that there is a significant positive residual at $\sim (4-5)\kev$ in the bottom-most best-fit panels of 2023, 2024, and 2025. We did not model that line in this work. See Table \ref{Table:stack_fit} and Sections \ref{subsubsec:emission_line_stacked} and \ref{subsection:iron_line_emission_analysis} for details.}
    \label{Figure:pn_emis_stacked_data_by_model_ratio}
\end{figure*}

We fitted PN, RGS1, and RGS2 stacked spectra simultaneously for each of the years 2022, 2023, 2024, and 2025.  We used the same baseline model for the continuum as we used in Sec \ref{subsubsection:emission_line_indiv_obs} for individual observations, and the emission lines were modeled using Gaussian profiles. The stacked spectra give us a better view of the emission lines at $\sim 0.56 \kev$ and $\sim 1 \kev$ with an increased SNR ratio, as shown in Figure \ref{fig:rgs1_rgs2_only}. We have used stacked RGS1 and RGS2 for 2024 spectra to visualize the emission features. The emission lines are also clearly visible in PN-only spectra, as shown in Figure \ref{Figure:pn_emis_stacked_data_by_model_ratio}. The fit results are presented in Table \ref{Table:stack_fit}. The Gaussian lines at $\sim 0.56 \kev$ are narrow with a line width of $\sim 5 \ev$ in 2023, 2024, and 2025, whereas the 2022 stack has an emission line at $\sim 0.47 \kev$ with a line width of $\sim 60 \ev$. In all the cases, the changes in the fit statistics were significant. $1 \kev$ line, on the other hand, is significant in 2024 and 2025 with $\Delta C \ge 122$, whereas it is less significant in 2022 and 2023 with $\Delta C \sim 12-16$. The line width of $1 \kev$ the line was significantly broader, ranging from $\sim 90-120 \kev$. We have also presented some unfolded spectra and models and the line contours to show how well the lines at $\sim 0.45-0.56 \kev$ and $\sim 1 \kev$ are constrained in Figures \ref{fig:pn_rgs_2022_stacked}, \ref{fig:pn_rgs_2023_stacked}, \ref{fig:pn_rgs_2024_stacked} and \ref{fig:pn_rgs_2025_stacked} for 2022, 2023, 2024, and 2025, respectively, in the Appendix. The line at $\sim 0.56 \kev$ and $\sim 1 \kev$ are constrained with $99\%$ confidence for 2023, 2024 and 2025. For stacked spectra 2022, lines appear at $\sim 0.47 \kev$ and $\sim 1 \kev$. The line at $\sim 0.47 \kev$ is constrained within $99\%$ confidence, whereas $\sim 1 \kev$ line is constrained within the $90\%$ confidence as shown in Figure \ref{fig:pn_rgs_2022_stacked}.


\begin{table*}
\centering
\setlength{\tabcolsep}{2pt}
\caption{The best-fit parameters, obtained from simultaneous fitting of the \xmm{} EPIC-pn and RGS spectra for all the individual observations, including Gaussian emission lines at $\sim 0.45-0.6 \kev$, $\sim 1.0 \kev$ and $\sim 6.0-7.2 \kev$. The date is in the DD/MM/YY format.} 
\label{Table:indiv_fit_2022_2023}
\begin{tabular}{llccccccc}
\hline\hline
Model & Parameter & \textbf{20/05/11} & \textbf{26/07/22} & \textbf{28/07/22} & \textbf{30/07/22} & \textbf{01/08/22} & \textbf{21/02/23} & \textbf{07/08/23} \\ \hline
ztbabs & $N_{\mathrm{H}}$ ($10^{20}$ cm$^{-2}$) & $3.21^{+0.79}_{-0.78}$ & $1.97^{+0.74}_{-0.62}$ & $2.62^{+0.76}_{-0.83}$ & $2.18^{+0.76}_{-0.72}$ & $1.13^{+0.80}_{-0.73}$ & $4.27^{+0.66}_{-0.69}$ & $5.16^{+0.56}_{-0.58}$ \\ \hline
bbody & $kT$ (keV) & $0.164^{+0.010}_{-0.010}$ & $0.149^{+0.010}_{-0.013}$ & $0.137^{+0.013}_{-(\mathrm{unc})}$ & $0.148^{+0.011}_{-0.012}$ & $0.146^{+0.008}_{-0.009}$ & $0.150^{+0.005}_{-0.005}$ & $0.150^{+0.003}_{-0.003}$ \\
& norm ($10^{-4}$) & $0.42^{+0.05}_{-0.05}$ & $0.62^{+0.08}_{-0.08}$ & $0.73^{+0.14}_{-0.13}$ & $0.63^{+0.09}_{-0.09}$ & $0.76^{+0.08}_{-0.07}$ & $2.64^{+0.25}_{-0.28}$ & $4.76^{+0.17}_{-0.17}$ \\ \hline
powerlaw & $\Gamma$ & $2.39^{+0.04}_{-0.04}$ & $2.63^{+0.06}_{-0.05}$ & $2.70^{+0.06}_{-0.05}$ & $2.70^{+0.06}_{-0.05}$ & $2.52^{+0.06}_{-0.06}$ & $3.10^{+0.06}_{-0.06}$ & $3.17^{+0.05}_{-0.05}$ \\
& norm ($10^{-2}$) & $0.27^{+0.01}_{-0.01}$ & $0.46^{+0.03}_{-0.02}$ & $0.51^{+0.02}_{-0.03}$ & $0.52^{+0.03}_{-0.03}$ & $0.44^{+0.03}_{-0.02}$ & $1.09^{+0.06}_{-0.06}$ & $1.33^{+0.06}_{-0.06}$ \\ \hline
zgauss & $E$ (keV) & $0.49^{+0.02}_{-0.02}$ & $0.47^{+0.01}_{-0.01}$ & $0.44^{+0.02}_{-0.03}$ & $0.48^{+0.02}_{-0.02}$ & $0.46^{+0.01}_{-0.01}$ & $0.54^{+0.03}_{-0.03}$ & $0.55^{+0.02}_{-0.02}$ \\
& $\sigma$ (eV) & $25.8$ (f) & $60.0$ (f) & $59.9$ (f) & $60.0$ (f) & $45.9$ (f) & $56.7$ (f) & $35.3$ (f) \\
& norm ($10^{-4}$) & $1.30^{+0.82}_{-0.78}$ & $13.38^{+3.05}_{-2.98}$ & $11.35^{+6.35}_{-4.76}$ & $8.66^{+3.31}_{-3.11}$ & $9.63^{+2.38}_{-2.23}$ & $9.40^{+4.76}_{-4.44}$ & $8.52^{+3.03}_{-3.05}$ \\
& EW (eV) & $5.78$ & $26.70$ & $17.25$ & $15.68$ & $20.36$ & $8.37$ & $5.24$ \\ \hline
$\Delta C/\Delta$ dof & & 10/3 & 97/3 & 44/3 & 47/3 & 78/3 & 14/3 & 30/3 \\ \hline
zgauss & $E$ (keV) & $0.89^{+0.07}_{-0.08}$ & $0.98^{+0.14}_{-(\mathrm{unc})}$ & $1.00^{+0.08}_{-0.08}$ & $1.02^{+0.05}_{-0.05}$ & $1.07^{+0.06}_{-0.07}$ & $0.85^{+0.06}_{-(\mathrm{unc})}$ & $0.99^{+0.04}_{-0.05}$ \\
& $\sigma$ (eV) & $120.0$ (f) & $120.0$ (f) & $120.0$ (f) & $71.5$ (f) & $60.0$ (f) & $113.4$ (f) & $91.5$ (f) \\
& norm ($10^{-4}$) & $0.82^{+0.48}_{-0.43}$ & $0.48^{+0.60}_{-(\mathrm{unc})}$ & $0.98^{+0.55}_{-0.65}$ & $0.62^{+0.39}_{-0.36}$ & $0.31^{+0.25}_{-0.25}$ & $3.26^{+2.23}_{-1.82}$ & $2.11^{+0.96}_{-0.92}$ \\
& EW (eV) & $15.46$ & $7.50$ & $15.08$ & $9.64$ & $6.00$ & $10.76$ & $8.33$ \\ \hline
$\Delta C/\Delta$ dof & & 10/3 & 2/3 & 0/3 & 7/3 & 3/3 & 8/3 & 13/3 \\ \hline
zgauss & $E$ (keV) & $7.50$ (f) & $6.00$ (f) & $6.00$ (f) & $7.50$ (f) & $7.50^{+(\mathrm{unc})}_{-0.40}$ & $7.15^{+(\mathrm{unc})}_{-0.98}$ & $7.50^{+(\mathrm{unc})}_{-0.82}$ \\
& $\sigma$ (eV) & $1036.2$ (f) & $1084.4$ (f) & $1032.7$ (f) & $1100.0$ (f) & $1099.7$ (f) & $1052.8$ (f) & $1100.0$ (f) \\
& norm ($10^{-4}$) & $0.00^{+0.04}_{-(\mathrm{unc})}$ & $0.00^{+0.10}_{-(\mathrm{unc})}$ & $0.02^{+0.11}_{-(\mathrm{unc})}$ & $0.05^{+0.09}_{-(\mathrm{unc})}$ & $0.38^{+0.19}_{-0.21}$ & $0.18^{+0.12}_{-0.12}$ & $0.19^{+0.12}_{-0.12}$ \\
& EW (eV) & $0.00$ & $0.01$ & $45.16$ & $206.47$ & $1294.65$ & $660.66$ & $808.95$ \\ \hline
$\Delta C/\Delta$ dof & & 0/3 & 0/3 & 6/3 & 1/3 & 9/3 & 6/3 & 7/3 \\ \hline
$C$/dof & & 3956/4243 & 4495/4576 & 3905/4252 & 4037/4312 & 4197/4560 & 4725/4675 & 4894/4720 \\ \hline
\end{tabular}

\vspace{0.5cm}
\textbf{Note}: For the individual spectral fits, we constructed the final model incrementally, starting from the continuum. We first fitted each spectrum with the baseline model, {\tt const$\times$tbabs$\times$ztbabs(bbody + pow)}. We then introduced additional Gaussian components one at a time and refitted. A Gaussian emission line at $\sim 0.5\kev$ was added first and fitted with its width free. After obtaining its best-fit width, we froze the line width for $0.5\kev$ and and added a second Gaussian line at $\sim 1\kev$. This second line was fitted in the same way: the width was initially allowed to vary, then frozen at its best‐fit value. Finally, we added the third Gaussian component in the $\sim 6$–$7.5\kev$ range and fitted for the remaining free parameters. The width of the line was needed to freeze because they were weak or moved to unphysical values while finding the error.  In all cases, the line widths were frozen at their respective best‐fit values, not arbitrary choices. We then computed the uncertainties on the free parameters. We have put (f) next to parameter values to indicate that they were frozen when the error was estimated; it does not mean frozen when finding the best fit. The cross-calibration constants were within $\sim (5-10)\%$.

\tablenotetext{(unc)}{\ \ \ \ \ \ The upper or lower limit for that parameters are not constrained within the bound of the model parameter we set up.}
\end{table*}

\begin{table*}
\centering
\setlength{\tabcolsep}{2pt}
\caption{Table \ref{Table:indiv_fit_2022_2023} continued..}
\label{Table:indiv_fit_2024}

\begin{tabular}{llcccccc}
\hline\hline
Model & Parameter & \textbf{04/03/24} & \textbf{12/03/24} & \textbf{19/07/24} & \textbf{27/07/24} & \textbf{13/10/24} & \textbf{21/10/24} \\ \hline
ztbabs & $N_{\mathrm{H}}$ ($10^{20}$ cm$^{-2}$) & $4.91^{+0.61}_{-0.64}$ & $5.32^{+0.47}_{-0.48}$ & $5.63^{+0.50}_{-0.49}$ & $5.72^{+0.48}_{-0.48}$ & $4.40^{+0.75}_{-0.76}$ & $5.22^{+0.51}_{-0.51}$ \\ \hline
bbody & $kT$ (keV) & $0.154^{+0.003}_{-0.003}$ & $0.153^{+0.002}_{-0.002}$ & $0.148^{+0.003}_{-0.003}$ & $0.150^{+0.002}_{-0.002}$ & $0.155^{+0.003}_{-0.003}$ & $0.150^{+0.002}_{-0.002}$ \\
& norm ($10^{-4}$) & $5.61^{+0.20}_{-0.19}$ & $5.51^{+0.17}_{-0.16}$ & $5.94^{+0.23}_{-0.22}$ & $5.86^{+0.19}_{-0.18}$ & $5.56^{+0.26}_{-0.26}$ & $5.94^{+0.21}_{-0.19}$ \\ \hline
powerlaw & $\Gamma$ & $3.17^{+0.05}_{-0.06}$ & $3.22^{+0.04}_{-0.04}$ & $3.22^{+0.04}_{-0.04}$ & $3.24^{+0.04}_{-0.04}$ & $3.13^{+0.07}_{-0.07}$ & $3.16^{+0.04}_{-0.04}$ \\
& norm ($10^{-2}$) & $1.39^{+0.07}_{-0.07}$ & $1.41^{+0.05}_{-0.05}$ & $1.51^{+0.06}_{-0.06}$ & $1.49^{+0.06}_{-0.05}$ & $1.34^{+0.09}_{-0.09}$ & $1.49^{+0.06}_{-0.06}$ \\ \hline
zgauss & $E$ (keV) & $0.56^{+0.01}_{-0.02}$ & $0.56^{+0.00}_{-0.00}$ & $0.56^{+0.00}_{-0.00}$ & $0.56^{+0.00}_{-0.00}$ & $0.56^{+0.02}_{-0.02}$ & $0.57^{+0.00}_{-0.00}$ \\
& $\sigma$ (eV) & $32.8$ (f) & $5.6$ (f) & $4.8$ (f) & $3.1$ (f) & $55.7$ (f) & $3.6$ (f) \\
& norm ($10^{-4}$) & $9.73^{+3.24}_{-3.22}$ & $9.24^{+2.19}_{-2.02}$ & $7.06^{+2.27}_{-2.29}$ & $5.49^{+1.88}_{-1.73}$ & $16.74^{+5.41}_{-5.29}$ & $7.22^{+1.99}_{-1.74}$ \\
& EW (eV) & $5.46$ & $4.87$ & $3.43$ & $2.64$ & $10.10$ & $4.02$ \\ \hline
$\Delta C/\Delta$ dof & & 25/3 & 60/3 & 29/3 & 24/3 & 27/3 & 50/3 \\ \hline
zgauss & $E$ (keV) & $0.96^{+0.04}_{-0.03}$ & $0.98^{+0.03}_{-0.03}$ & $1.00^{+0.03}_{-0.03}$ & $1.00^{+0.02}_{-0.03}$ & $0.97^{+0.03}_{-0.03}$ & $1.01^{+0.03}_{-0.03}$ \\
& $\sigma$ (eV) & $80.5$ (f) & $80.7$ (f) & $114.0$ (f) & $75.5$ (f) & $70.7$ (f) & $84.5$ (f) \\
& norm ($10^{-4}$) & $2.85^{+1.09}_{-1.05}$ & $2.92^{+0.89}_{-0.88}$ & $5.31^{+1.40}_{-1.34}$ & $2.95^{+0.86}_{-0.84}$ & $3.74^{+1.29}_{-1.24}$ & $3.21^{+0.98}_{-0.97}$ \\
& EW (eV) & $8.95$ & $9.75$ & $18.92$ & $10.20$ & $12.05$ & $11.63$ \\ \hline
$\Delta C/\Delta$ dof & & 20/3 & 32/3 & 46/3 & 30/3 & 26/3 & 32/3 \\ \hline
zgauss & $E$ (keV) & $6.00$ (f) & $6.95^{+0.19}_{-0.18}$ & $7.02^{+0.31}_{-0.28}$ & $7.20$ (f) & $6.40$ (f) & $6.69^{+0.31}_{-0.22}$ \\
& $\sigma$ (eV) & $438.1$ (f) & $200.8$ (f) & $200.0$ (f) & $1100.0$ (f) & $200.1$ (f) & $200.0$ (f) \\
& norm ($10^{-4}$) & $0.08^{+0.11}_{-(\mathrm{unc})}$ & $0.10^{+0.06}_{-0.06}$ & $0.05^{+0.05}_{-0.05}$ & $0.14^{+0.11}_{-0.11}$ & $0.02^{+0.10}_{-(\mathrm{unc})}$ & $0.09^{+0.07}_{-0.07}$ \\
& EW (eV) & $159.00$ & $355.89$ & $173.01$ & $533.63$ & $36.23$ & $228.72$ \\ \hline
$\Delta C/\Delta$ dof & & 1/3 & 9/3 & 3/3 & 5/3 & 0/3 & 5/3 \\ \hline
$C$/dof & & 4993/4691 & 4894/4753 & 4863/4660 & 4865/4674 & 4758/4581 & 5006/4736 \\ \hline
\end{tabular}

\end{table*}

\begin{table*}
\centering
\setlength{\tabcolsep}{2pt}
\caption{Table \ref{Table:indiv_fit_2022_2023} continued..}
\label{Table:indiv_fit_2025}

\begin{tabular}{llccccc}
\hline\hline
Model & Parameter & \textbf{19/01/25} & \textbf{25/01/25} & \textbf{30/04/25} & \textbf{02/05/25} & \textbf{05/05/25} \\ \hline
ztbabs & $N_{\mathrm{H}}$ ($10^{20}$ cm$^{-2}$) & $5.02^{+0.62}_{-0.63}$ & $5.23^{+0.60}_{-0.63}$ & $5.24^{+0.54}_{-0.51}$ & $4.13^{+0.62}_{-0.64}$ & $4.57^{+0.73}_{-0.78}$ \\ \hline
bbody & $kT$ (keV) & $0.154^{+0.003}_{-0.003}$ & $0.153^{+0.003}_{-0.003}$ & $0.151^{+0.003}_{-0.003}$ & $0.155^{+0.003}_{-0.003}$ & $0.156^{+0.003}_{-0.003}$ \\
& norm ($10^{-4}$) & $5.68^{+0.24}_{-0.24}$ & $5.59^{+0.23}_{-0.24}$ & $5.77^{+0.21}_{-0.21}$ & $5.76^{+0.24}_{-0.26}$ & $5.95^{+0.19}_{-0.18}$ \\ \hline
powerlaw & $\Gamma$ & $3.19^{+0.05}_{-0.05}$ & $3.21^{+0.05}_{-0.06}$ & $3.24^{+0.04}_{-0.04}$ & $3.13^{+0.05}_{-0.06}$ & $3.16^{+0.06}_{-0.07}$ \\
& norm ($10^{-2}$) & $1.45^{+0.07}_{-0.07}$ & $1.38^{+0.07}_{-0.07}$ & $1.43^{+0.06}_{-0.06}$ & $1.29^{+0.07}_{-0.07}$ & $1.40^{+0.09}_{-0.09}$ \\ \hline
zgauss & $E$ (keV) & $0.56^{+0.01}_{-0.01}$ & $0.55^{+0.02}_{-0.02}$ & $0.56^{+0.00}_{-0.00}$ & $0.56^{+0.02}_{-0.02}$ & $0.54^{+0.02}_{-0.02}$ \\
& $\sigma$ (eV) & $33.2$ (f) & $43.6$ (f) & $5.8$ (f) & $46.3$ (f) & $26.5$ (f) \\
& norm ($10^{-4}$) & $13.82^{+3.67}_{-3.60}$ & $10.37^{+4.05}_{-3.96}$ & $9.44^{+2.55}_{-2.39}$ & $12.71^{+4.09}_{-4.00}$ & $9.23^{+3.38}_{-3.43}$ \\
& EW (eV) & $7.48$ & $5.83$ & $4.72$ & $7.66$ & $5.05$ \\ \hline
$\Delta C/\Delta$ dof & & 40/3 & 19/3 & 44/3 & 26/3 & 23/3 \\ \hline
zgauss & $E$ (keV) & $0.96^{+0.04}_{-0.04}$ & $0.94^{+0.04}_{-0.04}$ & $0.97^{+0.03}_{-0.04}$ & $0.95^{+0.05}_{-0.05}$ & $1.02^{+0.06}_{-0.08}$ \\
& $\sigma$ (eV) & $120.0$ (f) & $119.8$ (f) & $120.0$ (f) & $119.9$ (f) & $60.0$ (f) \\
& norm ($10^{-4}$) & $5.31^{+1.80}_{-1.69}$ & $4.56^{+1.79}_{-1.65}$ & $4.97^{+1.56}_{-1.50}$ & $4.19^{+1.91}_{-1.73}$ & $1.09^{+0.86}_{-0.83}$ \\
& EW (eV) & $16.33$ & $13.64$ & $15.79$ & $12.80$ & $4.00$ \\ \hline
$\Delta C/\Delta$ dof & & 29/3 & 22/3 & 32/3 & 16/3 & 4/3 \\ \hline
zgauss & $E$ (keV) & $7.50^{+(\mathrm{unc})}_{-0.67}$ & $6.90^{+0.47}_{-0.87}$ & $7.50$ (f) & $7.50^{+(\mathrm{unc})}_{-0.20}$ & $6.27^{+1.02}_{-(\mathrm{unc})}$ \\
& $\sigma$ (eV) & $964.6$ (f) & $300.8$ (f) & $526.9$ (f) & $375.3$ (f) & $573.9$ (f) \\
& norm ($10^{-4}$) & $0.28^{+0.18}_{-0.18}$ & $0.09^{+0.08}_{-0.08}$ & $0.03^{+0.10}_{-(\mathrm{unc})}$ & $0.12^{+0.09}_{-0.09}$ & $0.12^{+0.12}_{-0.06}$ \\
& EW (eV) & $1101.85$ & $283.76$ & $148.37$ & $465.56$ & $256.93$ \\ \hline
$\Delta C/\Delta$ dof & & 7/3 & 3/3 & 0/3 & 5/3 & 3/3 \\ \hline
$C$/dof & & 4941/4675 & 5139/4693 & 4821/4674 & 4888/4643 & 4951/4639 \\ \hline
\end{tabular}

\end{table*}

\begin{table*}
\centering
\setlength{\tabcolsep}{4pt}
\caption{Best fit parameters obtained by simultaneous fitting of stacked RGS1, RGS2, and PN spectra for 2022–2025 observations including Gaussian emission lines.}
\label{Table:stack_fit}
\begin{tabular}{llcccc}
\hline\hline
Model & Parameter & \textbf{2022} & \textbf{2023} & \textbf{2024} & \textbf{2025} \\ \hline
ztbabs & $N_{\mathrm{H}}$ ($10^{20}$ cm$^{-2}$) & $1.91^{+0.45}_{-0.48}$ & $5.23^{+0.36}_{-0.36}$ & $5.54^{+0.23}_{-0.23}$ & $5.17^{+0.25}_{-0.25}$ \\ \hline
bbody & $kT$ (keV) & $0.145^{+0.006}_{-0.006}$ & $0.146^{+0.002}_{-0.002}$ & $0.150^{+0.001}_{-0.001}$ & $0.152^{+0.001}_{-0.001}$ \\
& norm ($10^{-4}$) & $0.68^{+0.05}_{-0.05}$ & $3.83^{+0.13}_{-0.13}$ & $5.73^{+0.09}_{-0.09}$ & $5.89^{+0.10}_{-0.10}$ \\ \hline
powerlaw & $\Gamma$ & $2.63^{+0.03}_{-0.03}$ & $3.17^{+0.03}_{-0.03}$ & $3.22^{+0.02}_{-0.02}$ & $3.21^{+0.02}_{-0.02}$ \\
& norm ($10^{-2}$) & $0.48^{+0.02}_{-0.01}$ & $1.25^{+0.03}_{-0.03}$ & $1.48^{+0.03}_{-0.03}$ & $1.42^{+0.03}_{-0.03}$ \\ \hline
zgauss & $E$ (keV) & $0.46^{+0.02}_{-0.01}$ & $0.56^{+0.00}_{-0.00}$ & $0.56^{+0.00}_{-0.00}$ & $0.56^{+0.00}_{-0.00}$ \\
& $\sigma$ (eV) & $60.0^{+(\mathrm{unc})}_{-16.3}$ & $4.6^{+1.9}_{-(\mathrm{unc})}$ & $4.9^{+1.0}_{-0.9}$ & $6.1^{+1.1}_{-1.0}$ \\
& norm ($10^{-4}$) & $11.53^{+1.90}_{-4.11}$ & $5.49^{+1.41}_{-1.49}$ & $7.95^{+1.12}_{-1.13}$ & $8.31^{+1.19}_{-1.16}$ \\
& EW (eV) & $21.43$ & $3.50$ & $3.89$ & $4.18$ \\ \hline
$\Delta C/\Delta$ dof & & 258/3 & 40/3 & 177/3 & 122/3 \\ \hline
zgauss & $E$ (keV) & $1.00^{+0.06}_{-0.05}$ & $0.95^{+0.05}_{-0.04}$ & $1.00^{+0.01}_{-0.02}$ & $0.99^{+0.02}_{-0.02}$ \\
& $\sigma$ (eV) & $114.3$ (f) & $115.8^{+(\mathrm{unc})}_{-40.1}$ & $88.9^{+15.5}_{-13.5}$ & $106.2^{+(\mathrm{unc})}_{-18.0}$ \\
& norm ($10^{-4}$) & $0.64^{+0.36}_{-0.40}$ & $2.37^{+0.97}_{-1.12}$ & $3.22^{+0.72}_{-0.60}$ & $3.48^{+0.98}_{-0.83}$ \\
& EW (eV) & $10.05$ & $9.58$ & $11.30$ & $11.63$ \\ \hline
$\Delta C/\Delta$ dof & & 12/3 & 16/3 & 150/3 & 142/3 \\ \hline
zgauss & $E$ (keV) & $7.28^{+(\mathrm{unc})}_{-0.60}$ & $7.17^{+(\mathrm{unc})}_{-0.57}$ & $7.00^{+(\mathrm{unc})}_{-0.68}$ & $7.22^{+(\mathrm{unc})}_{-0.77}$ \\
& $\sigma$ (eV) & $400.0$ (f) & $800.0$ (f) & $800.0$ (f) & $800.0$ (f) \\
& norm ($10^{-4}$) & $0.07^{+0.06}_{-0.06}$ & $0.16^{+0.07}_{-0.07}$ & $0.08^{+0.05}_{-0.05}$ & $0.09^{+0.06}_{-0.06}$ \\
& EW (eV) & $292.88$ & $604.94$ & $261.10$ & $314.53$ \\ \hline
$\Delta C/\Delta$ dof & & 5/2 & 36/2 & 18/2 & 22/2 \\ \hline
$C$/dof & & 5621/5405 & 5111/4917 & 6152/5264 & 6255/5159 \\ \hline
\end{tabular}

\vspace{0.5cm}
\textbf{Note}: Our final spectral model was constructed in a stepwise manner, starting from the simplest continuum, as we did for the individual fit. We first fitted each spectrum with the baseline model, {\tt const$\times$tbabs$\times$ztbabs(bbody + pow)}. We then added Gaussian components one by one and refitted the data at each stage. A Gaussian emission line at $\sim 0.5\kev$ was added first, followed by a second line at $\sim 1\kev$, and finally a third line in the $\sim 6$–$7.5\kev$ range. The width of the $\sim 6$–$7.5\kev$ line was kept fixed at $0.8 \kev$. Importantly, the width was frozen at the \emph{best-fit} value obtained from the initial free-width fit, not at an arbitrary value. All other free parameters were allowed to vary within the bounds we set up when we performed the fitting and computing uncertainties. For the stacked spectra, the widths of the $\sim 0.5\kev$ and $\sim 1\kev$ Gaussian components were kept free, whereas in the individual observations, these widths were fixed during error calculations. In most of the cases, the line energies and widths were constrained within physically motivated parameter ranges. We have put (f) next to parameter values to indicate that they were frozen when the error was estimated; it does not mean frozen when finding the best fit.

\tablenotetext{(unc)}{\ \ \ \ \ \ The upper or lower limit for that parameters are not constrained within the bound of the model parameter we set up.}
\end{table*}

\subsubsection{Fe K$\alpha$ emission line analysis}
\label{subsection:iron_line_emission_analysis}
After obtaining a best fit of simultaneous RGS+PN individual observation with the continuum and two emission lines at the $\sim 0.45-0.56 \kev$, we added a Gaussian line at $\sim 6-7 \kev$. The Fe K$\alpha$ emission line features at $6-7\kev$ were not detected significantly in any of the individual \xmm{} spectra, as shown in Tables \ref{Table:indiv_fit_2022_2023}, \ref{Table:indiv_fit_2024} and \ref{Table:indiv_fit_2025}. All the Gaussian parameters were free while performing the fitting. However, the line energy and line width were only allowed to vary within specific bounds (for the Fe K$\alpha$ emission line the following ranges were used: the line energy $6 \kev <E< 7.5\kev$, and the line width $0.2<\sigma<1.1\kev$).

We further probed the presence of the Fe K$\alpha$ emission line in the stacked PN spectra. We fitted the stacked spectra (from 2022, 2023, 2024, and 2025) with a continuum baseline model and added a Gaussian line. Initially we kept the line width frozen at $\sigma=0.2\kev$ to investigate if we at all detect a line with any significance. We found that the line energy of the Gaussian could be very well constrained, as demonstrated in the 1-D contours in Figure \ref{fig:ironline_detection}. This confirms that the Fe lines are robustly detected in the stacked spectra in 2023, 2024, and 2025, but only marginal in 2022.

The Fe emission lines being weak, the line energy and the $\sigma$ cannot be simultaneously constrained in any of the cases. So, we carried out the following test: We repeated the above fitting (addition of a Gaussian) technique with fixed line widths at $0.4 \kev$, $0.6 \kev$ and $0.8 \kev$. We find that the most improvement in statistics happen when $\sigma=0.8\kev$, as shown in Table \ref{Table:varying_FeK_widths} for all the stacked observations in 2023, 2024, and 2025. This demonstrates that a broader Gaussian line-width $\sigma=0.8 \kev$ is preferred in stacked observations of 2023, 2024, and 2025 (with an improvement in the fit statistics greater than 17). The improvement of the fit statistics in 2022 stacked was insignificant, with $\Delta C/ \Delta \mathrm{dof} \sim 5/2$ suggesting perhaps no line is required by the data.  The best-fit parameters are presented in the lowest panel of Table \ref{Table:stack_fit}. As a caveat, we note that the Fe K$\alpha$ emission line is very weak in this source, and it lies in the energy range ($\ge 6.5\kev$) where background dominates; hence, even though we detect the line clearly in stacked spectra, we refrain from commenting about any variability of this line. The details of FeK line modeling will be carried out in our future work with extensive \nustar{} observations.

\subsubsection{Statistical Significance of the FeK Emission Features}

To test the statistical significance of the broad FeK line complex, we employed a model (\texttt{tbabs$\times$ztbabs(zgauss + zgauss + zgauss + bbody + powerlaw)} as per XSPEC notation) to fit a PN stack spectrum for 2025 ($\sim 70$ ks). The two zgauss components represent the lines at $\sim0.56 \kev$ and $\sim 1.0 \kev$.   A broad emission line feature with a width of $\sigma \approx 1.5\kev$ is observed at $\sim 6.5\kev$.

\begin{figure}[h!]
    \centering
    \includegraphics[width=0.99\linewidth]{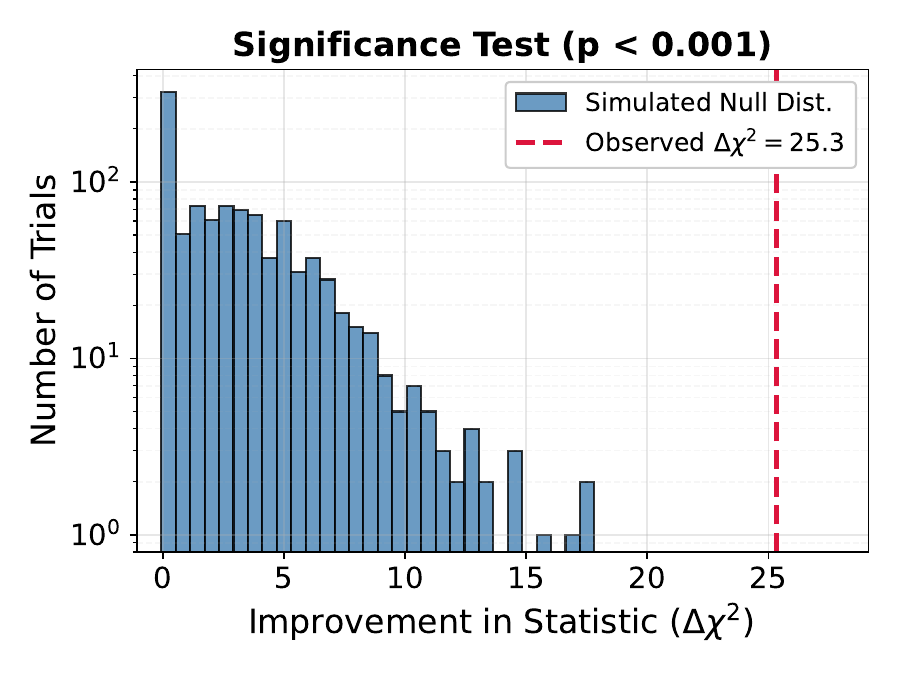}
    \caption{Distribution of the fit statistic improvement ($\Delta\chi^2$) from 1,000 Monte Carlo simulations evaluating the significance of the broad $6.46$\,keV emission feature using a 2025 stack PN spectrum. The histogram represents the expected improvements under the null hypothesis. The observed improvement of $\Delta\chi^2 \approx 25$ (indicated by the vertical dashed red line) falls well outside the simulated null distribution, establishing a line detection significance of $>99.9\%$.}
    \label{fig:fek_simulation}
\end{figure}

We performed Monte Carlo simulations to robustly evaluate the statistical significance of this feature at $\sim~6.5\kev$. We utilized the \texttt{simftest} routine in XSPEC to compare the test model (the full model including the $\sim~6.5$\,keV Gaussian) against a null hypothesis model (the identical model with the $6.5 \kev$ Gaussian component removed).

We generated 1,000 fake spectra based on the best-fitting parameters using the null hypothesis. Each simulated spectrum was then fitted with both the null model and the test model to empirically map the probability distribution of the fit statistic improvement, $\Delta\chi^2$, expected purely from statistical fluctuations. The probability distribution is given in Figure \ref{fig:fek_simulation}.

In our actual observed data, the addition of the $6.46$\,keV Gaussian component resulted in a substantial improvement to the fit, yielding a statistical reduction of $\Delta\chi^2 \approx 25$ (improving from $\chi^2 = 1070$ to 1045). Across all 1,000 simulated iterations, none of the mock spectra yielded a $\Delta\chi^2$ improvement equal to or greater than the observed value of $25$ (the maximum improvement was $\Delta\chi^2 \approx 18$). Hence, we establish the significance of the $6.46$\,keV emission feature to be $>99.9\%$ ($>3\sigma$), confirming that the broad iron line detection is statistically favored.

\section{Discussion}
\label{section:discussion}

We carried out a multi-wavelength study focusing on a detailed analysis of the X-ray spectra of 1ES~1927+654 using \xmm{} PN and RGS datasets. We detected newly emerging emission lines in $\sim 0.56\kev$, $\sim 1\kev$ and $(6-7) \kev$ energies. We detected a strong WA in 2022, and it's getting weaker in the subsequent years. A UFO, however, is not detected in any of these observations.  During this period of 2022-2025, the source is accreting at a rate $\lambdaedd\sim 0.3$ \citep{ghosh2023}. In the context of the simultaneous jet activity, soft X-ray rise by a factor of $\sim 10$ times, a QPO detection, and minimal changes in the UV or $(2-10)\kev$ fluxes \citep{meyer2025, laha2025, masterson2025}, we discuss the following scientific question in this section.

\subsection{The Origin of X-ray Emission Features}
\label{sec:reflection_discussion_combined}

Our spectral analysis reveals the emergence of several emission features in the 2023--2025 epochs that were notably absent in historical data \citep[May 2011;][]{Gallo2013}. These include soft X-ray lines at $\sim 0.56\kev$ and $\sim 1\kev$, as well as a broad ($\sigma \approx 800\ev$) Fe K emission feature at $\sim 6-7\kev$ detected in the stacked EPIC-pn spectra. The simultaneous appearance of these features alongside the onset of radio-jet formation suggests an evolution in the inner accretion flow. However, we can not say anything conclusively given the current statistics and the phenomenological method we used.

\subsubsection{Reflection Scenario}
The emergence of the $1\kev$ feature and the broad Fe K$\alpha$ complex ($6.4-7.2\kev$) suggests a shared origin in the reprocessing of radiation within the immediate vicinity of the central engine \citep{fabian2000, garcia2014}. An outflow was invoked to explain transient emissions during the 2018 super-Eddington phase \citep{masterson2022} in this source. However, given the current sub-Eddington state of the source ($\lambdaedd \sim 0.3$, \citealt{ghosh2023}), the broad $1\kev$ and FeK lines may arise from enhanced reflection column density, possibly driven by a reconfiguration of the inner disk or magnetic field realignment during jet formation \citep{Fabian2009, meyer2025}.

Given that the $(2-10)\kev$ illuminating continuum has remained relatively stable (varying by less than a factor of 2) during the 2022--2025 epoch, the appearance of these features likely points to a structural change in the reflecting medium itself. An increase in the inner disk accretion rate within a few tens of gravitational radii ($R_{\rm G}$) could provide the required column density to enhance the reflection spectrum. Furthermore, the Fe K$\alpha$ feature is centered at the higher-energy end of the iron complex (above $6.7\kev$), suggesting significant contributions from highly ionized species such as Fe~XXV ($\sim 6.7\kev$) or Fe~XXVI ($\sim 6.97\kev$). However, we note that the centroid of the line is dependent on the line width we choose during the fitting. In case of $\sim 0.8\kev$ line width, the centroids are $\sim (6.7-7.0)\kev$ in stacked spectra. Such high-ionization features are characteristic of reflection off the innermost accretion disk rather than a distant material such as a neutral torus.

While the qualitative alignment of these features points toward a disk-reflection origin, we note that our current interpretation is based on phenomenological Gaussian modeling. The current SNR and spectral resolution preclude robust constraints using self-consistent reflection models (e.g., \texttt{relxill}). Higher-sensitivity observations are required to break the degeneracies between the continuum shape and blurred reflection components to confirm if the evolution is driven by a change in geometry or the primary X-ray illuminator.

\subsubsection{BLR gas illumination}
Another possibility would be ionized gas. Even though this source is known as a "true type 2" AGN \citep{Boller2003,Gallo2013}, the broad line region (BLR) in \onees{} may always be present but remains weakly ionized or under-illuminated \citep{trak19, Li_broad_line_1es}. The order-of-magnitude increase in soft X-ray flux could sufficiently illuminate these pre-existing clouds in the BLR and emit these emission lines. While the narrowness of the $0.56\kev$ line ($\sim 5\ev$ in RGS) supports a photoionized gas origin \citep{Matteo2007, Piconcelli2008, Kinkhabwala2002}, the significant width of the $1\kev$ line ($\sim 100\ev$) and Fe K features ($\sigma \approx 0.8\kev$) suggests higher-velocity gas moving near the accretion disk. We will investigate these features and their underlying reflection morphology in detail in a following study utilizing long \nustar{} observations with higher SNR.

\subsection{Ionized Gas Absorption it's Implications}
\label{subsec:absorber_discussion_revised}

\onees{} was reported to have a transient ionized absorber with $\nh \sim 10^{20}\cmsqi$ and $\log(\xi/\xiunit) \sim 2.9$ in 2018 \citep{Ricci2021}. During this period of changing look phase, the source was accreting at the super-Eddington limit. However, the source transitioned to a sub-Eddington state ($\lambdaedd \sim 0.3$; \citealt{ghosh2023}) in the post-changing-look phase. 

\begin{figure}
    \centering
    \includegraphics[width=0.99\linewidth]{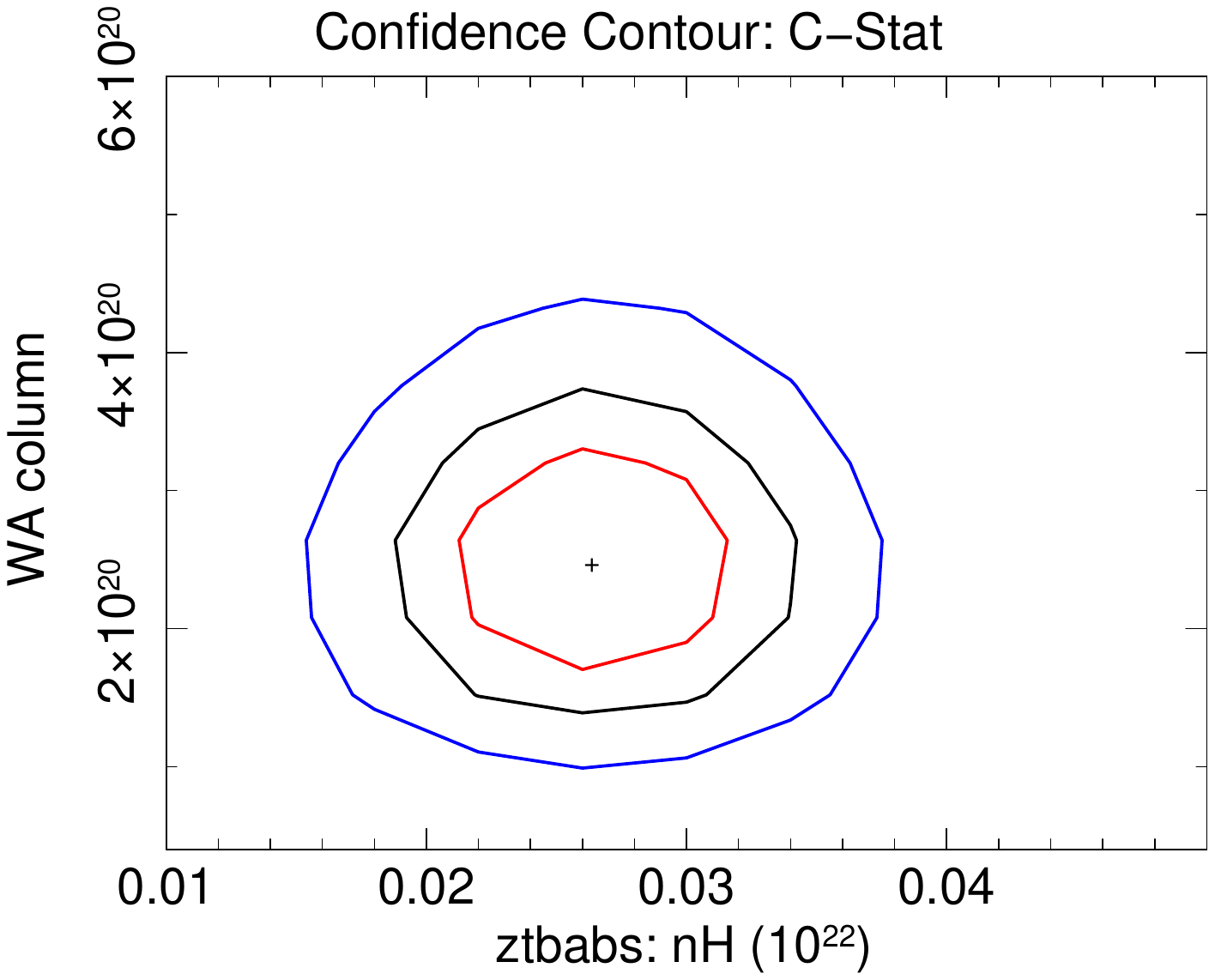}
    \caption{Confidence contours (68\%, 90\%, and 99\%) for the column density of the host galaxy neutral absorber ($n_{\text{H,ztbabs}}$) versus the column density of the ionized warm absorber (WA$_{\text{column}}$) for the July 2022 epoch. The closed, nearly circular contours indicate a lack of significant correlation or degeneracy between the neutral and ionized absorption components. This independence demonstrates that both a neutral foreground and an evolving ionized outflow are statistically required.}
    \label{fig:cont_tbabs_wcol}
\end{figure}

A critical consideration in our absorption analysis is the potential for degeneracy between the intrinsic neutral absorber (\texttt{ztbabs}) and the ionized warm absorber (WA), particularly in the soft X-ray band ($< 2 \kev$). This source shows a host galactic absorption with a variable column since its detection \citep{Boller2003,Gallo2013, trak19, ricci2022, laha2022,masterson2022}. To address this, we created a contour for the column densities of both components for the observation on July 26, 2022. As illustrated in Figure \ref{fig:cont_tbabs_wcol}, the confidence contours are well-constrained and show no evidence of a parameter degeneracy. This lack of correlation suggests that the \texttt{ztbabs} and the WA are physically and statistically distinct.

The presence of strong ionized absorbers in 2022, which subsequently weakened as a nascent jet emerged, \citep{laha2025, meyer2025} is very interesting. AGNs are thought to switch between two primary states: a radiatively efficient "wind mode" at high accretion rates ($\lambdaedd$) that drives wide-angle outflows and a "jet mode" at lower rates that tends to suppress them \citep{fabian2012}. Even though the observations hint that \onees{} may have undergone a similar transition, we cannot rule out the coincidence. The test of physical link between different modes of outflows is beyond the scope of this study.

An anti-correlation between WA column density and radio loudness $R$ has been proposed in the past literature. This anti-correlation is potentially linked to a reconfiguration of the inner accretion flow or magnetic field geometry that preferentially favors one type of outflow\citep{mehd2019}. While the temporal alignment between the onset of the jet and the weakening of the X-ray wind in \onees{} is suggestive of a mode transition, this does not necessarily imply that they are mutually exclusive. As observed in broader AGN samples, jets and winds can also coexist \citep{Tombesi2013, Tombesi2014}. The current data indicate a likely transition, though the precise relationship between the weakening of wind and the formation of a jet requires further investigation. Future high-resolution observations, such as those from the \textit{XRISM} micro-calorimeter, will be instrumental in placing tighter constraints on weak or highly ionized gases such as WA and UFO and robustly mapping the disk-wind-jet connection.

\subsection{The soft X-ray and 5 GHz radio plateau}
 Figure \ref{fig:swift_lc_zoom} shows the multi-band light curve of the source from May 2022 to Aug 2025. 
 We note that the $2-10 \kev$ (hard X-ray) flux is almost at the same level during this period except for some variability at $<2$ times. On the other hand, the $0.3-2 \kev$ flux has steadily increased since May 2022 and reached a peak value (about 10 times the preflare value) somewhere in April 2024, after which it has plateaued. It is interesting to note from panel 6 of Figure \ref{fig:swift_lc_zoom} that the radio emission at $5$ GHz has also plateaued (for details, see \citealt{meyer2024}), indicating a close connection between the soft X-ray emission and radio emission. On the contrary, the optical and UV flux densities show random variability within $<30\%$ of the mean value, indicating perhaps a steady accretion in the source. We note that the source is currently in a radio-loud state \citep{meyer2024} with a newly formed jet emerging from the central engine at a speed of $\sim 0.2c$. As discussed in our earlier work, \citep{laha2025} the soft X-ray excess of this source can be explained neither from a reflection scenario \citep{garcia2014} nor from a warm corona perspective \citep{Done2012}. The soft X-rays are closely linked with the jet formation and may originate from regions located as close as the coronal plasma. This is because of the following reasons: (1) The shortest variability time scale in soft X-rays is $\sim 200$ sec which is similar to that of the hard X-rays \citep{laha2025,masterson2025}, (2) The QPO \citep{masterson2025} was detected in both the soft and the hard X-rays, (3) The soft X-rays started to rise when the radio flare happened and, subsequently, the radio jet was launched, (4) The soft X-rays plateaued near about the same time as the radio, (5) The soft X-rays share no long-term correlations with the UV and the hard X-rays, implying a complex interplay between several parameters such as the jet launch, magnetic field changes, steady accretion, etc., producing this emission. Existing literature established that jet formation can follow tidal disruption events (TDEs; e.g., \citealt{Bloom2011, Pasham2023}). A similar evolution is observed in \onees{}, which underwent a violent UV/optical outburst in late 2017 \citep{trak19, Ricci2020, Ricci2021, laha2022}, followed by reported jet formation in 2023 \citep{meyer2025, laha2025}. However, the gap between the UV/optical outburst and radio outburst is $\sim 5$ years, which is much greater than the dynamical timescale of accretion in AGN \citep{Kara2025}.

\section{Conclusion}
\label{sec:conclusion}

In this study, we conducted a multi-wavelength campaign of the changing-look AGN 1ES 1927+654, centered on a detailed X-ray spectral analysis using \xmm{} RGS and EPIC-pn observations from 2022 to 2025. The data reveal significant spectral evolution in X-ray and radio and relatively stable in optical/UV. Given the current SNR ratio and phenomenological modeling, the results, however, are indicative rather than definitive physical conclusions. The key results are summarized as follows:

\begin{enumerate}

\item \textbf{Emergence of Soft X-ray Emission Lines}: 
Our analysis reveals two soft X-ray emission lines at $\sim0.56$~keV and $\sim1$~keV during the 2023--2025 epochs. These features, absent in historical and early flare-state data, could be associated with the evolution of the accretion disk. Specifically, the $\sim 1\kev$ line is consistent with an increase in column density in the inner accretion disk region, \citep{Fabian2009}, possibly linked to magnetic field reconfiguration during jet formation. The narrower $\sim 0.56\kev$ feature, however, likely originates from O~VII ionic transitions in a photoionized gas, tracing a change in the illumination of the surrounding gases in the context of a rise in the soft X-ray by nearly an order of magnitude.

\item \textbf{Detection of a Broad Iron Feature}: 
For the first time in this source, we report the detection of a statistically significant, broad ($\sigma \approx 800\ev$) FeK emission feature in yearly stacked spectra from 2023 to 2025. The broadness and high rest-frame energy (suggestive of Fe~XXV or Fe~XXVI) point toward a reflecting medium that is highly ionized and possibly in the innermost regions of the disk. Given the relative stability of the $(2-10)\kev$ continuum, this feature indicates a potential structural change in the inner disk geometry coincident with jet emergence.

\item \textbf{Evolution of Ionized Absorption}: 
The data show the presence of an ionized warm absorber (WA) in 2022 and weakening in between 2023 and 2025, a period coinciding with the formation of a nascent jet \citep{laha2025,meyer2025}. While these observations are qualitatively consistent with a transition from a  "wind-dominated" to a "jet-dominated" mode \citep{fabian2012, mehd2019} of outflow based on the accretion rate and radio loudness, we cannot definitively claim the total quenching of the wind. Instead, the results indicate a shift in the mode of outflow as the source remains in a sub-Eddington state ($\lambdaedd \sim 0.3$) . Future high-resolution observations with \textit{XRISM} are required to conclusively rule out or constrain highly ionized outflows in the $7-9\kev$ band.

\item \textbf{Co-evolution of Soft X-ray and Radio Emission}: 
The quasi-simultaneous rise and subsequent plateaus observed in both the soft X-ray and 5 GHz radio bands suggest a potential physical link between these emission components. This is further supported by the detection and persistence of Quasi-Periodic Oscillations (QPOs) since 2022, which have now reached a saturated state \citep{masterson2025, Masterson2026}. Collectively, these phenomena imply that the soft X-ray emission likely originates from regions in close proximity to the corona or jet base. While a detailed establishment of this relationship is beyond the scope of this study, the dramatic spectral evolution in X-ray and radio alongside a nearly constant optical flux during the post-changing-look phase points toward a complex interplay between magnetic field evolution, steady accretion, and jet-launching dynamics. 

\end{enumerate}

\section{Acknowledgements}
The material is based upon work supported by NASA under award number 80GSFC21M0002. We thank the annonymous referee for their highly constructive and insightful comments during the review process.

\bibliographystyle{aasjournal}
\bibliography{xmm_one_es}

\appendix

\section{Creating an ionized absorber model}
\label{sec:creating_ionized_absorber_model}
Previous studies on ionized absorption modeling in AGN X-ray spectra have pointed out the necessity of a realistic spectral energy distribution (SED) of the source as seen by the absorbers to generate the ionized absorption models \citep{laha2013IRAS}. The different regions of the SED: UV, soft X-ray ($0.3-2 \kev$) and hard X-ray ($2-100 \kev$),  impact the WA models differently, leading to very different inferences from the data (for details, see \citealt{laha2013IRAS}). Therefore, we create a realistic broadband SED to create the warm absorber and UFO models as enumerated in several previous studies \citep{laha2011Mrk704,laha2014,laha2021natAs,laha2013IRAS}. Below we briefly discuss the steps to create the realistic SED and refer the readers to previous work for details \citep[e.g.,][]{laha2014}.

\subsection*{Creating a realistic SED}
\begin{figure}[h!]
    \centering
    \includegraphics[width=0.5\textwidth]{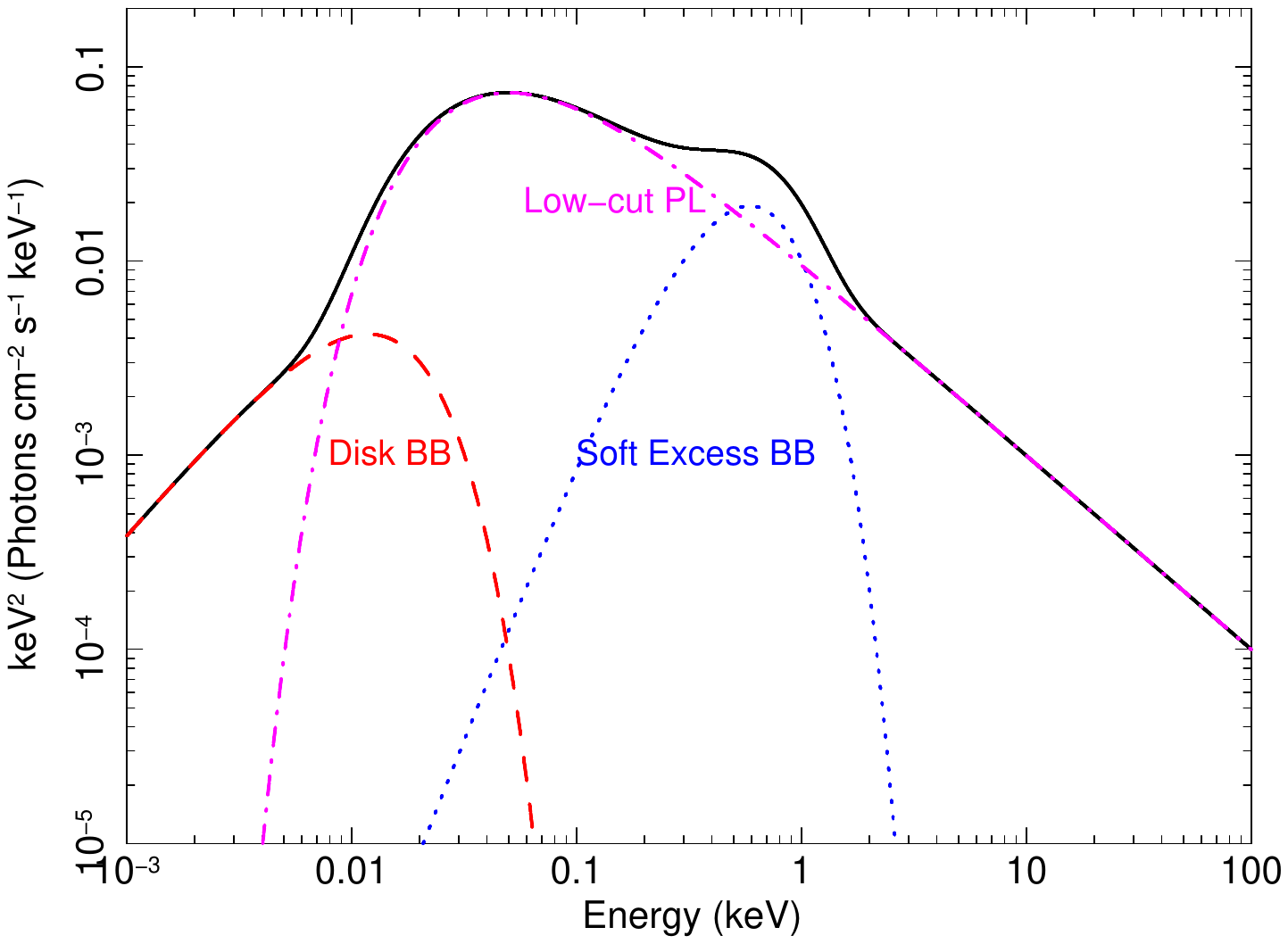}

            \caption{ A broadband SED for 1ES 1927+654 based on the recent observations in 2025. The red dotted curve is an accretion disk multi-color blackbody with a temperature of 5 eV at the innermost zone, the blue dotted curve is for a blackbody emission from the soft excess component with a temperature of 0.16 keV, the pink dotted curve is a power law emission with a lower cutoff energy of 50 eV, and the black solid line is the total continuum emission for this source.} 
    \label{fig:sed}
\end{figure}

A typical realistic SED as observed by the ionized clouds near the AGN can be a continuous spectrum from UV ($1\ev$) to hard X-rays ($100\kev$). However, we note that the energy band $13.6\ev-100\ev$ is not observable because of the Galactic absorption, and hence some approximations may have to be made \citep[see for e.g.,][]{laha2013IRAS}. For 1ES 1927+654, we constructed the SED using the model {\tt diskbb+bb+expabs$\times$pow}, with parameter values provided in Table \ref{table:model_parameters}. Where {\tt diskbb} describes the accretion disk blackbody, {\tt bbody} is the blackbody model describing the soft X-ray excess emission, {\tt expabs} is a low-energy exponential roll-off, and {\tt pow} is the power law emission from the corona. The {\tt bbody} and {\tt pow} parameters were adopted from the best fit to the XMM-Newton PN spectrum as in March 2024 \citep{laha2025}, while the {\tt diskbb} (modeling the UV bump) component parameters were obtained from a previous work using \swift{}-UVOT \citep{ghosh2023}. We note here that the UV flux did not vary appreciably ($<30\%$) during this 3-year period (2022 - May, 2025), but the source flux in the soft X-ray band varies significantly with time, increasing by a factor of $\sim 10$ in the last 3 years. Even though creating a SED for each observation is ideal, that level of accuracy is not needed in this work because, as we will find out, we did not detect any ionized absorbers. Figure \ref{fig:sed} illustrates the SED and its various components. We assumed a lower energy cut-off for the power-law emission at $1\ev$ and higher energy at $100\kev$.

\begin{table}[h!]
\centering
\caption{Model parameters and values of SED for 1ES 1927+654.}
\begin{tabular}{cccc}
\hline
\textbf{Model Comp.} & \textbf{Parameter} & \textbf{Unit} & \textbf{Value} \\
\hline
diskbb & Tin & eV & 5 \\
diskbb & norm & -- & 1.00E+09 \\
bbody & kT & keV & 0.15 \\
bbody & norm & -- & 5.00E-04 \\
expabs & LowECut & eV & 50 \\
pow & PhoIndex & -- & 3 \\
pow & norm & -- & 1.00E-02 \\
\hline
\end{tabular}

\label{table:model_parameters}
\end{table}

\section{Swift fit table and light curves}

We constructed a comprehensive multi-wavelength light curve to track the long-term evolution of 1ES~1927+654. The full light curve, covering the period from May~2018 to August~2025, is shown in Figure~\ref{fig:swift_lc_main} in the Appendix, while a zoomed-in view highlighting the recent activities between May~2022 and August~2025 is presented in Figure~\ref{fig:swift_lc_zoom}. The X-ray and UV data are obtained from continued \swift{} monitoring, with the observations from April~2024 to August~2025 extending the light curves published in our earlier works (e.g., \citealt{laha2022, ghosh2023, laha2025}). These new observations are also discussed in Section~\ref{subsec:swift} and listed in Tables~\ref{tab:swift_obs_tab1} and \ref{tab:swift_obs_tab2}. The corresponding spectral fit parameters are provided in Table~\ref{tab:swift_fit} in the Appendix.

\newpage
\begin{longtable}{llllllll}
\caption[Swift spectral fits]{Swift spectral fit parameters for 1ES~1927+654 }\\
\label{tab:swift_fit}\\
\hline
Short-ID & Date & $kT$ (keV) & $\Gamma$ & $F_{(0.3-2)\,\mathrm{keV}}^{a}$ & $F_{(2-10)\,\mathrm{keV}}^{a}$ & UV Flux Density$^{b}$ & $\chi^2/\mathrm{dof}$ \\
\hline
S145 & 18/04/2024 & $0.18^{+0.02}_{-0.01}$ & $2.29^{+0.44}_{-0.80}$ & $61.33^{+3.85}_{-3.84}$ & $9.56^{+0.60}_{-0.60}$ & $1.25 \pm 0.10$ & $74/108$ \\
S146 & 25/04/2024 & $0.20^{+0.04}_{-0.03}$ & $2.80^{+0.76}_{-0.79}$ & $65.13^{+4.80}_{-4.79}$ & $7.08^{+0.52}_{-0.52}$ & $1.15 \pm 0.10$ & $84/79$ \\
S146A & 25/04/2024 & $0.17^{+0.03}_{-0.02}$ & $2.31^{+0.82}_{-4.18}$ & $56.99^{+4.63}_{-4.63}$ & $8.87^{+0.72}_{-0.72}$ & $nan \pm nan$ & $59/68$ \\
S147 & 30/04/2024 & $0.19^{+0.02}_{-0.01}$ & $2.65^{+0.22}_{-0.29}$ & $59.47^{+2.76}_{-2.76}$ & $8.13^{+0.38}_{-0.38}$ & $1.17 \pm 0.14$ & $149/140$ \\
S148 & 02/05/2024 & $0.16^{+0.02}_{-0.01}$ & $2.23^{+0.67}_{-1.57}$ & $58.10^{+4.34}_{-4.34}$ & $10.62^{+0.79}_{-0.79}$ & $1.14 \pm 0.08$ & $89/81$ \\
S149 & 09/05/2024 & $0.16^{+0.04}_{-0.02}$ & $2.40^{+0.52}_{-1.06}$ & $58.01^{+4.73}_{-4.72}$ & $11.85^{+0.97}_{-0.96}$ & $1.16 \pm 0.12$ & $67/70$ \\
S150 & 24/05/2024 & $0.23^{+0.05}_{-0.04}$ & $2.93^{+0.34}_{-0.31}$ & $69.73^{+5.08}_{-5.08}$ & $7.86^{+0.57}_{-0.57}$ & $1.15 \pm 0.13$ & $48/81$ \\
S151 & 30/05/2024 & $0.19^{+0.04}_{-0.02}$ & $2.71^{+0.27}_{-0.34}$ & $59.46^{+3.96}_{-3.95}$ & $8.84^{+0.59}_{-0.59}$ & $1.12 \pm 0.11$ & $77/89$ \\
S152 & 10/06/2024 & $0.18^{+0.01}_{-0.01}$ & $2.31^{+0.30}_{-0.43}$ & $59.05^{+2.63}_{-2.63}$ & $10.13^{+0.45}_{-0.45}$ & $nan \pm nan$ & $143/146$ \\
S153 & 20/06/2024 & $0.19^{+0.04}_{-0.02}$ & $2.65^{+0.42}_{-0.66}$ & $58.86^{+5.16}_{-5.15}$ & $7.84^{+0.69}_{-0.69}$ & $1.19 \pm 0.12$ & $50/63$ \\
S154 & 27/06/2024 & $0.17^{+0.02}_{-0.02}$ & $2.06^{+0.63}_{-1.30}$ & $55.47^{+5.02}_{-5.01}$ & $13.90^{+1.26}_{-1.25}$ & $1.07 \pm 0.10$ & $51/57$ \\
S155 & 04/07/2024 & $0.18^{+0.02}_{-0.02}$ & $2.70^{+0.24}_{-0.30}$ & $67.79^{+3.98}_{-3.97}$ & $9.07^{+0.53}_{-0.53}$ & $1.29 \pm 0.12$ & $104/111$ \\
S156 & 14/07/2024 & $0.16^{+0.03}_{-0.02}$ & $2.67^{+0.37}_{-0.61}$ & $65.31^{+4.32}_{-4.32}$ & $8.35^{+0.55}_{-0.55}$ & $1.19 \pm 0.12$ & $72/87$ \\
S157 & 18/07/2024 & $0.18^{+0.03}_{-0.02}$ & $2.68^{+0.36}_{-0.58}$ & $62.58^{+4.05}_{-4.05}$ & $7.94^{+0.51}_{-0.51}$ & $1.12 \pm 0.11$ & $115/106$ \\
S158 & 25/07/2024 & $0.19^{+0.03}_{-0.02}$ & $2.70^{+0.27}_{-0.33}$ & $71.34^{+4.36}_{-4.19}$ & $10.00^{+0.61}_{-0.59}$ & $0.92 \pm 0.10$ & $92/99$ \\
S159 & 01/08/2024 & $0.19^{+0.03}_{-0.02}$ & $2.66^{+0.31}_{-0.52}$ & $68.03^{+4.47}_{-4.28}$ & $9.43^{+0.62}_{-0.59}$ & $1.16 \pm 0.13$ & $94/88$ \\
S160 & 08/08/2024 & $0.19^{+0.02}_{-0.02}$ & $2.52^{+0.26}_{-0.35}$ & $66.80^{+3.54}_{-3.42}$ & $10.93^{+0.58}_{-0.56}$ & $1.00 \pm 0.14$ & $125/122$ \\
S161 & 15/08/2024 & $0.24^{+0.04}_{-0.04}$ & $2.91^{+0.30}_{-0.30}$ & $62.26^{+4.47}_{-4.26}$ & $7.48^{+0.54}_{-0.51}$ & $0.77 \pm 0.10$ & $71/78$ \\
S162 & 22/08/2024 & $0.18^{+0.02}_{-0.01}$ & $2.39^{+0.30}_{-0.43}$ & $61.79^{+3.53}_{-3.40}$ & $10.55^{+0.60}_{-0.58}$ & $1.20 \pm 0.12$ & $93/112$ \\
S163 & 10/09/2024 & $0.21^{+0.05}_{-0.04}$ & $3.08^{+0.45}_{-0.41}$ & $75.72^{+6.04}_{-5.73}$ & $6.41^{+0.51}_{-0.49}$ & $1.10 \pm 0.10$ & $56/64$ \\
S164 & 12/09/2024 & $0.20^{+0.04}_{-0.03}$ & $2.90^{+0.39}_{-0.41}$ & $79.80^{+5.39}_{-5.15}$ & $8.32^{+0.56}_{-0.54}$ & $1.17 \pm 0.12$ & $74/87$ \\
S165 & 20/09/2024 & $0.18^{+0.03}_{-0.02}$ & $2.75^{+0.28}_{-0.38}$ & $65.98^{+5.32}_{-4.86}$ & $8.10^{+3.32}_{-2.41}$ & $2.09 \pm 0.07$ & $87/109$ \\
S166 & 26/09/2024 & $0.21^{+0.04}_{-0.03}$ & $3.05^{+0.46}_{-0.43}$ & $75.42^{+8.40}_{-7.54}$ & $6.25^{+4.10}_{-2.77}$ & $2.05 \pm 0.17$ & $71/71$ \\
S167 & 03/10/2024 & $0.19^{+0.04}_{-0.02}$ & $2.63^{+0.43}_{-0.88}$ & $68.44^{+8.32}_{-7.36}$ & $9.19^{+7.56}_{-3.83}$ & $2.05 \pm 0.17$ & $52/59$ \\
S168 & 10/10/2024 & $0.20^{+0.03}_{-0.02}$ & $2.44^{+0.31}_{-0.54}$ & $67.91^{+5.87}_{-5.36}$ & $12.72^{+5.84}_{-3.73}$ & $2.17 \pm 0.18$ & $97/102$ \\
S169 & 24/10/2024 & $0.18^{+0.04}_{-0.02}$ & $2.49^{+0.44}_{-0.84}$ & $67.37^{+7.81}_{-6.74}$ & $11.44^{+9.07}_{-4.48}$ & $2.07 \pm 0.20$ & $66/67$ \\
S170 & 31/10/2024 & $0.20^{+0.05}_{-0.03}$ & $2.92^{+0.45}_{-0.64}$ & $72.98^{+7.82}_{-9.54}$ & $6.46^{+4.68}_{-2.92}$ & $2.42 \pm 0.18$ & $56/45$ \\
\multicolumn{8}{c}{{\bfseries \tablename~\thetable{} -- continued}}\\
\hline
Short-ID & Date & $kT$ (keV) & $\Gamma$ & $F_{(0.3-2)\,\mathrm{keV}}^{a}$ & $F_{(2-10)\,\mathrm{keV}}^{a}$ & UV Flux Density$^{b}$ & $\chi^2/\mathrm{dof}$ \\
\hline
S171 & 07/11/2024 & $0.19^{+0.05}_{-0.03}$ & $2.83^{+0.48}_{-0.72}$ & $71.10^{+9.92}_{-8.58}$ & $7.17^{+5.58}_{-3.34}$ & $2.15 \pm 0.18$ & $51/46$ \\
S172 & 14/11/2024 & $0.19^{+0.05}_{-0.03}$ & $2.74^{+0.29}_{-0.41}$ & $71.07^{+7.28}_{-6.65}$ & $11.15^{+5.31}_{-3.71}$ & $2.24 \pm 0.18$ & $88/78$ \\

S173 & 21/11/2024 & $0.18^{+0.01}_{-0.01}$ & $2.02^{+0.48}_{-0.68}$ & $63.22^{+5.08}_{-4.54}$ & $13.48^{+7.87}_{-4.41}$ & $2.22 \pm 0.17$ & $88/96$ \\
S174 & 28/11/2024 & $0.20^{+0.04}_{-0.03}$ & $2.64^{+0.37}_{-0.48}$ & $69.31^{+7.10}_{-6.35}$ & $10.93^{+5.94}_{-4.10}$ & $2.02 \pm 0.17$ & $85/82$ \\
S175 & 12/12/2024 & $0.18^{+0.03}_{-0.02}$ & $2.58^{+0.31}_{-0.44}$ & $69.43^{+6.20}_{-5.59}$ & $10.59^{+4.38}_{-3.14}$ & $2.59 \pm 0.18$ & $104/98$ \\
S176 & 19/12/2024 & $0.20^{+0.04}_{-0.03}$ & $2.62^{+0.29}_{-0.39}$ & $64.93^{+6.19}_{-5.63}$ & $11.63^{+4.90}_{-3.57}$ & $1.98 \pm 0.17$ & $74/87$ \\

S177 & 26/12/2024 & $0.18^{+0.02}_{-0.01}$ & $2.52^{+0.34}_{-0.52}$ & $73.85^{+5.98}_{-5.41}$ & $11.32^{+5.61}_{-3.54}$ & $2.42 \pm 0.17$ & $99/108$ \\


S178 & 09/01/2025 & $0.16^{+0.01}_{-0.01}$ & $2.01^{+0.54}_{-0.74}$ & $83.49^{+9.25}_{-3.59}$ & $17.16^{+10.89}_{-5.94}$ & $2.05 \pm 0.17$ & $47/66$ \\

S179 & 16/01/2025 & $0.16^{+0.01}_{-0.01}$ & $2.06^{+0.46}_{-0.59}$ & $69.81^{+6.36}_{-5.82}$ & $16.78^{+8.59}_{-5.20}$ & $1.96 \pm 0.15$ & $85/78$ \\

S180 & 06/02/2025 & $0.19^{+0.02}_{-0.01}$ & $2.23^{+0.43}_{-0.79}$ & $73.76^{+4.36}_{-6.19}$ & $10.26^{+6.76}_{-3.48}$ & $2.15 \pm 0.17$ & $108/108$ \\

S181 & 13/02/2025 & $0.18^{+0.04}_{-0.02}$ & $2.69^{+0.33}_{-0.46}$ & $78.32^{+4.56}_{-8.75}$ & $10.30^{+5.17}_{-3.56}$ & $1.59 \pm 0.14$ & $102/82$ \\
S182 & 20/02/2025 & $0.20^{+0.03}_{-0.02}$ & $2.70^{+0.28}_{-0.37}$ & $71.83^{+5.75}_{-5.28}$ & $9.02^{+3.55}_{-2.64}$ & $1.83 \pm 0.15$ & $97/105$ \\
S183 & 27/02/2025 & $0.18^{+0.02}_{-0.01}$ & $2.08^{+0.47}_{-0.74}$ & $54.86^{+4.84}_{-4.27}$ & $13.69^{+8.59}_{-4.34}$ & $2.05 \pm 0.17$ & $74/91$ \\
S184 & 06/03/2025 & $0.18^{+0.02}_{-0.02}$ & $2.50^{+0.33}_{-0.50}$ & $72.47^{+6.84}_{-6.11}$ & $12.37^{+5.20}_{-3.53}$ & $1.83 \pm 0.15$ & $116/86$ \\
S185 & 22/03/2025 & $0.20^{+0.04}_{-0.03}$ & $2.74^{+0.25}_{-0.30}$ & $71.80^{+6.12}_{-5.63}$ & $11.01^{+4.12}_{-3.22}$ & $2.24 \pm 0.17$ & $87/100$ \\
S186 & 01/04/2025 & $0.16^{+0.03}_{-0.02}$ & $2.22^{+0.58}_{-0.98}$ & $83.36^{+9.28}_{-8.09}$ & $17.34^{+16.39}_{-6.52}$ & $1.89 \pm 0.15$ & $62/62$ \\
S187 & 02/04/2025 & $0.17^{+0.02}_{-0.02}$ & $2.61^{+0.35}_{-0.57}$ & $77.91^{+6.92}_{-6.21}$ & $10.74^{+5.65}_{-3.40}$ & $2.20 \pm 0.11$ & $81/94$ \\
S188 & 03/04/2025 & $0.19^{+0.03}_{-0.02}$ & $2.63^{+0.29}_{-0.42}$ & $65.58^{+6.12}_{-5.54}$ & $10.06^{+4.17}_{-3.00}$ & $2.28 \pm 0.13$ & $77/94$ \\
S189 & 04/04/2025 & $0.22^{+0.07}_{-0.04}$ & $3.02^{+0.41}_{-0.39}$ & $74.47^{+9.44}_{-8.32}$ & $7.44^{+3.80}_{-2.70}$ & $2.15 \pm 0.13$ & $87/74$ \\

S190 & 05/04/2025 & $0.17^{+0.02}_{-0.01}$ & $2.14^{+0.66}_{-1.16}$ & $62.42^{+6.80}_{-5.77}$ & $9.64^{+10.94}_{-4.19}$ & $2.22 \pm 0.13$ & $59/64$ \\
S191 & 07/04/2025 & $0.22^{+0.04}_{-0.03}$ & $3.01^{+0.39}_{-0.42}$ & $70.96^{+8.87}_{-7.92}$ & $6.40^{+3.71}_{-2.53}$ & $nan \pm nan$ & $52/58$ \\
S192 & 08/04/2025 & $0.17^{+0.02}_{-0.01}$ & $2.39^{+0.34}_{-0.48}$ & $67.65^{+5.63}_{-5.10}$ & $12.29^{+5.21}_{-3.50}$ & $2.28 \pm 0.13$ & $85/99$ \\
S192A & 08/04/2025 & $0.19^{+0.03}_{-0.02}$ & $2.63^{+0.44}_{-0.73}$ & $71.39^{+7.75}_{-6.86}$ & $8.57^{+6.11}_{-3.56}$ & $2.28 \pm 0.09$ & $70/66$ \\
S193 & 09/04/2025 & $0.18^{+0.03}_{-0.02}$ & $2.66^{+0.36}_{-0.52}$ & $63.83^{+6.67}_{-5.80}$ & $7.97^{+3.69}_{-2.60}$ & $2.17 \pm 0.13$ & $99/83$ \\
S194 & 11/04/2025 & $0.18^{+0.02}_{-0.01}$ & $2.40^{+0.36}_{-0.58}$ & $67.58^{+5.37}_{-4.88}$ & $10.62^{+4.92}_{-3.12}$ & $2.20 \pm 0.11$ & $92/106$ \\
S195 & 12/04/2025 & $0.17^{+0.02}_{-0.01}$ & $2.47^{+0.36}_{-0.50}$ & $64.40^{+5.50}_{-4.98}$ & $9.52^{+4.38}_{-2.93}$ & $2.07 \pm 0.11$ & $98/94$ \\
S196 & 15/04/2025 & $0.18^{+0.03}_{-0.02}$ & $2.60^{+0.30}_{-0.41}$ & $67.91^{+6.20}_{-5.60}$ & $11.02^{+4.48}_{-3.29}$ & $2.28 \pm 0.07$ & $85/85$ \\
S197 & 22/04/2025 & $0.16^{+0.02}_{-0.01}$ & $2.42^{+0.40}_{-0.61}$ & $68.22^{+5.92}_{-5.26}$ & $10.65^{+5.37}_{-3.20}$ & $2.18 \pm 0.07$ & $101/93$ \\
S198 & 13/05/2025 & $0.17^{+0.01}_{-0.01}$ & $2.26^{+0.35}_{-0.50}$ & $65.76^{+5.10}_{-4.71}$ & $12.70^{+5.10}_{-3.50}$ & $2.07 \pm 0.07$ & $96/101$ \\
S199 & 24/05/2025 & $0.23^{+0.10}_{-0.08}$ & $3.03^{+1.09}_{-0.63}$ & $85.81^{+15.19}_{-12.77}$ & $9.05^{+9.34}_{-5.67}$ & $2.26 \pm 0.11$ & $44/35$ \\
S200 & 27/05/2025 & $0.19^{+0.04}_{-0.03}$ & $2.98^{+0.51}_{-0.94}$ & $68.01^{+10.64}_{-9.27}$ & $5.60^{+5.17}_{-2.75}$ & $2.13 \pm 0.07$ & $44/49$ \\
S201 & 29/05/2025 & $0.19^{+0.09}_{-0.05}$ & $2.97^{+0.55}_{-0.66}$ & $76.52^{+12.37}_{-10.63}$ & $7.77^{+6.64}_{-4.01}$ & $2.13 \pm 0.09$ & $19/38$ \\
S202 & 03/06/2025 & $0.19^{+0.06}_{-0.04}$ & $2.89^{+0.27}_{-0.32}$ & $82.17^{+8.17}_{-7.46}$ & $10.34^{+4.18}_{-3.31}$ & $1.94 \pm 0.07$ & $74/82$ \\
S203 & 10/06/2025 & $0.22^{+0.04}_{-0.03}$ & $3.02^{+0.28}_{-0.28}$ & $66.76^{+6.25}_{-5.75}$ & $6.40^{+2.71}_{-2.02}$ & $2.22 \pm 0.07$ & $95/87$ \\
S204 & 26/06/2025 & $0.18^{+0.03}_{-0.02}$ & $2.26^{+0.53}_{-0.92}$ & $53.31^{+6.92}_{-5.77}$ & $10.44^{+7.93}_{-4.06}$ & $2.20 \pm 0.11$ & $57/54$ \\
S205 & 01/07/2025 & $0.18^{+0.02}_{-0.01}$ & $2.50^{+0.36}_{-0.61}$ & $68.16^{+5.52}_{-5.01}$ & $9.48^{+4.93}_{-2.93}$ & $2.20 \pm 0.07$ & $85/106$ \\
S206 & 08/07/2025 & $0.18^{+0.02}_{-0.01}$ & $2.24^{+0.40}_{-0.73}$ & $63.80^{+5.03}_{-4.51}$ & $12.32^{+6.45}_{-3.45}$ & $2.09 \pm 0.07$ & $102/111$ \\
S207 & 15/07/2025 & $0.19^{+0.03}_{-0.02}$ & $2.87^{+0.30}_{-0.36}$ & $67.25^{+5.91}_{-5.38}$ & $6.65^{+2.78}_{-2.11}$ & $2.22 \pm 0.07$ & $75/97$ \\
S208 & 22/07/2025 & $0.20^{+0.03}_{-0.02}$ & $2.58^{+0.25}_{-0.32}$ & $60.52^{+4.60}_{-4.25}$ & $9.77^{+3.12}_{-2.44}$ & $2.13 \pm 0.07$ & $104/116$ \\
S209 & 29/07/2025 & $0.18^{+0.04}_{-0.03}$ & $2.83^{+0.26}_{-0.40}$ & $72.68^{+6.67}_{-6.16}$ & $9.63^{+4.18}_{-2.89}$ & $2.15 \pm 0.07$ & $77/91$ \\
S210 & 03/08/2025 & $0.18^{+0.04}_{-0.02}$ & $2.71^{+0.40}_{-0.66}$ & $72.07^{+7.86}_{-7.00}$ & $9.36^{+6.75}_{-3.81}$ & $2.18 \pm 0.11$ & $73/73$ \\
S211 & 14/08/2025 & $0.18^{+0.03}_{-0.02}$ & $2.66^{+0.40}_{-0.78}$ & $69.09^{+7.26}_{-6.44}$ & $8.73^{+6.48}_{-3.28}$ & $1.91 \pm 0.07$ & $76/76$ \\
S212 & 19/08/2025 & $0.17^{+0.03}_{-0.02}$ & $2.67^{+0.29}_{-0.44}$ & $68.40^{+6.84}_{-6.09}$ & $10.19^{+3.87}_{-2.75}$ & $1.44 \pm 0.06$ & $84/90$ \\
S213 & 26/08/2025 & $0.17^{+0.01}_{-0.01}$ & $2.05^{+0.62}_{-0.93}$ & $64.02^{+5.76}_{-5.14}$ & $11.60^{+11.89}_{-4.76}$ & $2.04 \pm 0.07$ & $78/77$ \\
\hline

\multicolumn{8}{l}{\textbf{Notes.}}\\
\multicolumn{8}{l}{$^{a}$ Unabsorbed X-ray flux in units of $10^{-12}\,\mathrm{erg\,s^{-1}\,cm^{-2}}$.}\\
\multicolumn{8}{l}{$^{b}$ Swift UVOT UVW2 flux density in units of $10^{-15}\,\mathrm{erg\,s^{-1}\,cm^{-2}\,\AA^{-1}}$.}\\
\multicolumn{8}{l}{UVW2 central wavelength is 1928\,\AA.} \\
\multicolumn{8}{l}{The UV flux density was corrected for Galactic absorption using the correction magnitude of $A_{\lambda}$ = 0.690 obtained from NED.}\\
\end{longtable}

\begin{figure*}
    \centering
    \includegraphics[width=0.8\linewidth]{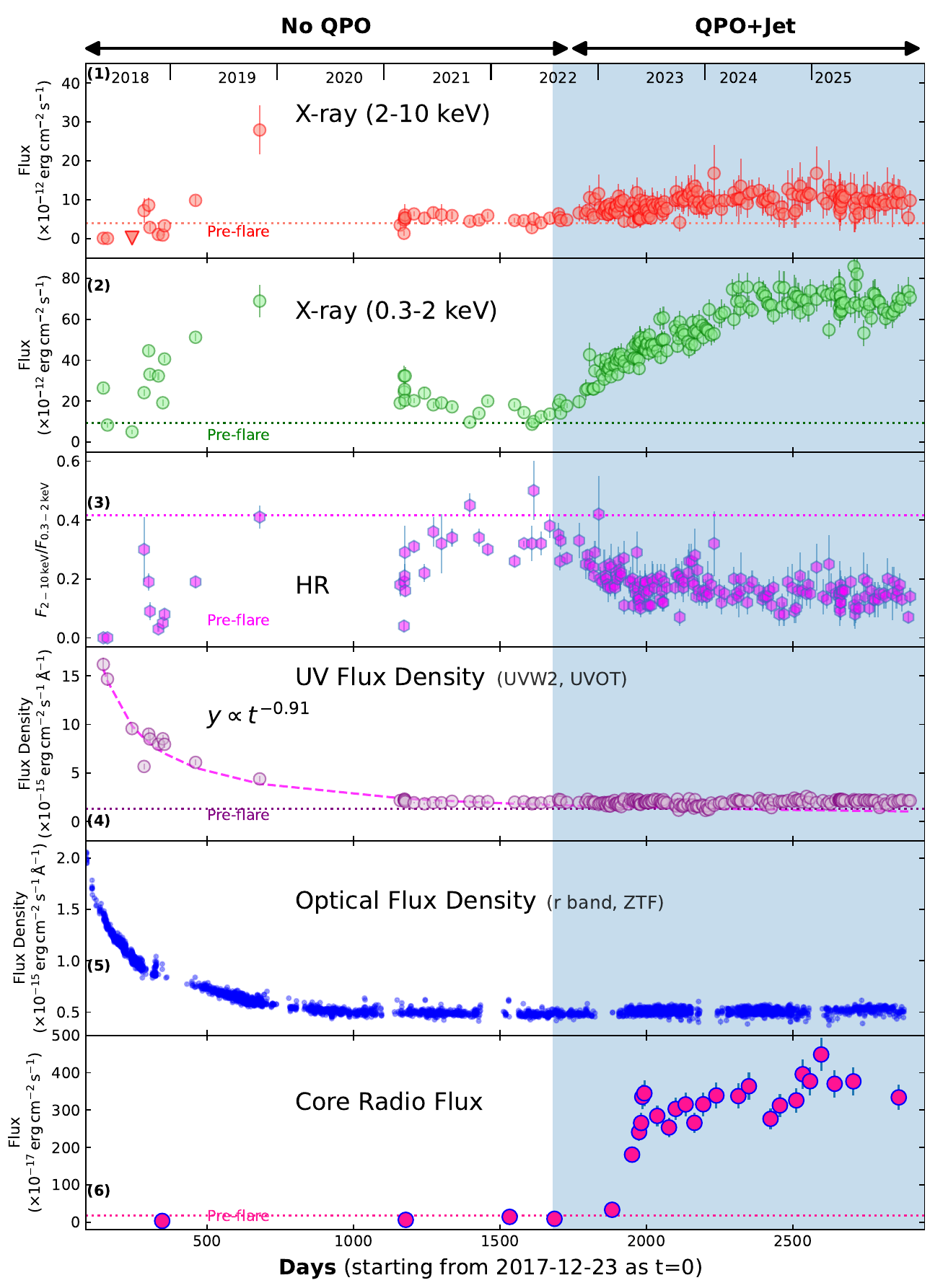}
    \caption{The X-ray, UV, optical, and radio light curves of 1ES1927+654 from May 2018 (when \swift{} monitoring started) to August 2025. We note the optical UV outburst in December 2017 in this source. The QPO+Jet+Soft X-ray rise phase is shaded in blue. We refer \citet{laha2025} to the details. The descriptions of the different panels are the same as in Figure \ref{fig:swift_lc_zoom}
    \label{fig:swift_lc_main}.}
\end{figure*}

\newpage

\section{Warm absorber simulation for RGS detector }\label{app:sim}

\begin{figure}[h!]
\centering
\hbox{
\includegraphics[width=0.5\textwidth]{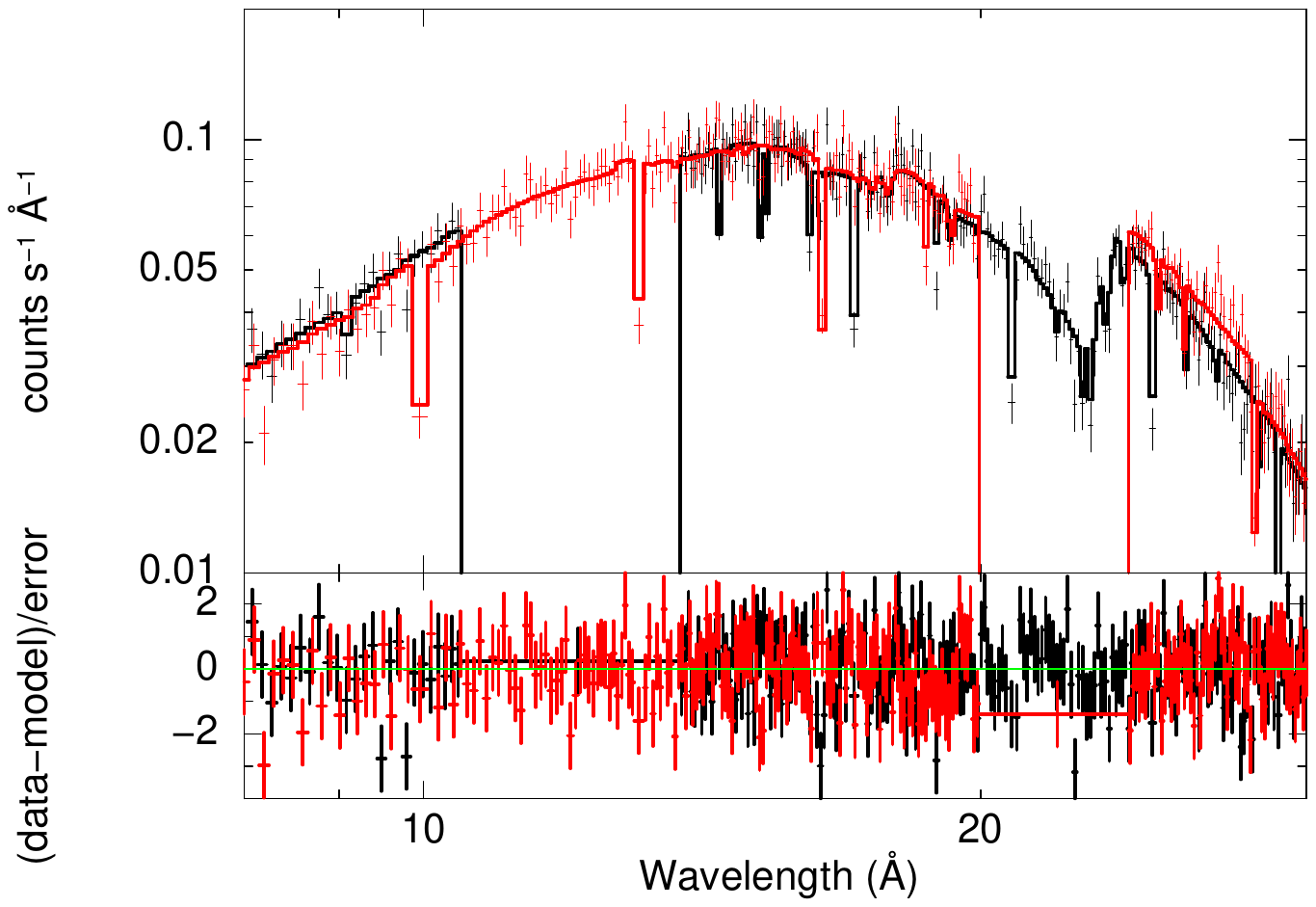 }
\includegraphics[width=0.5\textwidth]{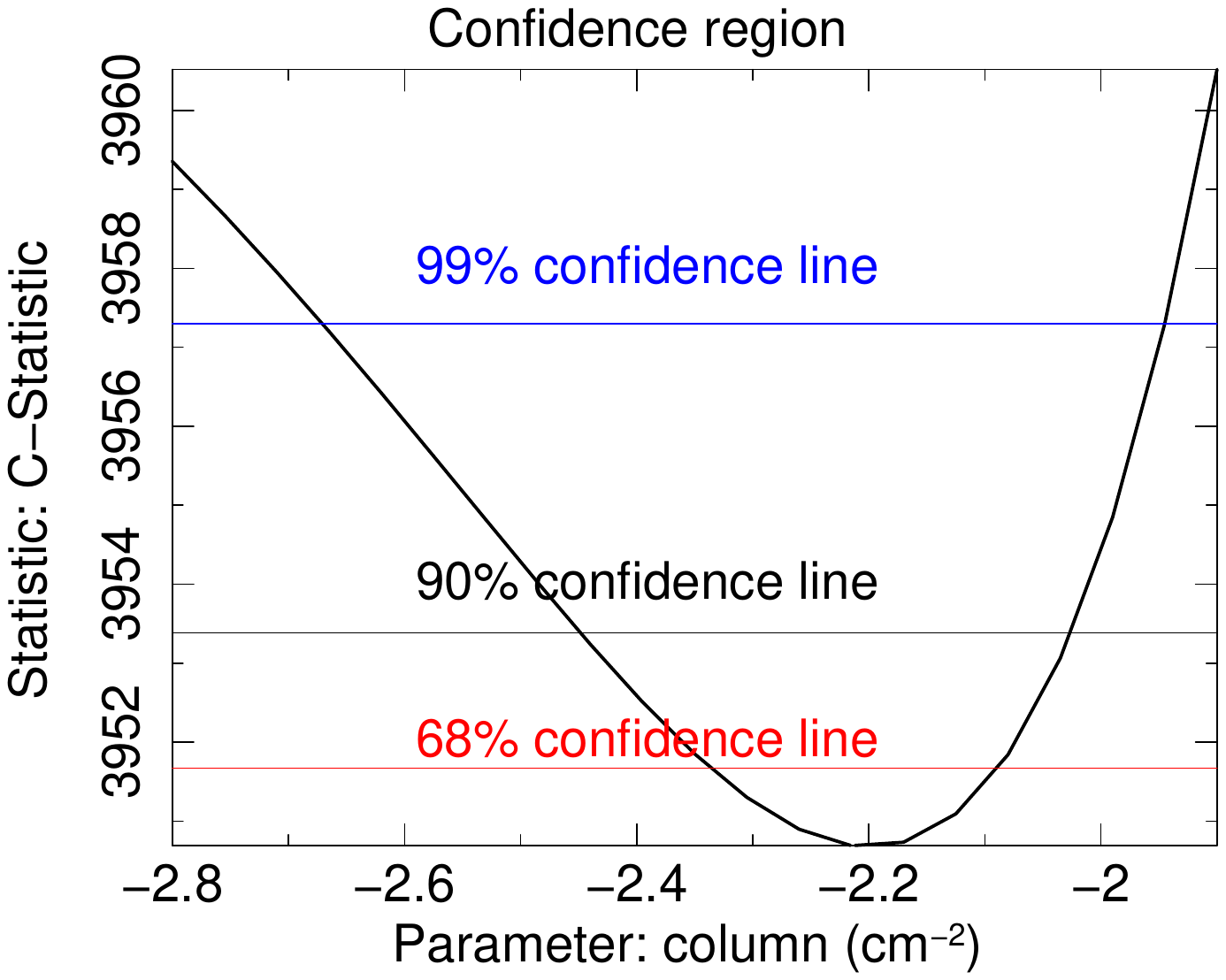}
    }
\caption{Simulated RGS1 and RGS2 \xmm{} spectra for the warm absorber parameters column of $5 \times 10^{19}$ cm$^{-2}$, log$(\xi) \sim 2$ and $v_{turb}=500 \ \mathrm{km \ s}^{-1}$. Right: X axes are converted to a column density as $10^{22-a}$, where a is the number in the x-axis. The value, -2.2, corresponds to $10^{22-2.2}$. This is equivalent to $ \sim 5 \times 10^{19}$ cm$^{-2}$.} 

\label{fig:sim_spec_rgs_wa}
\end{figure}

We have performed a simulation of WA in RGS spectra \ref{fig:sim_spec_rgs_wa} using a model we generated as discussed in Section \ref{sec:creating_ionized_absorber_model}. The goal of this simulation is to demonstrate that a warm absorber with a minimum of $5\times 10^{19}\cmsqi$ for a given log$(\xi) \sim 2/\xiunit$ and $v_{turb}=500 \ \mathrm{km \ s}^{-1}$ can be detected and constrained using RGS spectra. Any detection above this column is likely physical. 

\newpage
\section{UFO}

We have also tested the presence of the UFO, which is described in detail in Section \ref{subsubsec:ufo}. The fit parameters, including the improvement in the fit statistics, are presented here in Table \ref{tab:ufo_fits} in the Appendix.

\begin{table*}[h]
\centering
\caption{Best-fit parameters obtained from simultaneous fitting of the RGS1, RGS2, and PN spectra
for individual observations for the model \texttt{tbabs*ztbabs*warmabs*(bbody+pow)}.
The last column shows the improvement in fit statistic ($\Delta C$) upon addition of a UFO component in the 
model for each individual observational epoch for the loss of two degrees of freedom.}
\label{tab:ufo_fits}
\begin{tabular}{cccccccccc}
\hline
\textbf{Date} & \textbf{Short ID} & \textbf{ztbabs} & \textbf{kT} & \textbf{kT Norm} & \textbf{$\Gamma$}
& \textbf{$\Gamma$ Norm} & \textbf{Cstat/dof} & \textbf{Cstat/dof} & \textbf{$\Delta C$} \\
 & & $(\times 10^{20}$ cm$^{-2})$ & (keV) & $(\times 10^{-4})$ & & $(\times 10^{-2})$ & (w/o WA) & (with WA) & \\
\hline
2011-05-20 & x0 & $3.33 \pm 0.67$ & $0.163 \pm 0.006$ & $0.46 \pm 0.04$ & $2.40 \pm 0.04$ & $0.27 \pm 0.01$ & 3607/3806 & 3618/3805 & -11 \\
2022-07-26 & x1 & $2.64 \pm 0.65$ & $0.126 \pm 0.008$ & $0.59 \pm 0.10$ & $2.76 \pm 0.04$ & $0.52 \pm 0.02$ & 4129/4049 & 4122/4048 & 7 \\
2022-07-28 & x2 & $1.98 \pm 0.62$ & $0.138 \pm 0.008$ & $0.59 \pm 0.08$ & $2.73 \pm 0.04$ & $0.53 \pm 0.02$ & 3546/3784 & 3545/3783 & 1 \\
2022-07-30 & x3 & $2.09 \pm 0.65$ & $0.143 \pm 0.009$ & $0.59 \pm 0.08$ & $2.74 \pm 0.04$ & $0.54 \pm 0.02$ & 3682/3818 & 3691/3817 & -10 \\
2022-08-01 & x4 & $1.19 \pm 1.02$ & $0.135 \pm 0.010$ & $0.69 \pm 0.09$ & $2.60 \pm 0.07$ & $0.48 \pm 0.03$ & 3879/4089 & 3904/4088 & -25 \\
2023-02-21 & x5 & $4.22 \pm 0.50$ & $0.148 \pm 0.003$ & $2.94 \pm 0.13$ & $3.08 \pm 0.04$ & $1.08 \pm 0.04$ & 4148/4035 & 4166/4034 & -18 \\
2023-08-07 & x6 & $5.03 \pm 0.46$ & $0.150 \pm 0.002$ & $4.86 \pm 0.15$ & $3.17 \pm 0.04$ & $1.32 \pm 0.05$ & 4289/4073 & 4303/4072 & -14 \\
2024-03-04 & x7 & $4.79 \pm 0.53$ & $0.154 \pm 0.002$ & $5.77 \pm 0.18$ & $3.17 \pm 0.04$ & $1.38 \pm 0.06$ & 4304/4045 & 4300/4044 & 4 \\
2024-03-12 & x8 & $4.74 \pm 0.47$ & $0.155 \pm 0.002$ & $5.59 \pm 0.15$ & $3.19 \pm 0.04$ & $1.35 \pm 0.05$ & 4307/4103 & 4325/4102 & -18 \\
2024-07-19 & x9 & $5.06 \pm 0.48$ & $0.153 \pm 0.002$ & $5.96 \pm 0.18$ & $3.19 \pm 0.04$ & $1.44 \pm 0.05$ & 4228/4014 & 4208/4013 & 21 \\
2024-07-27 & x10 & $5.16 \pm 0.48$ & $0.153 \pm 0.002$ & $5.89 \pm 0.17$ & $3.21 \pm 0.04$ & $1.43 \pm 0.05$ & 4226/4025 & 4237/4024 & -11 \\
2024-10-13 & x11 & $4.35 \pm 0.67$ & $0.154 \pm 0.003$ & $5.85 \pm 0.23$ & $3.13 \pm 0.06$ & $1.35 \pm 0.07$ & 4212/3942 & 4211/3941 & 1 \\
2024-10-21 & x12 & $4.59 \pm 0.50$ & $0.154 \pm 0.002$ & $5.93 \pm 0.17$ & $3.13 \pm 0.04$ & $1.43 \pm 0.06$ & 4384/4090 & 4373/4089 & 11 \\
2025-01-19 & x13 & $4.52 \pm 0.56$ & $0.156 \pm 0.002$ & $5.98 \pm 0.18$ & $3.16 \pm 0.05$ & $1.39 \pm 0.06$ & 4312/4028 & 4303/4027 & 9 \\
2025-01-25 & x14 & $5.09 \pm 0.52$ & $0.154 \pm 0.002$ & $5.81 \pm 0.17$ & $3.21 \pm 0.04$ & $1.37 \pm 0.06$ & 4481/4043 & 4483/4042 & -2 \\
2025-04-30 & x15 & $4.73 \pm 0.52$ & $0.155 \pm 0.002$ & $5.88 \pm 0.17$ & $3.21 \pm 0.04$ & $1.37 \pm 0.06$ & 4227/4031 & 4218/4030 & 9 \\
2025-05-02 & x16 & $3.84 \pm 0.57$ & $0.155 \pm 0.002$ & $6.01 \pm 0.18$ & $3.11 \pm 0.05$ & $1.27 \pm 0.06$ & 4272/3995 & 4267/3994 & 5 \\
2025-05-05 & x17 & $4.70 \pm 0.58$ & $0.155 \pm 0.002$ & $5.99 \pm 0.19$ & $3.18 \pm 0.05$ & $1.43 \pm 0.07$ & 4294/3994 & 4308/3993 & -14 \\
\hline
\end{tabular}
\textbf{Note}: The UFO is created using the different set of parameters in warmabs model. The UFO model component used here had the ionization parameter and column density free to fit but bounded. The $log(\xi)$ had lower and upper bound $(2-4.5)/\xiunit$  whereas $N_H^{UFO}$ was bound within $(10^{22}-10^{24}) \cmsqi$. While velocity was frozen at 0.2c, where c is the speed of light. This is the typical velocity for a UFO. The initial parameters for $log(\xi)$ and $N_H^{UFO}$ were $3.4 /\xiunit$, $10^{23} \cmsqi$ respectively. The turbulent speeds were taken to be $10000 \kms$. The positive $\Delta C$ value in the last column shows the improvement in the fit after adding the WA component, while a negative $\Delta C$ means the fit worsened by adding the WA component. 
\end{table*}

\newpage
\section{Emission lines below 2 keV}\label{app:emission_line}
\begin{figure*}
\begin{minipage}{0.7\textwidth}
\includegraphics[width=\textwidth, height=10cm]{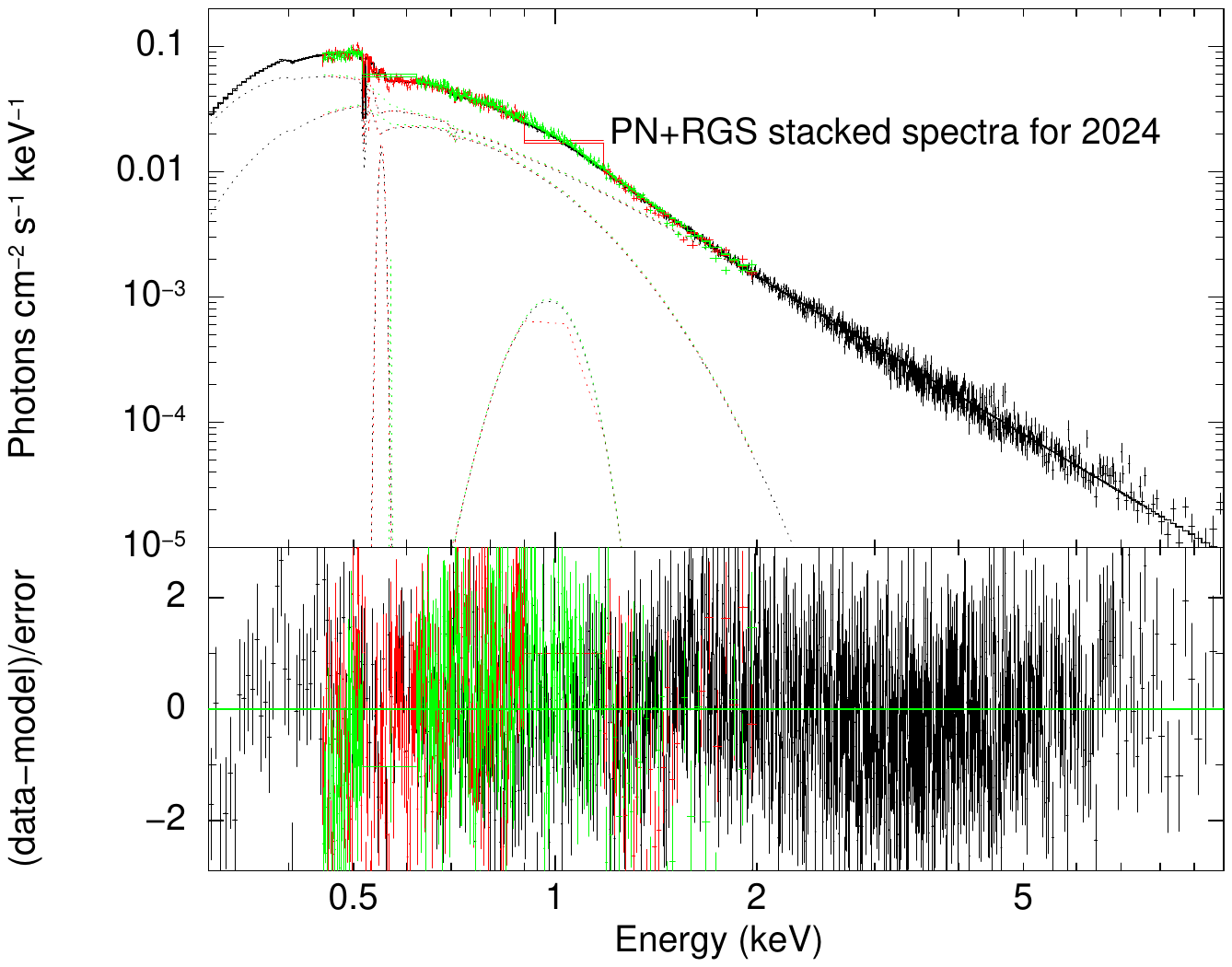}

\end{minipage}
\hspace{-1cm}
\begin{minipage}{0.3\textwidth}
\includegraphics[width=\textwidth,height=4cm]{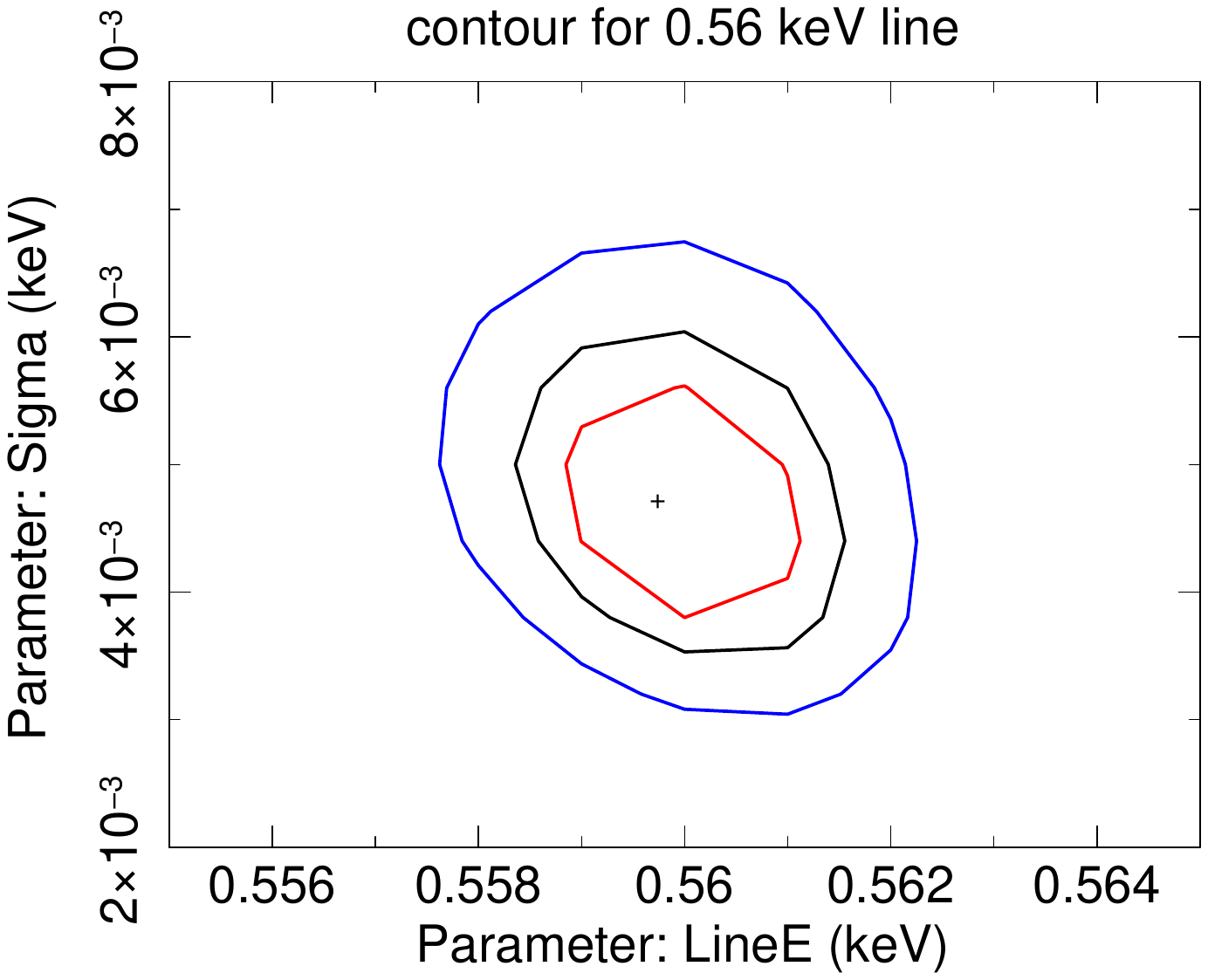}
\\[1mm]
\includegraphics[width=\textwidth,height=4cm]{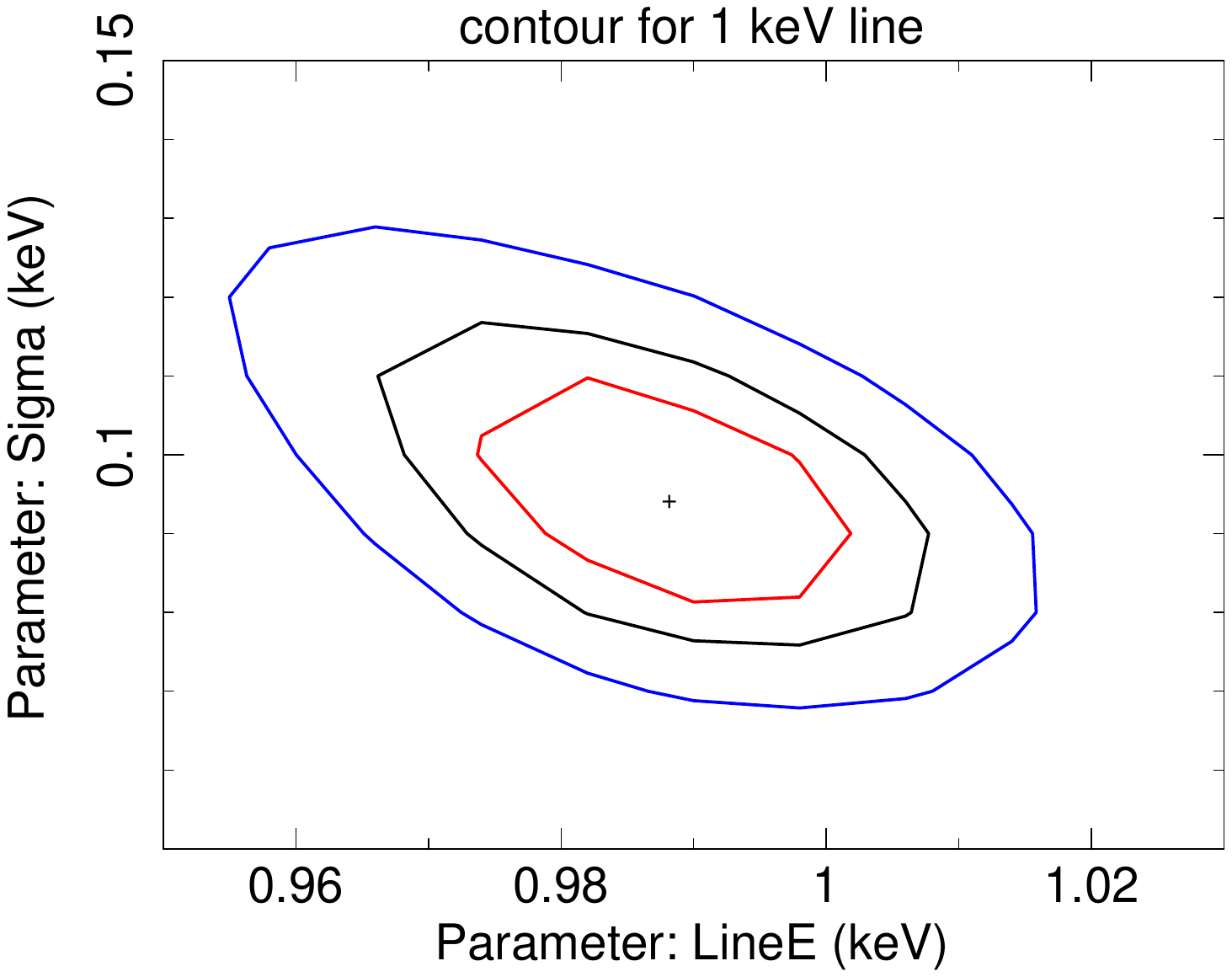}
\end{minipage}
\caption{Left: The best-fitted unfolded spectrum obtained from the simultaneous fit of PN, RGS1 and RGS2 of the stacked spectra for 2024. ${\tt tbabs\times ztbabs \times(zgauss+zgauss+bbody+pow)}$ model was used used to fit the spectra. Red and green colors correspond to RGS1 and RGS2 whereas black is for pn. The different model components are presented in the dotted lines. Top Right: The confidence contour using c-statistics for the line at $\sim 0.56$ keV. Line sigma and energy are constrained within the 99 percent of the confidence. Bottom right: The confidence contour using c-statistics for the line at $\sim 1$ keV. Line sigma and energy are constrained within the 99 percent of the confidence. Color code for contours: red, green, and blue correspond to 68, 90, and 99 percent confidence intervals.}
\label{fig:pn_rgs_2024_stacked}
\end{figure*}

\begin{figure*}
\begin{minipage}{0.7\textwidth}
\includegraphics[width=\textwidth, height=9cm]{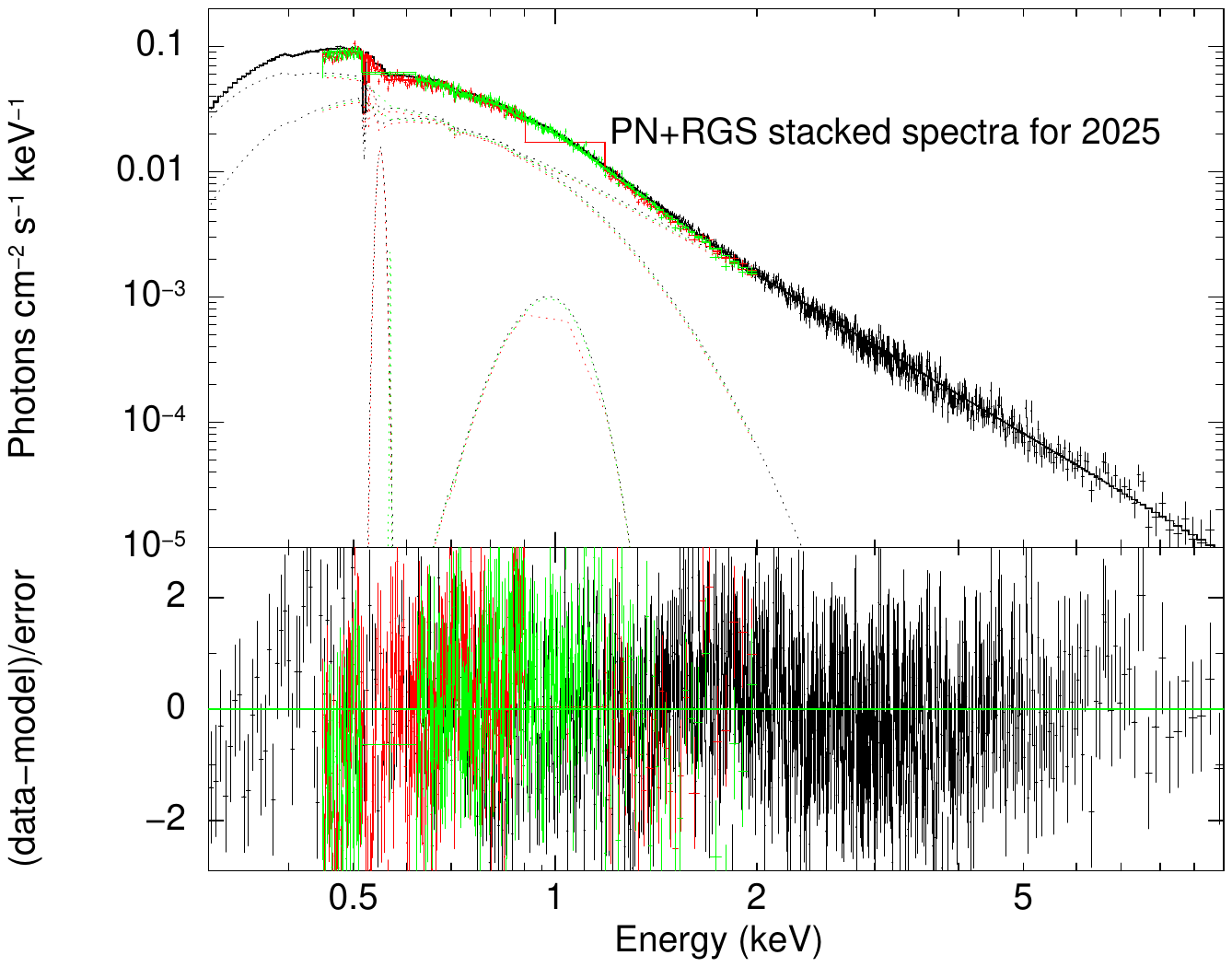}

\end{minipage}
\hspace{-2cm}
\begin{minipage}{0.3\textwidth}
\includegraphics[width=\textwidth,height=3.8cm]{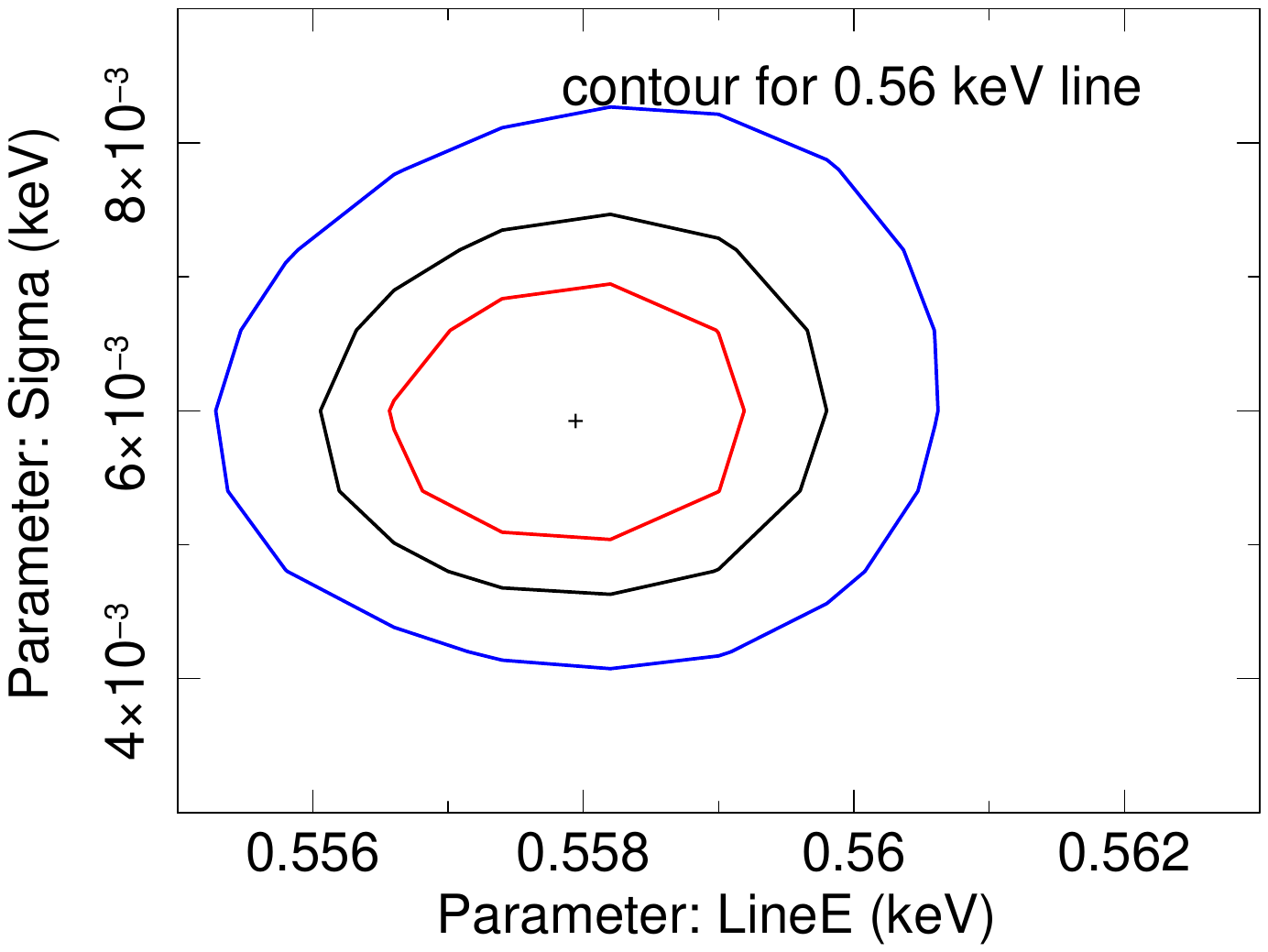}
\\[1mm]
\includegraphics[width=\textwidth,height=3.8cm]{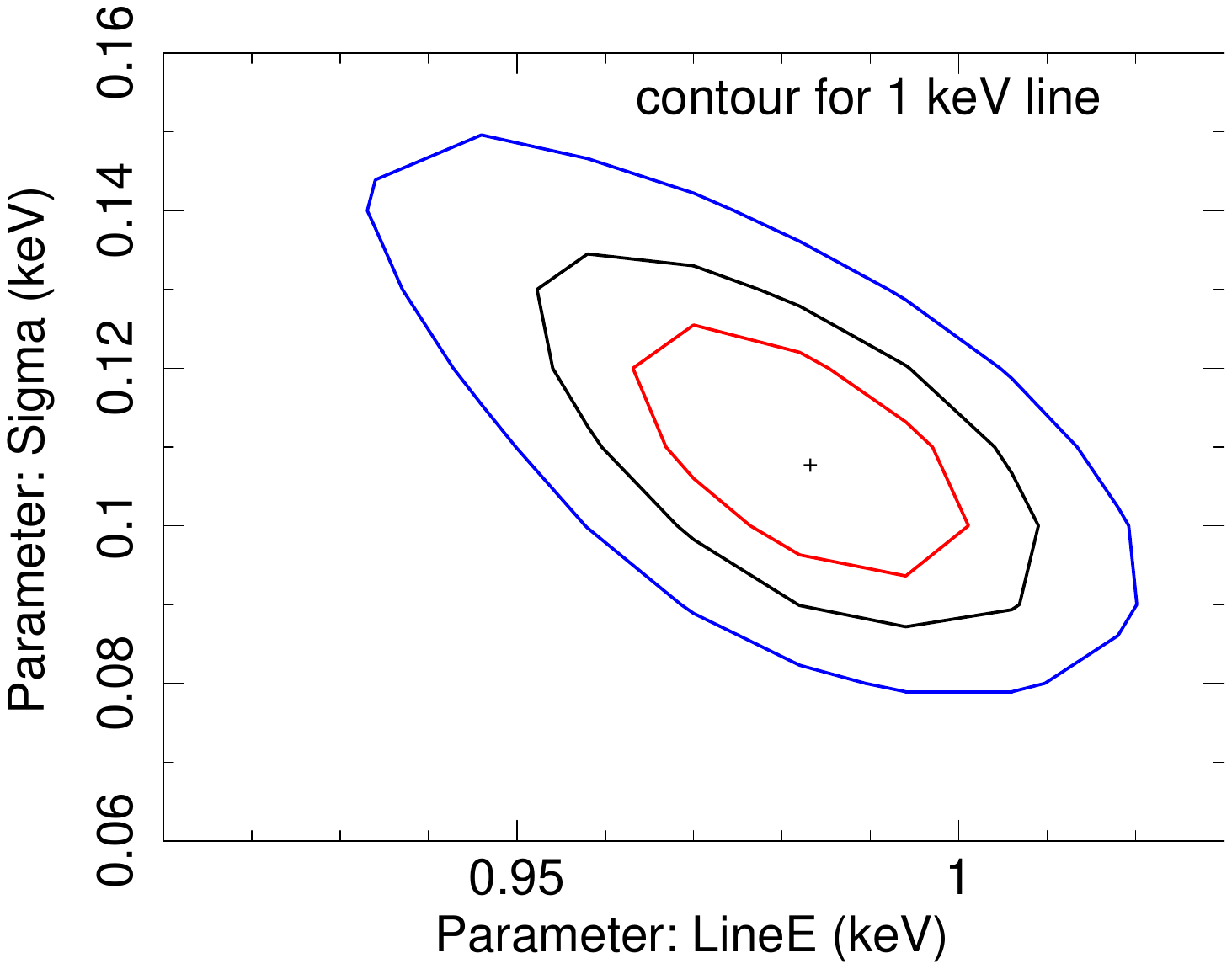}
\end{minipage}

\caption{Left: The best-fitted unfolded spectrum obtained from the simultaneous fit of PN, RGS1 and RGS2 of the stacked spectra from four epochs in 2025. Top Right: The confidence contour using c-statistics for the line at $\sim 0.56$ keV. Both line sigma and energy are constrained within the 99 percent confidence interval. Bottom right: The confidence contour using c-statistics for the line at $\sim 1$ keV. Both line sigma and energy are constrained within the 99 percent confidence interval. Color code: red, green, and blue correspond to 68, 90, and 99 percent confidence intervals.}
\label{fig:pn_rgs_2025_stacked}
\end{figure*}

We provide a detailed discussion of the soft X-ray emission lines at $\sim(0.47$–$0.57),\kev$ in Section~\ref{subsubsection:emission_line_indiv_obs}. For completeness, the contour plots illustrating the detection significance of the $\sim 0.47\kev$,  $\sim 0.57\kev$ and $\sim 0.57\kev$ features for each epoch are shown in Appendix Figures~\ref{fig:pn_rgs_2022_stacked}, \ref{fig:pn_rgs_2023_stacked}, \ref{fig:pn_rgs_2024_stacked}, and \ref{fig:pn_rgs_2025_stacked}. In addition, Figure~\ref{Figure:pn_emis_stacked_data_by_model_ratio} presents the corresponding ratio for data and models including only the continuum and for those incorporating the three Gaussian emission components, providing a clear visualization of the emission features and improvement in the fits.

\begin{figure*}
\begin{minipage}{0.7\textwidth}
\includegraphics[width=\textwidth, height=9cm]{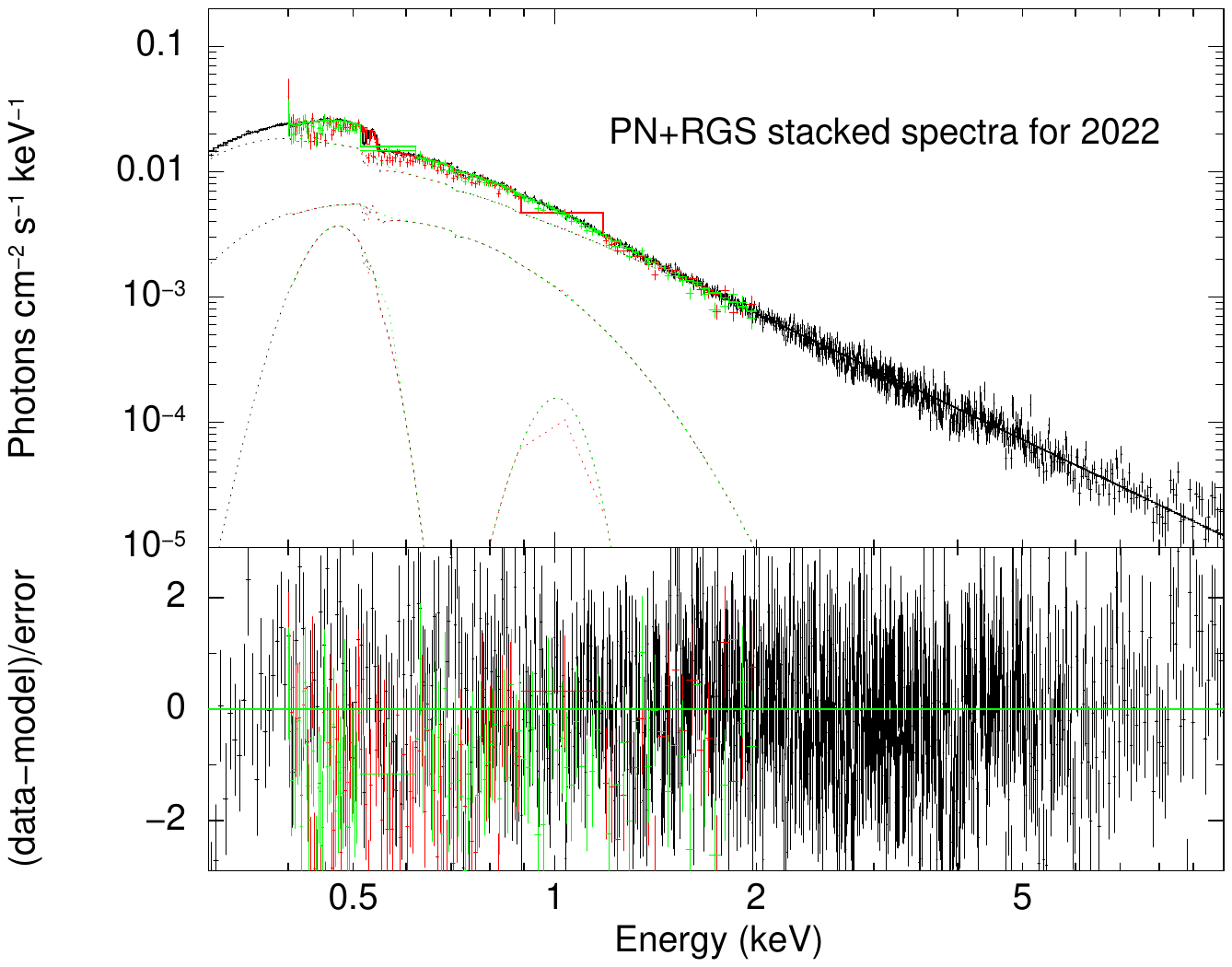}

\end{minipage}
\hspace{-2cm}
\begin{minipage}{0.3\textwidth}
\includegraphics[width=\textwidth,height=3.8cm]{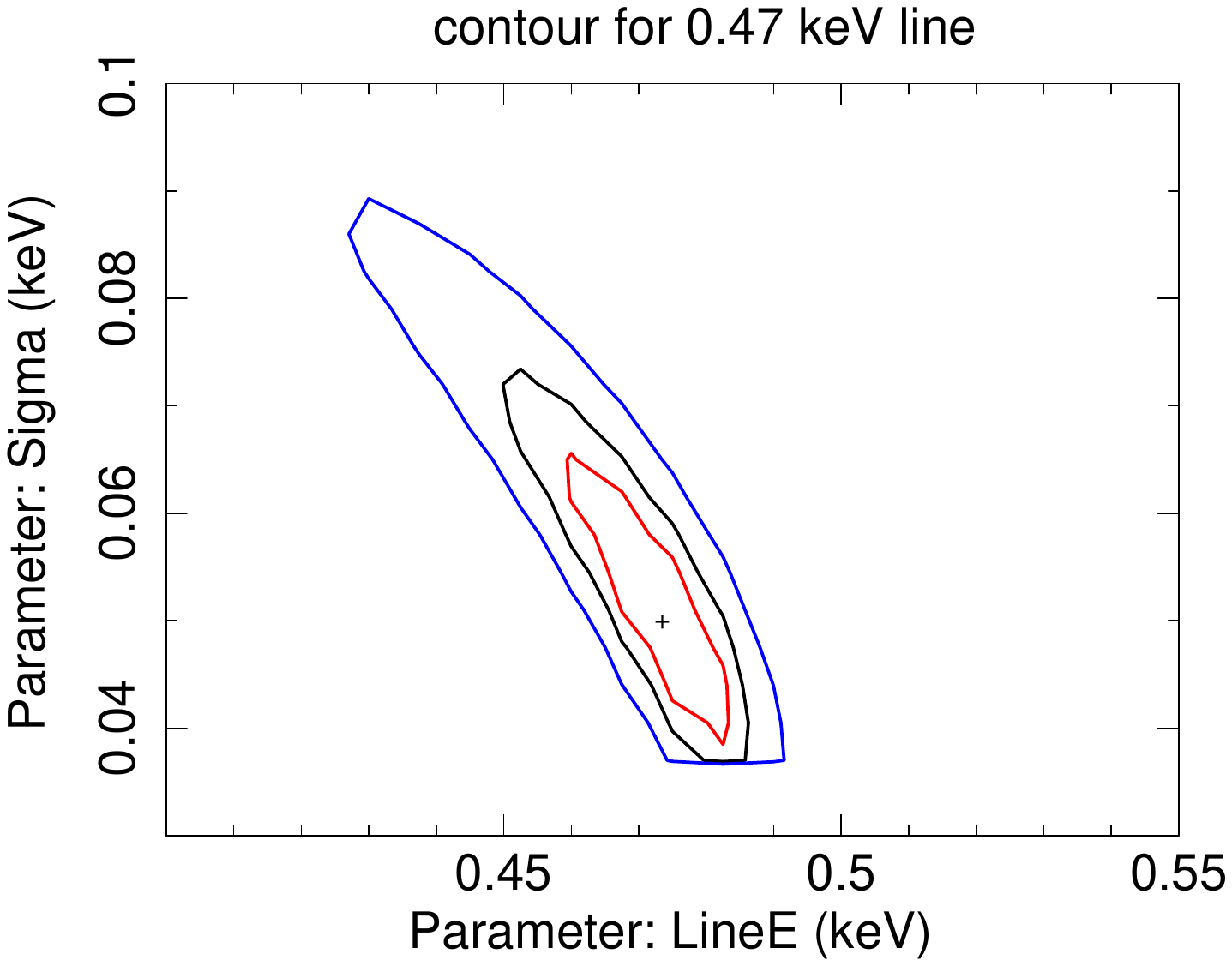}
\\[1mm]
\includegraphics[width=\textwidth,height=3.8cm]{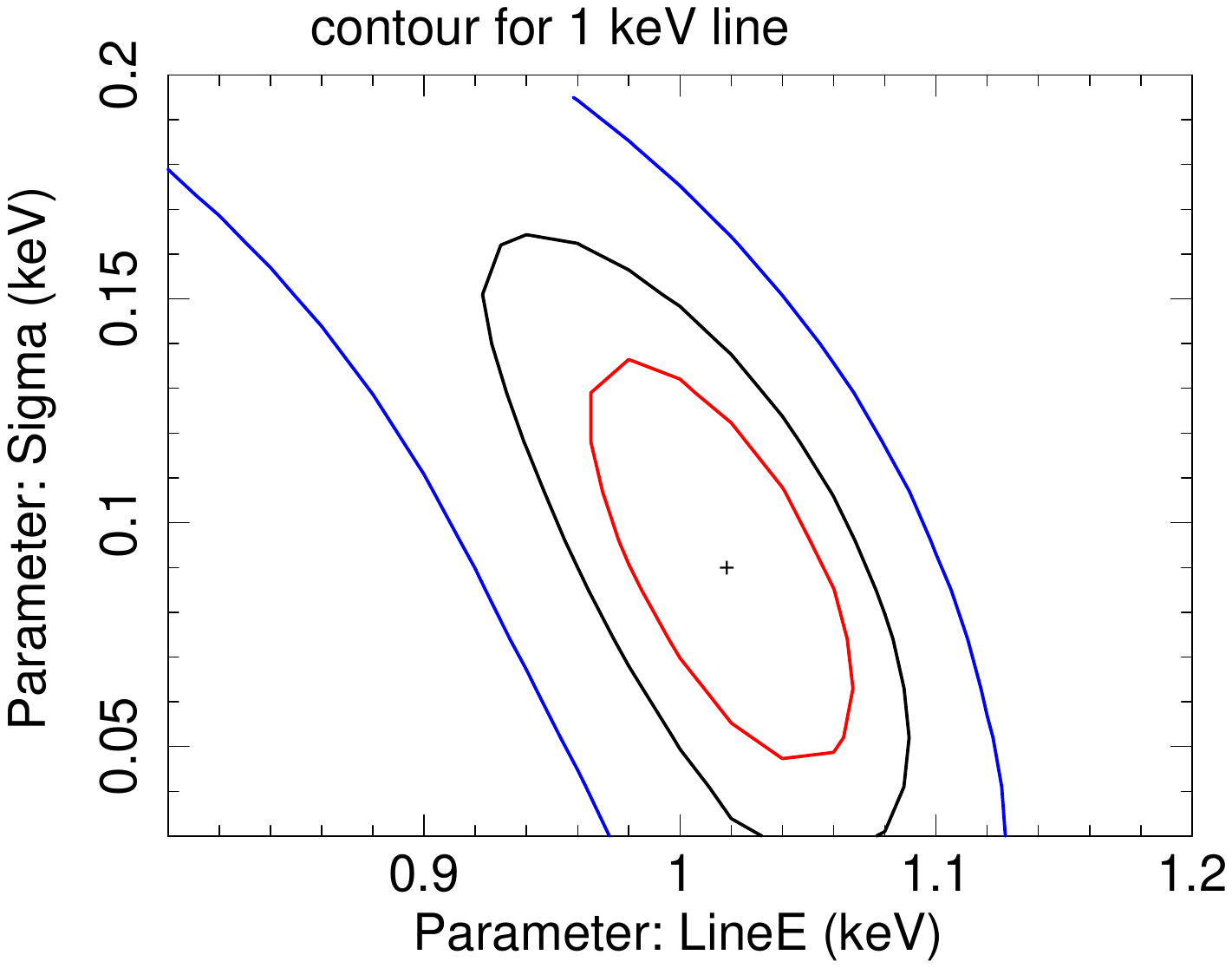}
\end{minipage}

\caption{Left: The best-fitted unfolded spectrum obtained from the simultaneous fit of PN, RGS1 and RGS2 of the stacked spectra from four epochs in 2022. Top Right: The confidence contour using c-statistics for the line at $\sim 0.47$ keV. Both line sigma and energy are constrained within the 99 percent of the confidence. Bottom right: The confidence contour using c-statistics for the line at $\sim 1$ keV. Both line sigma and energy are constrained within the 90 percent of the confidence. Color code: red, green, and blue correspond to 68, 90, and 99 percent confidence intervals.}

\label{fig:pn_rgs_2022_stacked}
\end{figure*}

\begin{figure*}
\begin{minipage}{0.7\textwidth}
\includegraphics[width=\textwidth, height=9cm]{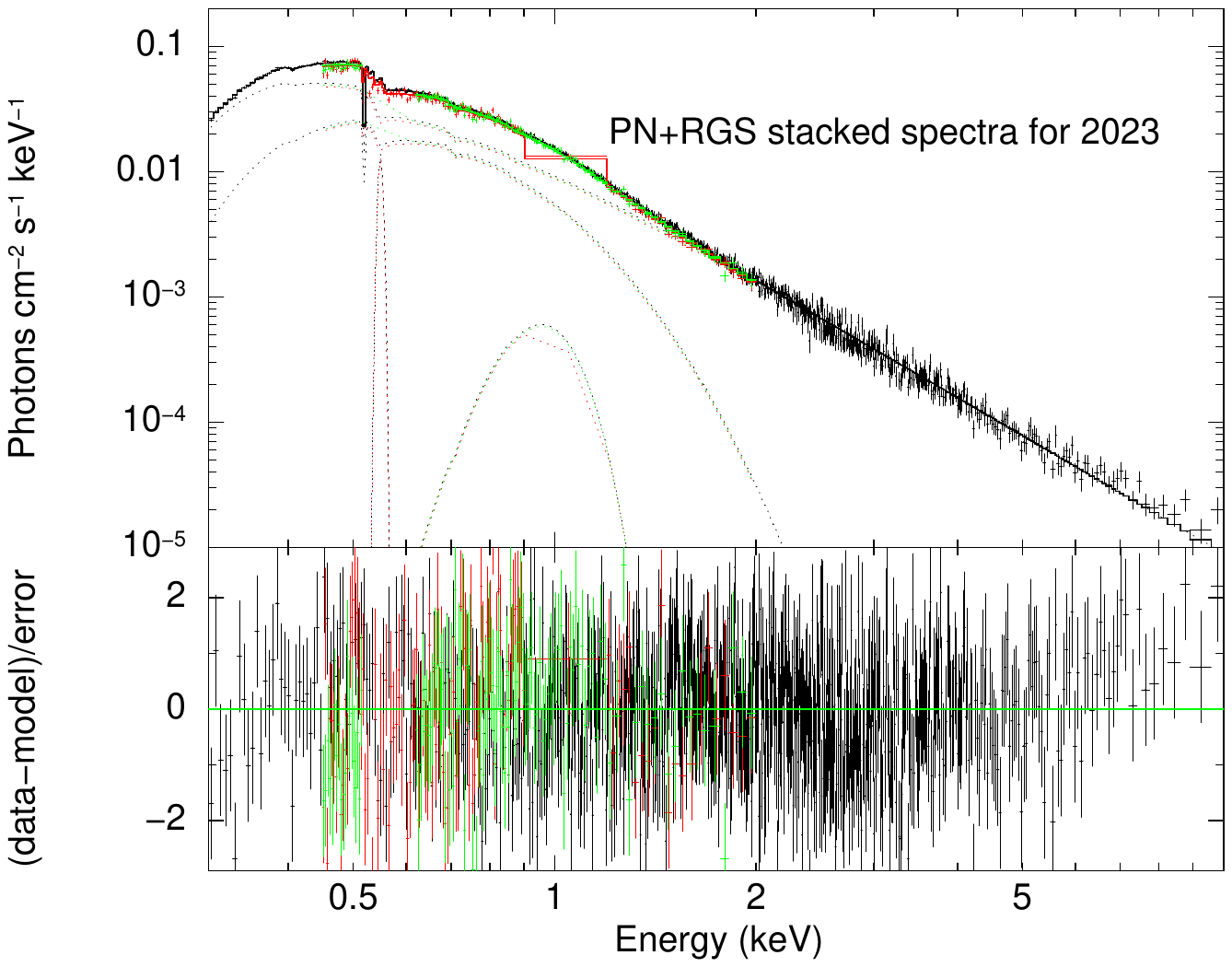}

\end{minipage}
\hspace{-2cm}
\begin{minipage}{0.3\textwidth}
\includegraphics[width=\textwidth,height=3.8cm]{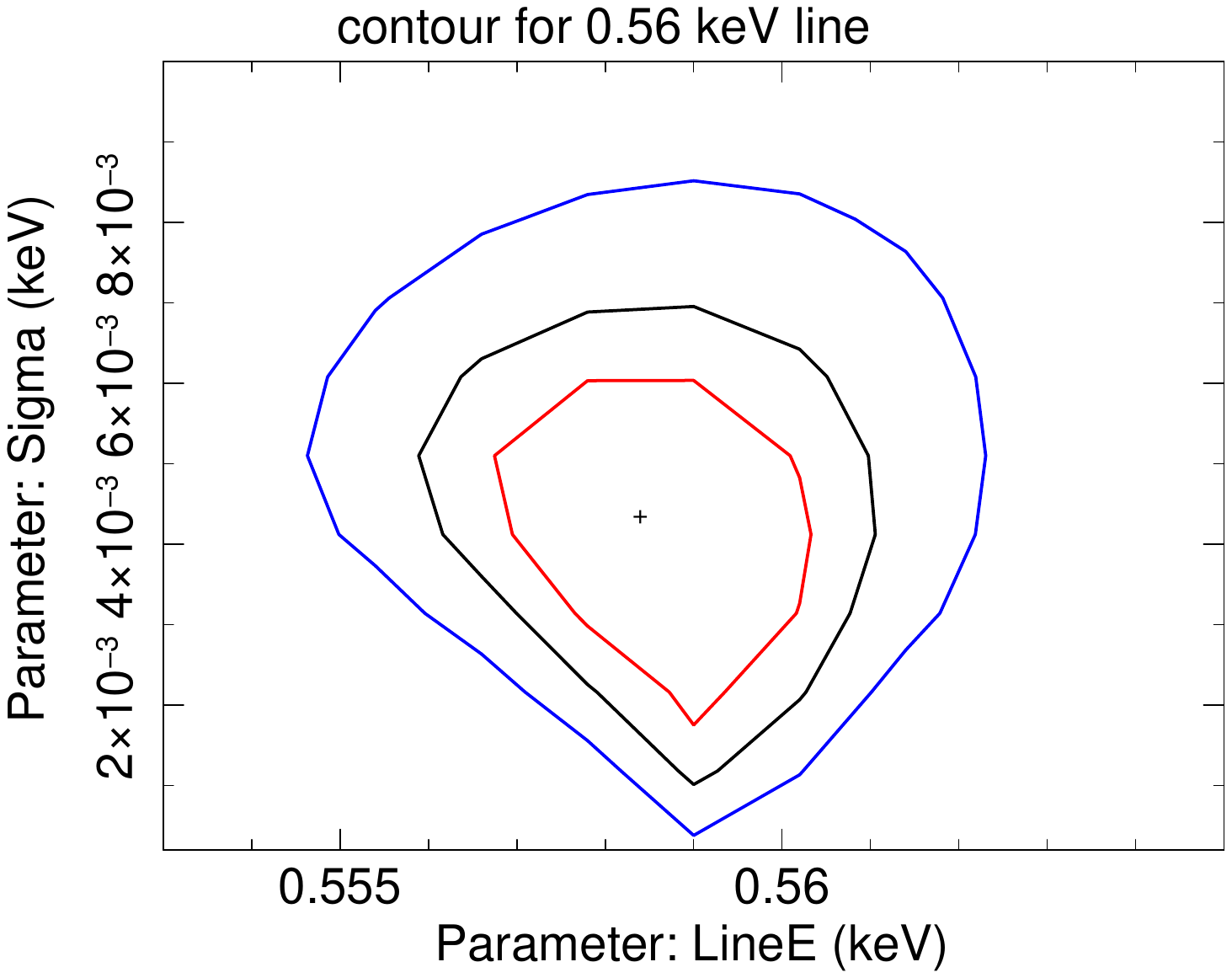}
\\[1mm]
\includegraphics[width=\textwidth,height=3.8cm]{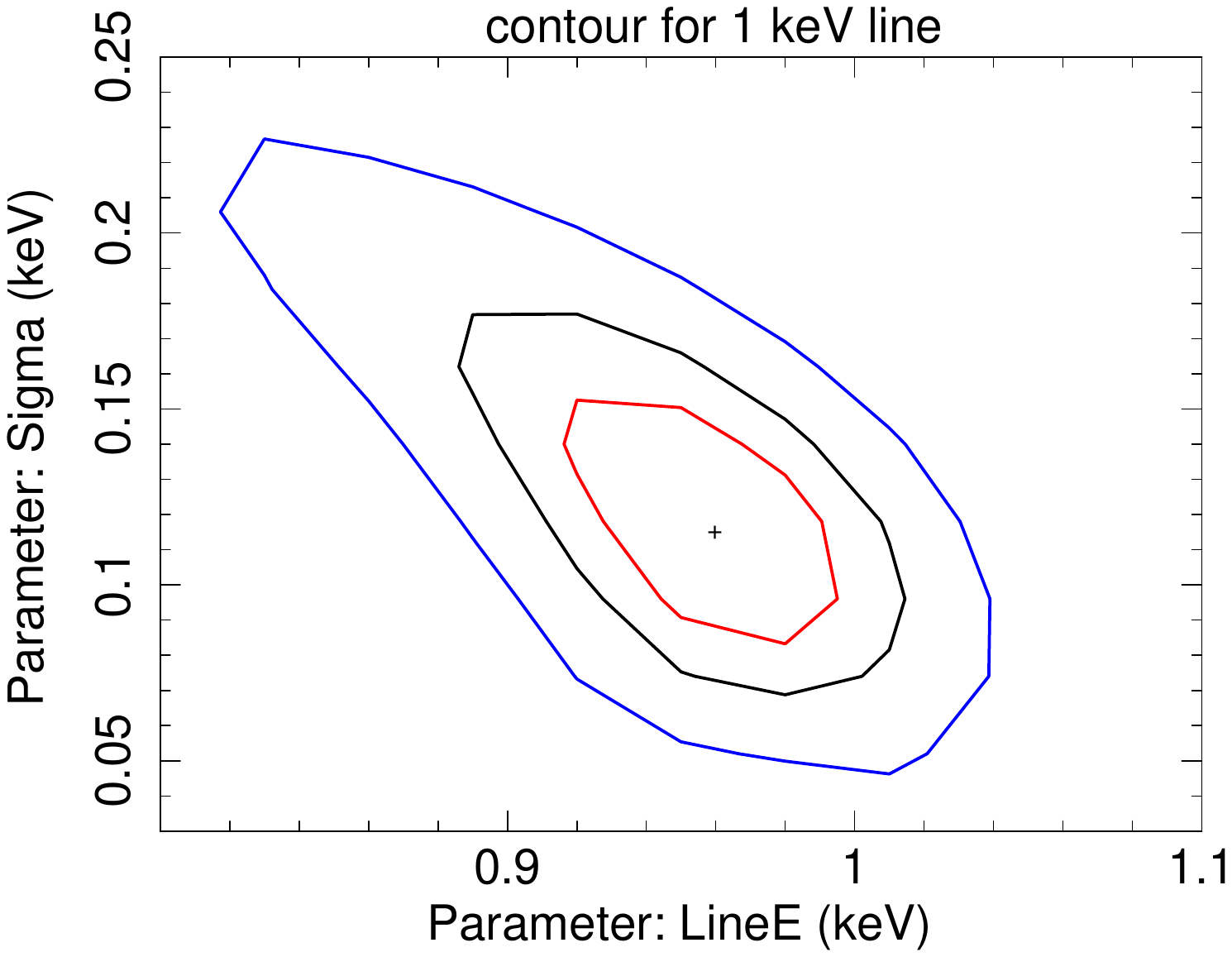}
\end{minipage}

\caption{Left: The best-fitted unfolded spectrum obtained from the simultaneous fit of PN, RGS1 and RGS2 of the stacked spectra from four epochs in 2023. Top Right: The confidence contour using c-statistics for the line at $\sim 0.56$ keV. Both line sigma and energy are constrained within the 99 percent of the confidence. Bottom right: The confidence contour using c-statistics for the line at $\sim 1$ keV. Both line sigma and energy are constrained within the 99 percent of the confidence. Color code: red, green and blue corresponds to 68, 90 and 99 percent confidence intervals.}
\label{fig:pn_rgs_2023_stacked}
\end{figure*}


\newpage

\section{Iron emission line}
\begin{figure*}
    \centering
    \vbox{
    \hbox{
    \includegraphics[width=0.32\linewidth]{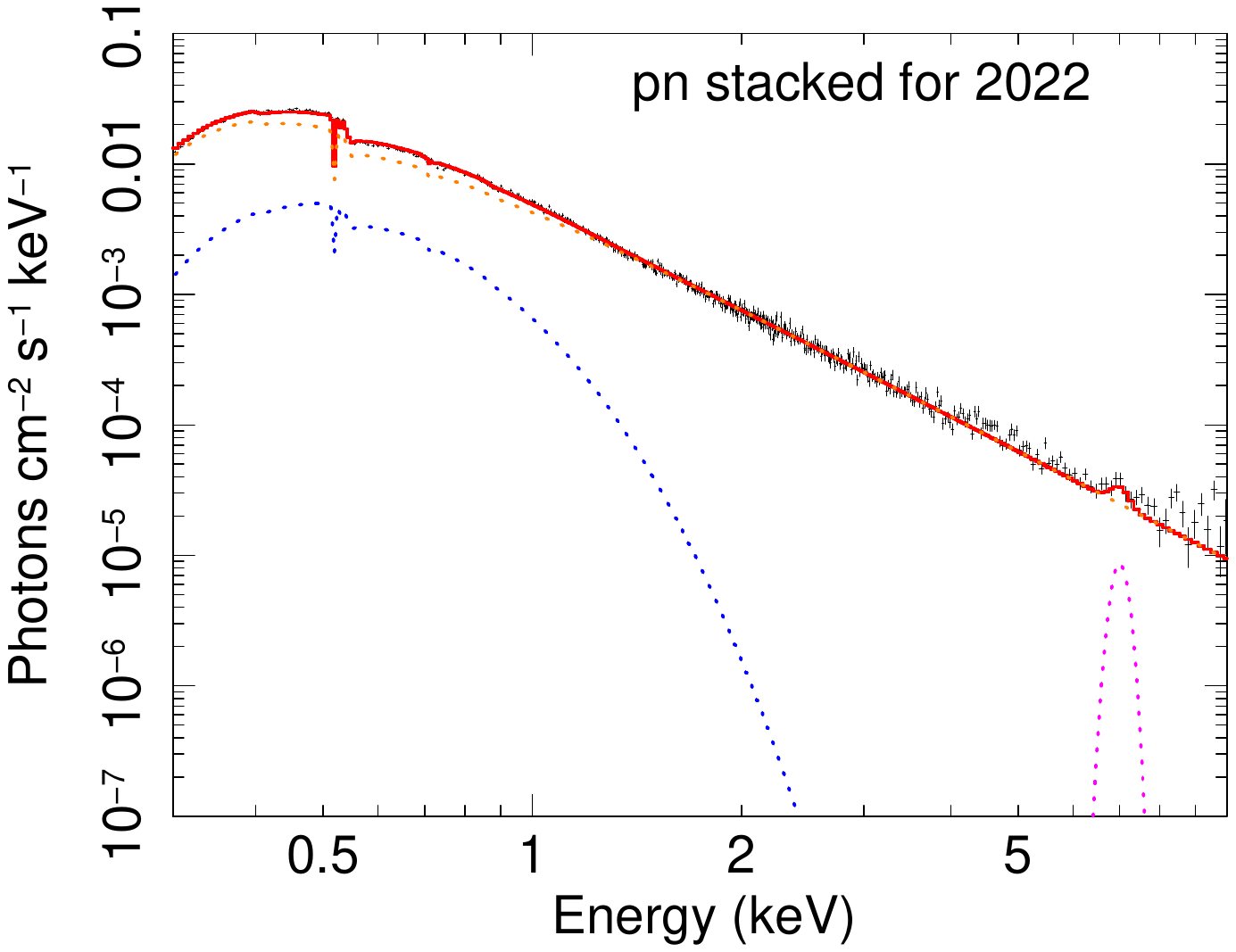}
    \hspace{-16mm}
    \includegraphics[width=0.32\linewidth]{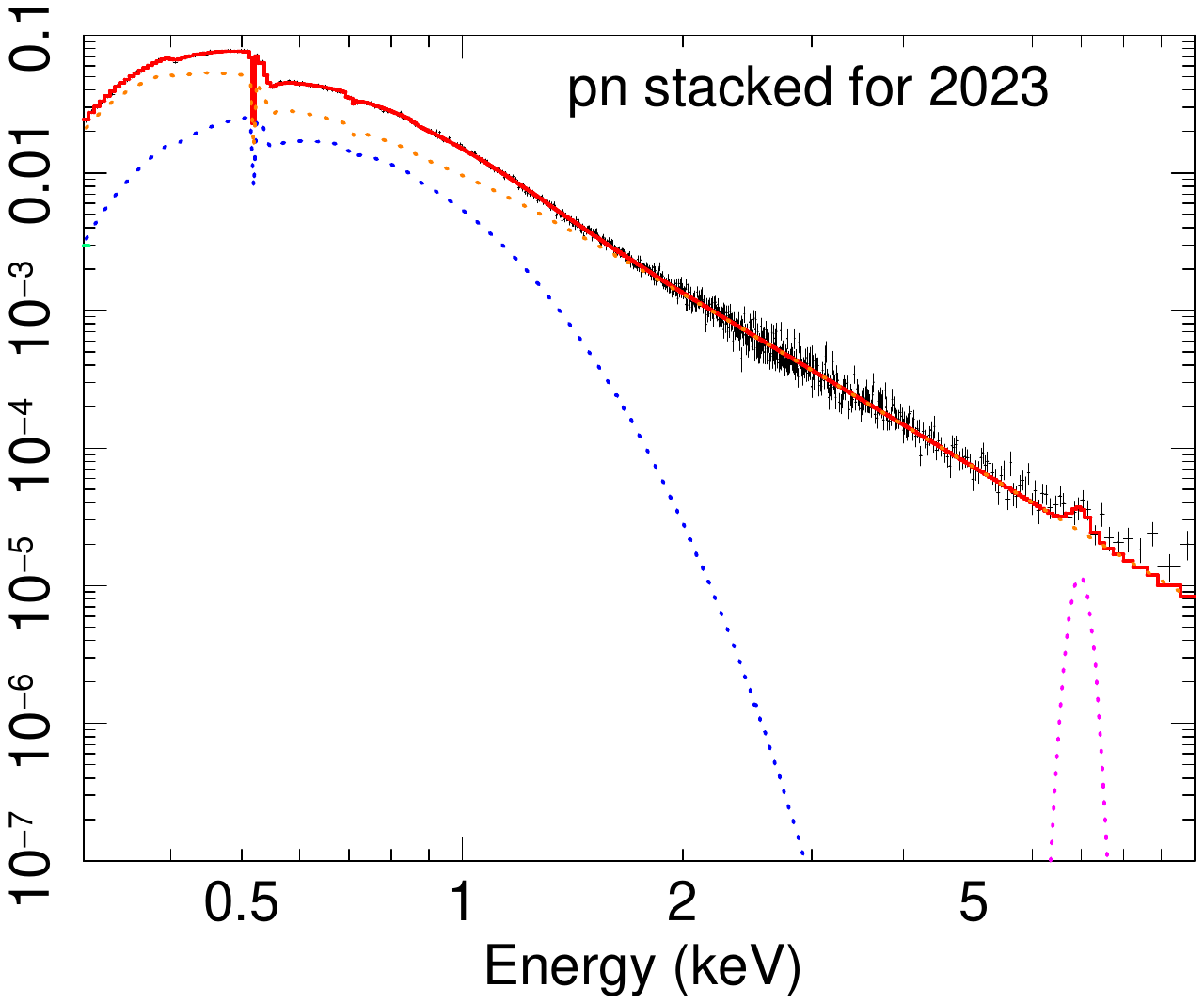}
    \hspace{-16mm}
    \includegraphics[width=0.32\linewidth]{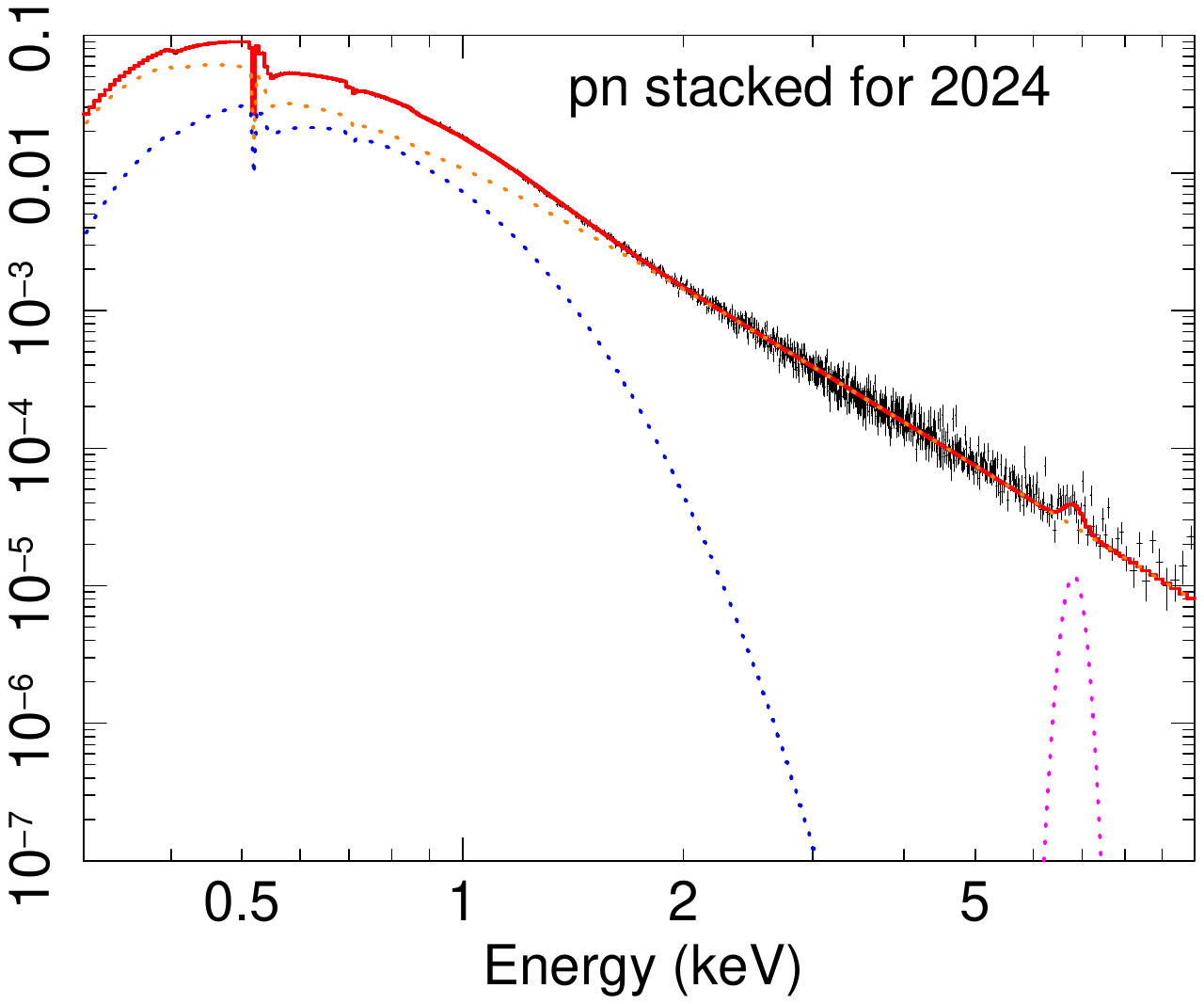}
    \hspace{-16mm}
    \includegraphics[width=0.32\linewidth]{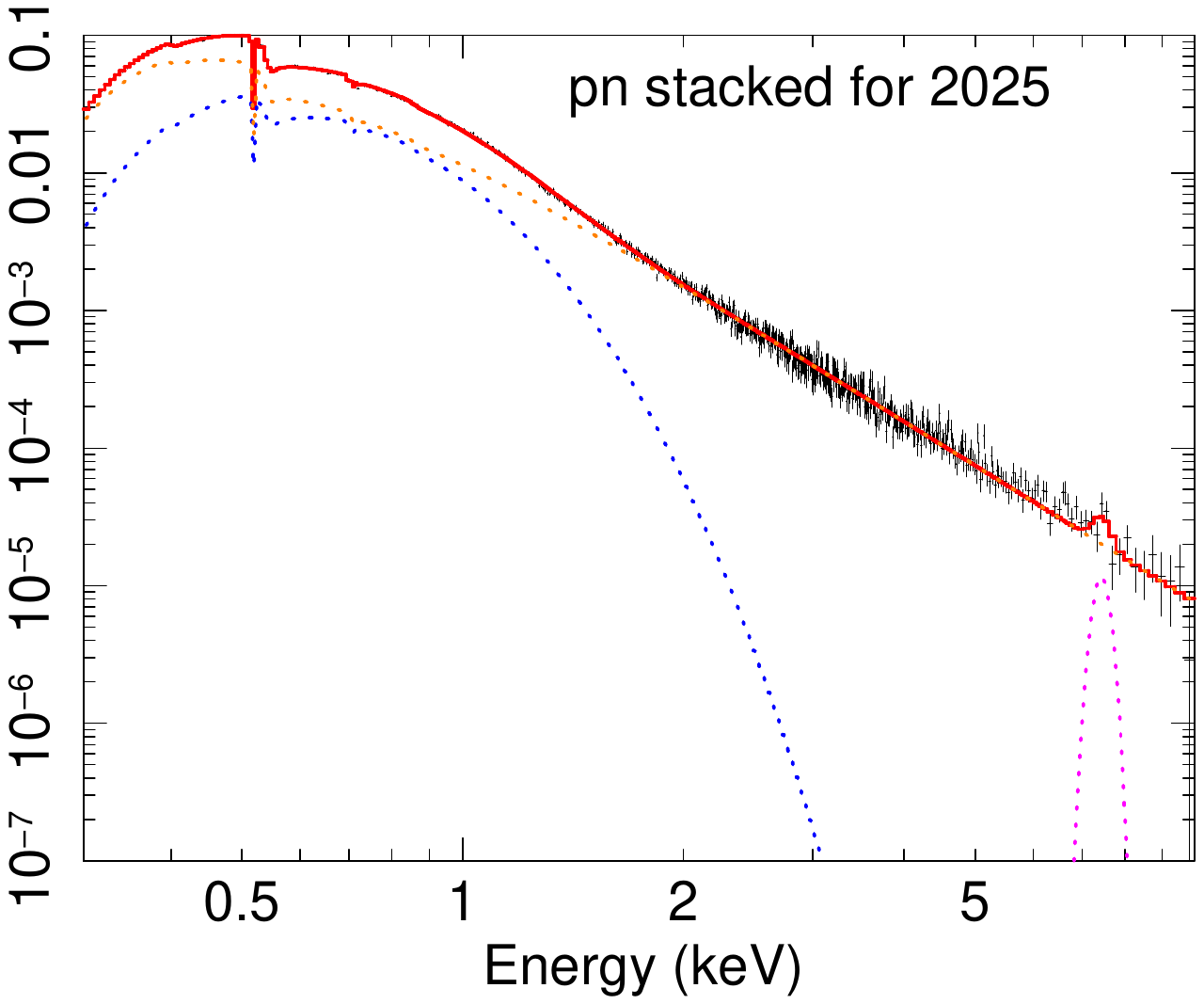}
    }

    \centering
    \hbox{
    \includegraphics[width=0.32\linewidth]{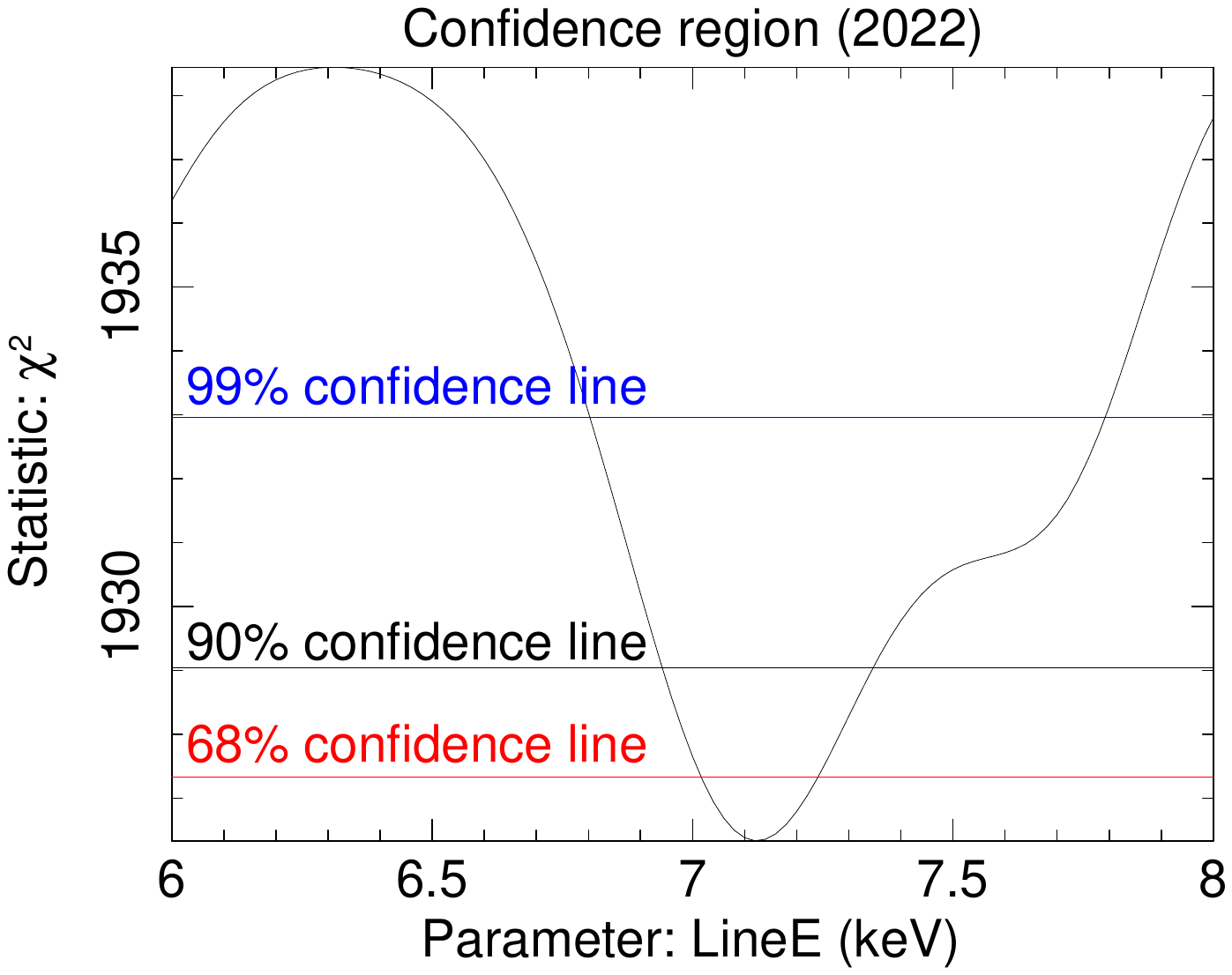}
    \hspace{-16mm}
    \includegraphics[width=0.32\linewidth]{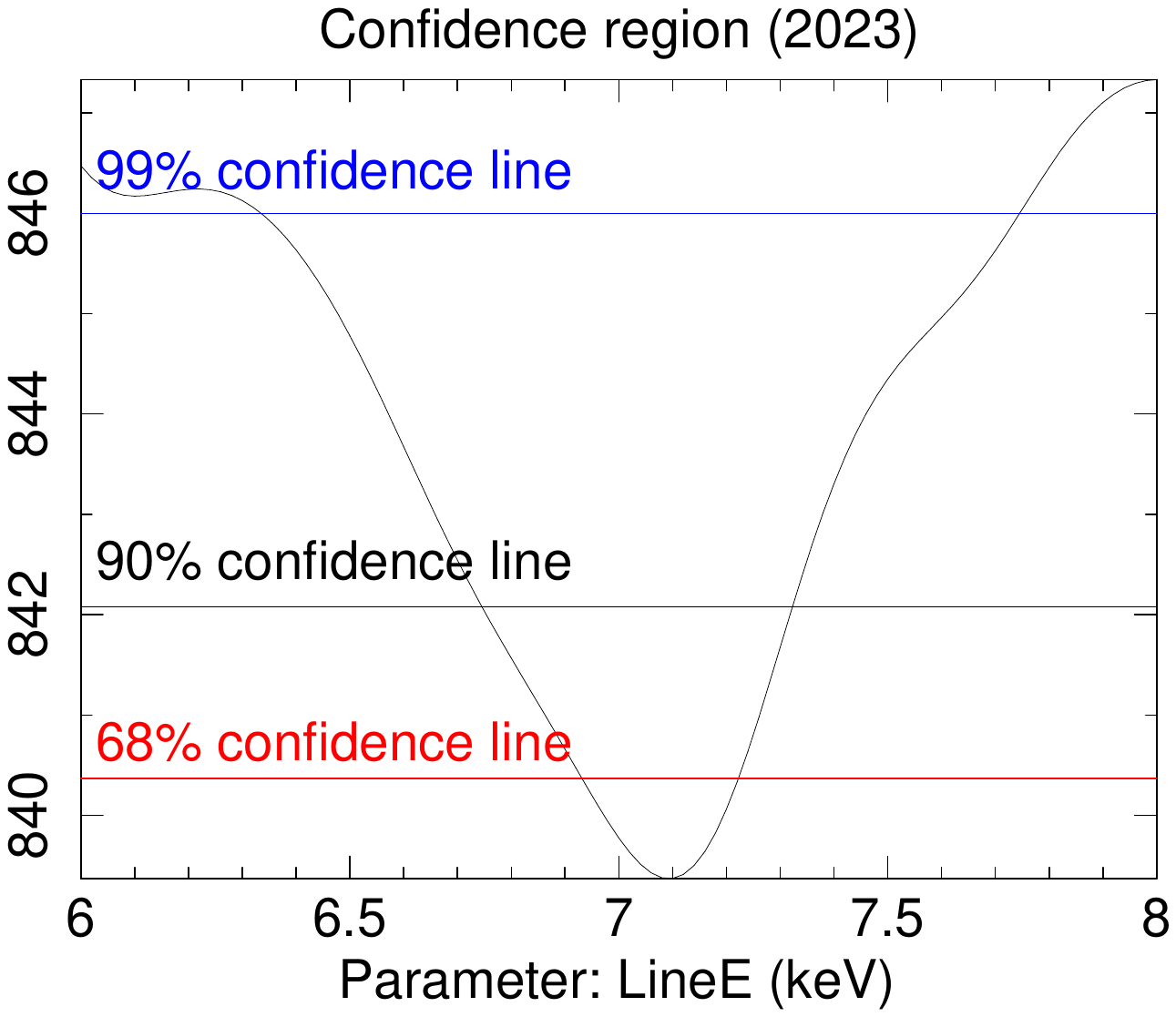}
    \hspace{-16mm}
    \includegraphics[width=0.32\linewidth]{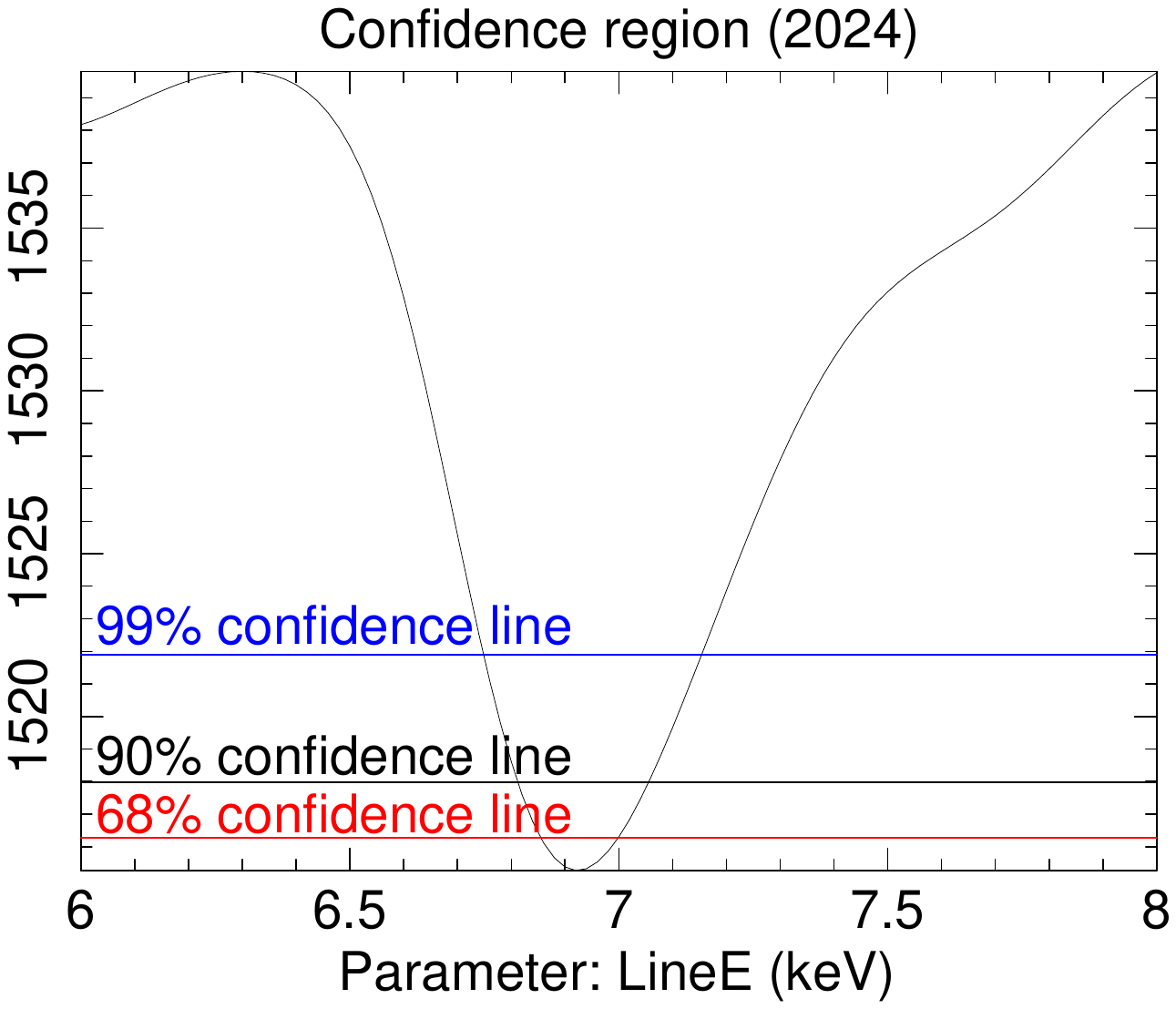}

    \hspace{-16mm}
    \includegraphics[width=0.32\linewidth]{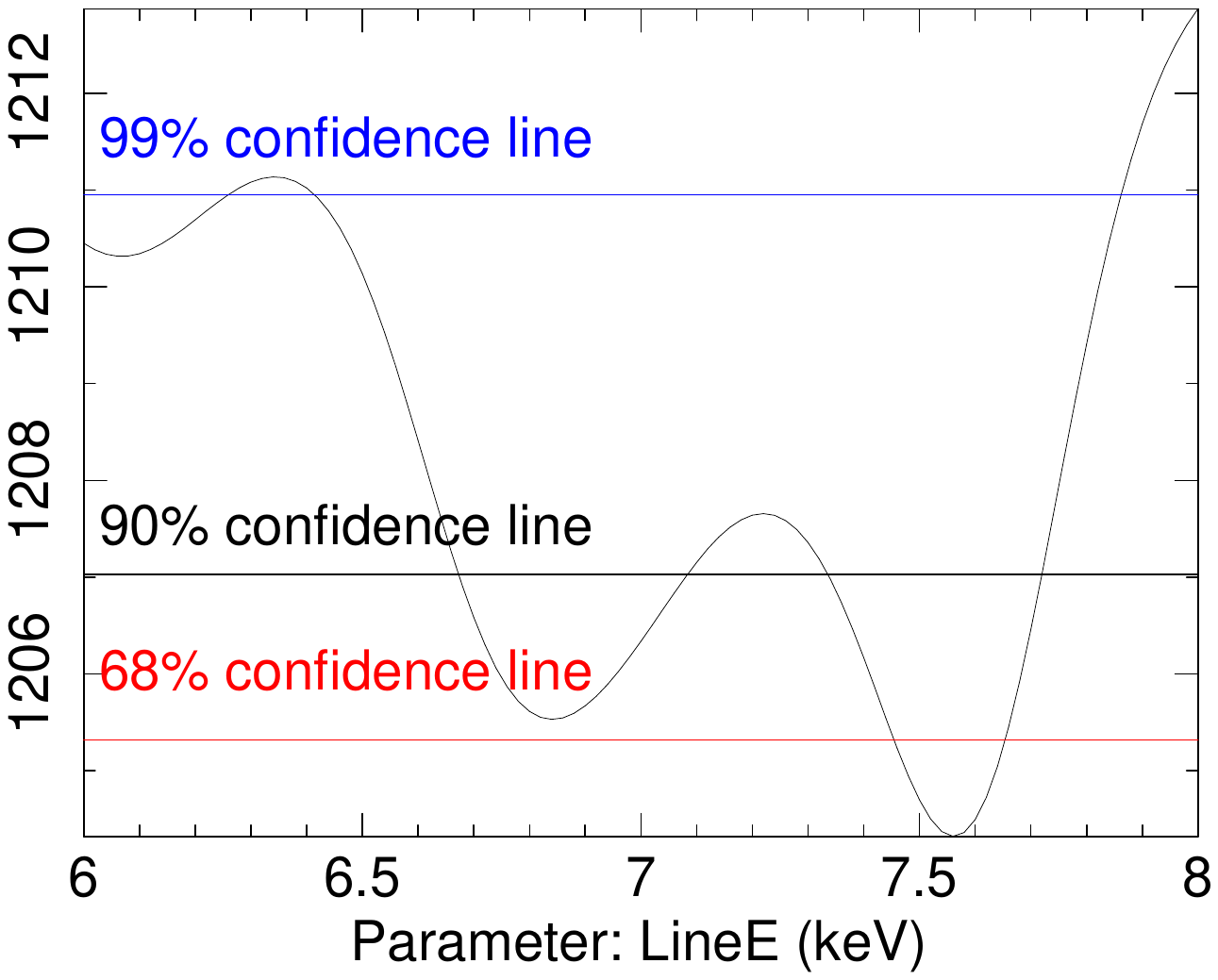}
    }
    }
    \caption{Upper column: Best-fitted unfolded stacked spectra for 2022, 2023, 2024, and 2025 from left to right. Here we added an iron line, with a width of 0.2 keV, which is typical in the case of AGN, froze the line width, and fitted the stacked spectra. In all the cases, the spectral fit improves significantly with the iron line.
    Lower Column: Confidence interval for the iron line detected in the stacked spectra for  2022, 2023, 2024, and 2025 from left to right. The green lines in each panel are the confidence levels for 68, 90, and 99 percent from the bottom to the top. All the lines are detected with 99 percent confidence given the fact that line width is fixed. In 2023, there were only two observations with a total exposure of $\sim 50 $ ks, and hence the S/N ratio is smaller compared to other years.}
    \label{fig:ironline_detection}
\end{figure*}

We have discussed the Fe line in depth in Sections \ref{subsection:iron_line_emission_analysis} and \ref{sec:reflection_discussion_combined} in the main text. Plot \ref{fig:ironline_detection} illustrates the level of detection of the Fe line, and Table \ref{Table:varying_FeK_widths}, which demonstrates the improvement of the fit statistics upon the addition of the Fe line in the stacked spectra with different line widths; they are shown here in the Appendix. The residual at $\sim 6-7.5 \kev$ can also be seen in Figure \ref{Figure:pn_emis_stacked_data_by_model_ratio}.

\begin{table*}[]
    \centering
    \caption{Fitted Fe\,K line parameters from the stacked \textit{XMM-Newton}/pn spectra for 2022 to 2025.}
    \begin{tabular}{lcccc}
    \hline
    \textbf{Parameter} & \textbf{2022} & \textbf{2023} & \textbf{2024} & \textbf{2025} \\ \hline
    \multicolumn{5}{l}{\textit{Gaussian line width = 0.2\,keV}} \\
LineE (keV) & 7.13$_{-0.17}^{+0.16}$ & 7.13$_{-0.25}^{+0.48}$ & 6.90$_{-0.13}^{+0.14}$ & 7.56$_{-0.16}^{+0.21}$ \\
Sigma (eV) & 200(f) & 200(f) & 200(f) & 200(f) \\
Norm ($10^{-6}$) & 6.12$_{-2.58}^{+2.60}$ & 6.44$_{-3.27}^{+3.42}$ & 6.66$_{-2.67}^{+2.72}$ & 6.11$_{-3.11}^{+3.16}$ \\
    \hline
    $\Delta \chi^2$ & 15 & 11 & 17 & 11 \\
    \hline
    \multicolumn{5}{l}{\textit{Gaussian line width = 0.4\,keV}} \\
LineE (keV) & 7.20$_{-0.26}^{+0.25}$ & 6.99$_{-0.36}^{+0.35}$ & 6.93$_{-0.24}^{+0.23}$ & 7.14$_{-0.44}^{+0.47}$ \\
Sigma (eV) & 400(f) & 400(f) & 400(f) & 400(f) \\
Norm ($10^{-6}$) & 9.27$_{-3.58}^{+3.61}$ & 11.06$_{-4.57}^{+4.70}$ & 9.04$_{-3.67}^{+3.71}$ & 9.47$_{-4.13}^{+4.21}$ \\
    \hline
    $\Delta \chi^2$ & 18 & 16 & 17 & 14 \\
    \hline
    \multicolumn{5}{l}{\textit{Gaussian line width = 0.6\,keV}} \\
LineE (keV) & 7.24$_{-0.44}^{+0.39}$ & 6.97$_{-0.45}^{+0.41}$ & 6.93$_{-0.35}^{+0.39}$ & 7.04$_{-0.44}^{+0.51}$ \\
Sigma (eV) & 600(f) & 600(f) & 600(f) & 600(f) \\
Norm ($10^{-6}$) & 12.17$_{-4.46}^{+4.49}$ & 15.22$_{-5.70}^{+5.82}$ & 11.13$_{-4.54}^{+4.57}$ & 13.10$_{-5.09}^{+5.15}$ \\
    \hline
    $\Delta \chi^2$ & 20 & 20 & 16 & 18 \\
    \hline
    \multicolumn{5}{l}{\textit{Gaussian line width = 0.8\,keV}} \\
LineE (keV) & 4.78$_{-0.27}^{+0.25}$ & 7.03$_{-0.58}^{+0.51}$ & 6.79$_{-0.49}^{+0.57}$ & 6.86$_{-0.53}^{+0.67}$ \\
Sigma (eV) & 800(f) & 800(f) & 800(f) & 800(f) \\
Norm ($10^{-6}$) & 29.50$_{-7.34}^{+7.28}$ & 19.23$_{-6.75}^{+6.87}$ & 13.36$_{-5.36}^{+5.39}$ & 16.47$_{-6.00}^{+6.04}$ \\
    \hline
    $\Delta \chi^2$ & 45 & 23 & 17 & 21 \\
    \hline
    \end{tabular}
    \label{Table:varying_FeK_widths}
\end{table*}

\newpage

\section{RGS spectra for individual observation}\label{app:rgs_spectra}
The individual RGS1 and RGS2 spectra are presented in here in the appendix for visualization purposes. We have included the delta C to show the quality of the fit. The delta C within $\pm 20$ is considered to be acceptable for the quality of the data we have in this work.

\begin{figure*}
    \hbox{
    \includegraphics[width=0.5\textwidth]{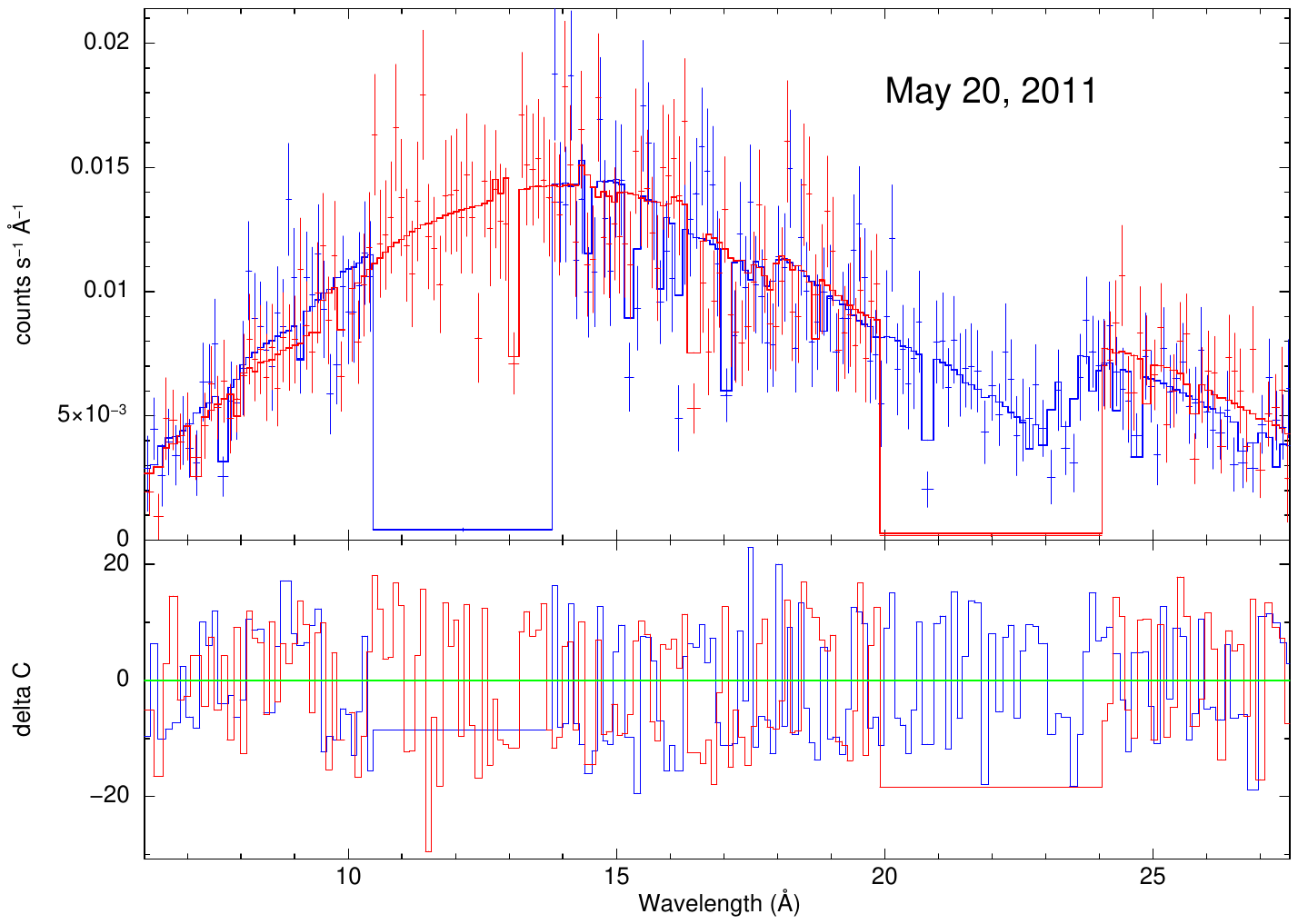}
    \includegraphics[width=0.5\textwidth]{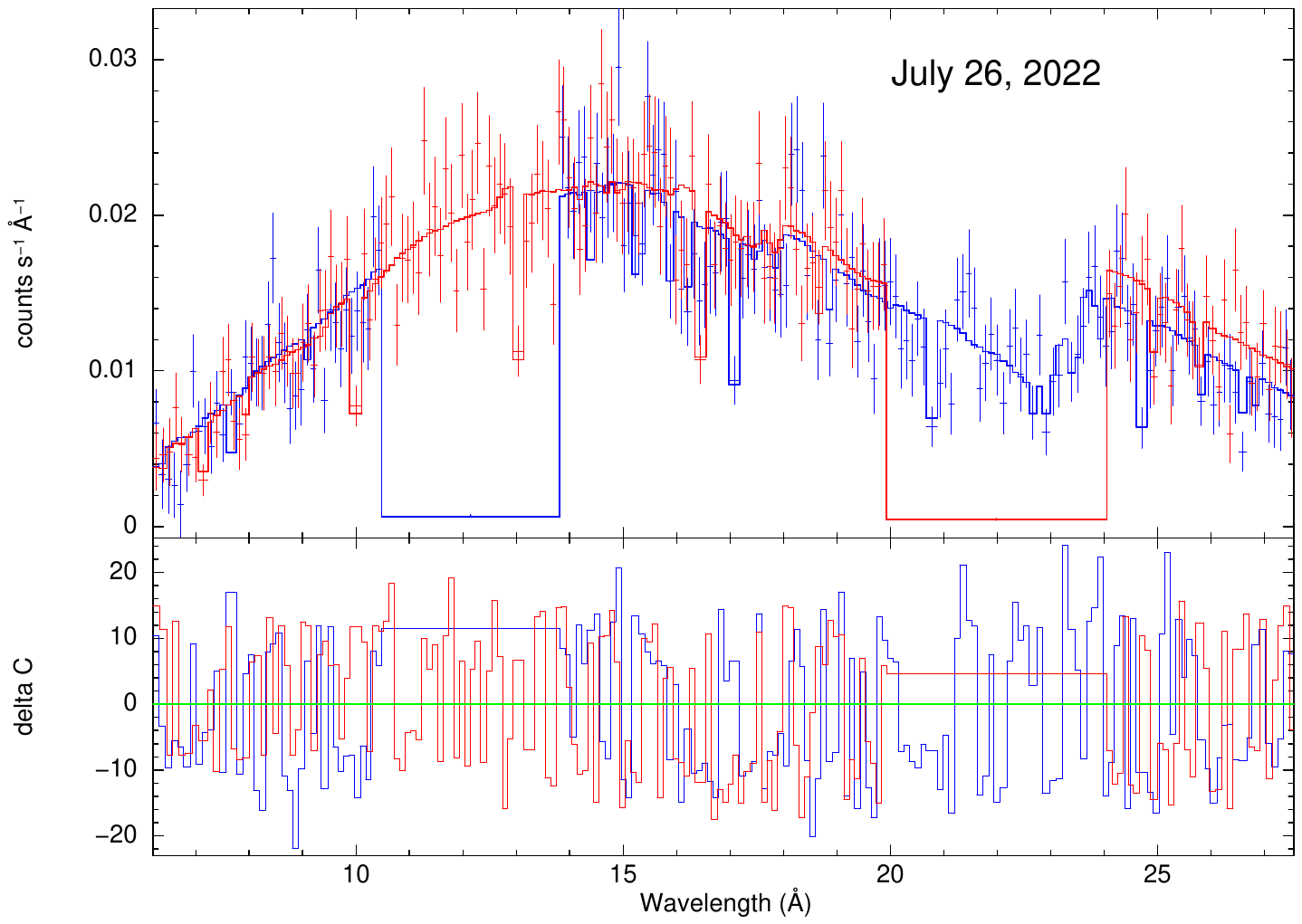}
    }
    \caption{Left: The RGS1 and RGS2 spectra (in blue and red, respectively) for May 20, 2011, observations. The upper panel shows the data and the best-fit models. The bottom panel shows the residual delta C. Right: Similar to the left for observation from July 26, 2022. } 
    \label{fig:rgs_set1}
\end{figure*}

\begin{figure*}
    \hbox{
    \includegraphics[width=0.5\textwidth]{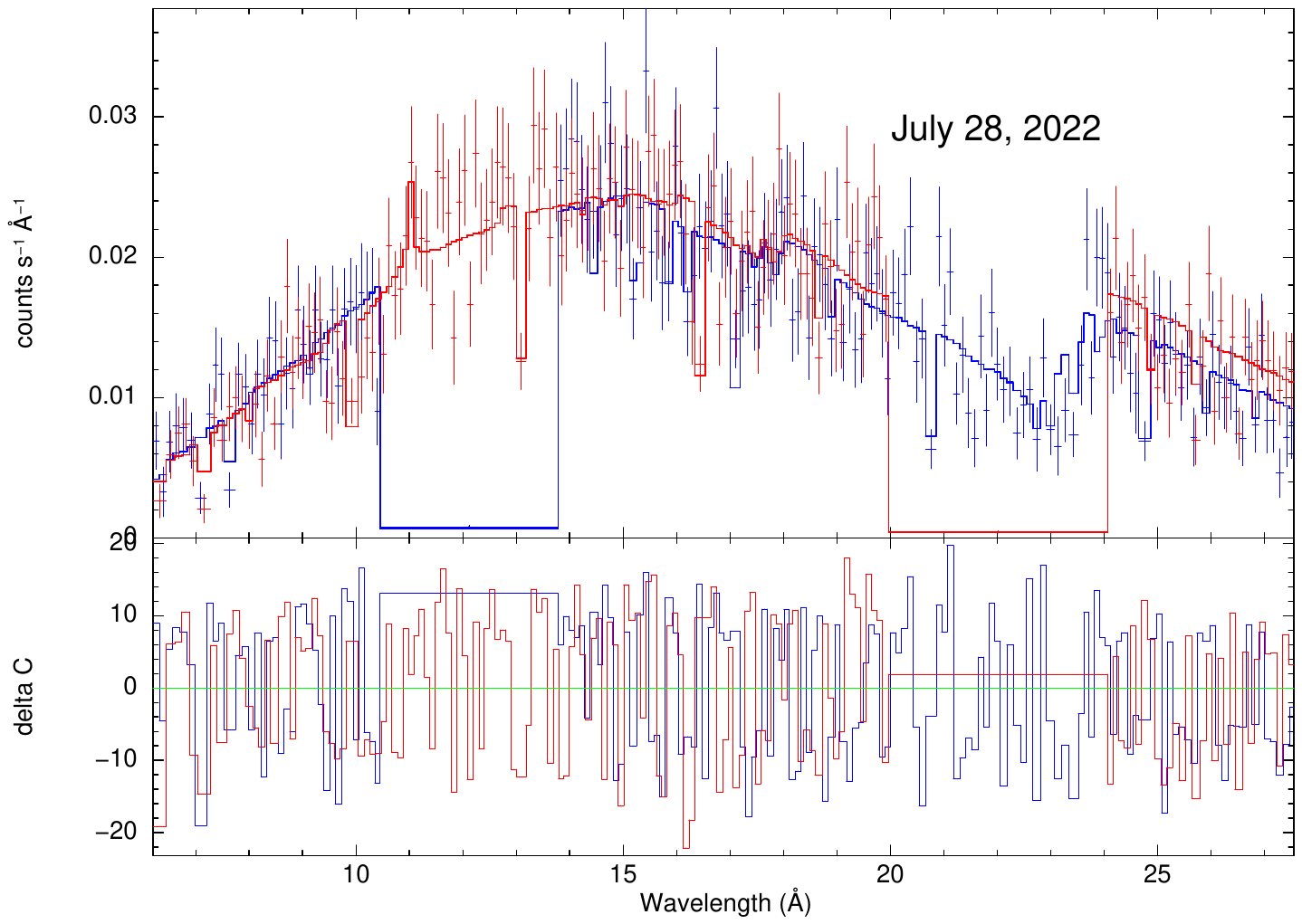}
    \includegraphics[width=0.5\textwidth]{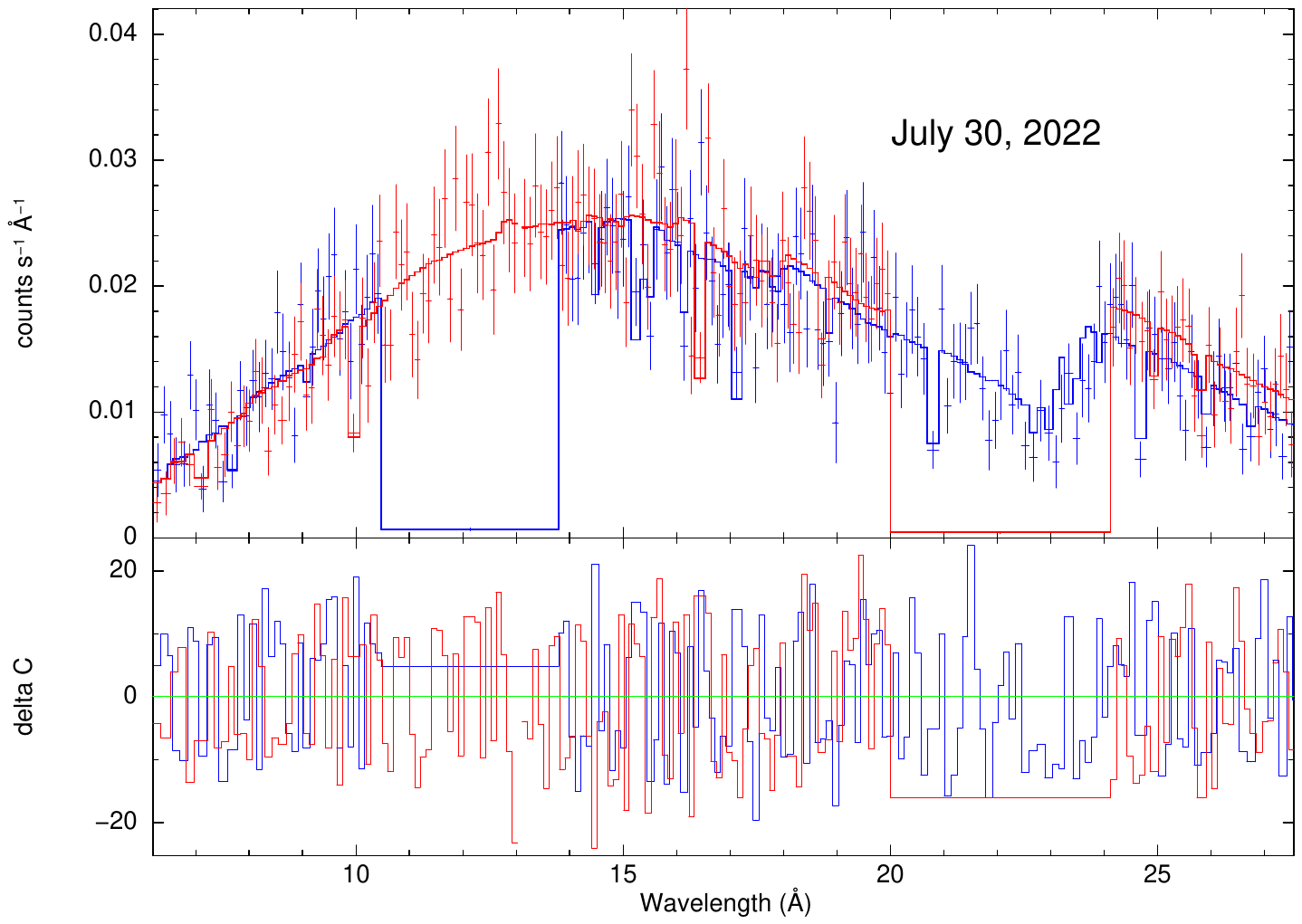}
    }
    \caption{Same as Fig. \ref{fig:rgs_set1} for the observations taken on July 28, 2022, and July 30, 2022. } 
    \label{fig:rgs_set2}
\end{figure*}

\begin{figure*}
    \hbox{
    \includegraphics[width=0.5\textwidth]{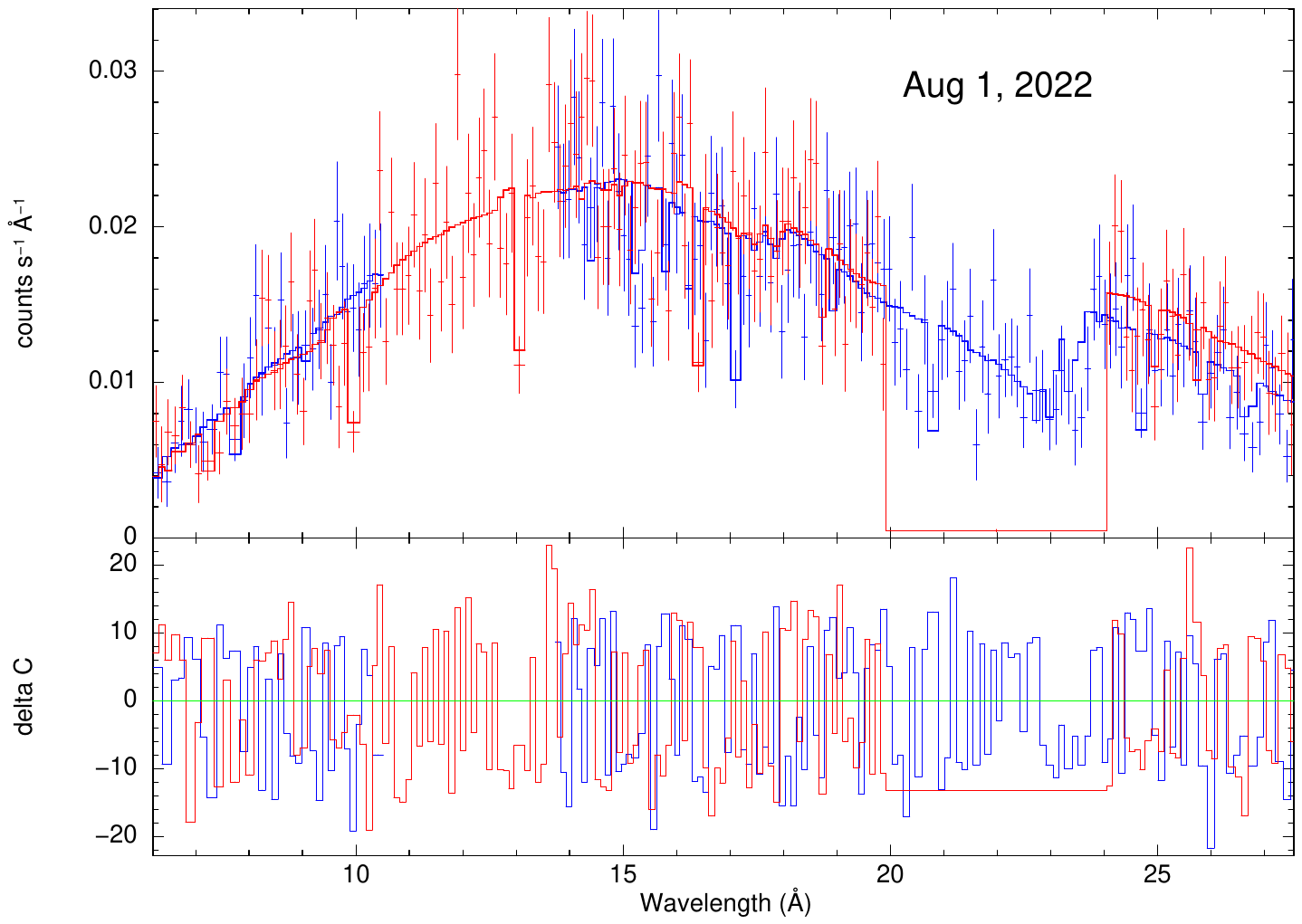}
    \includegraphics[width=0.5\textwidth]{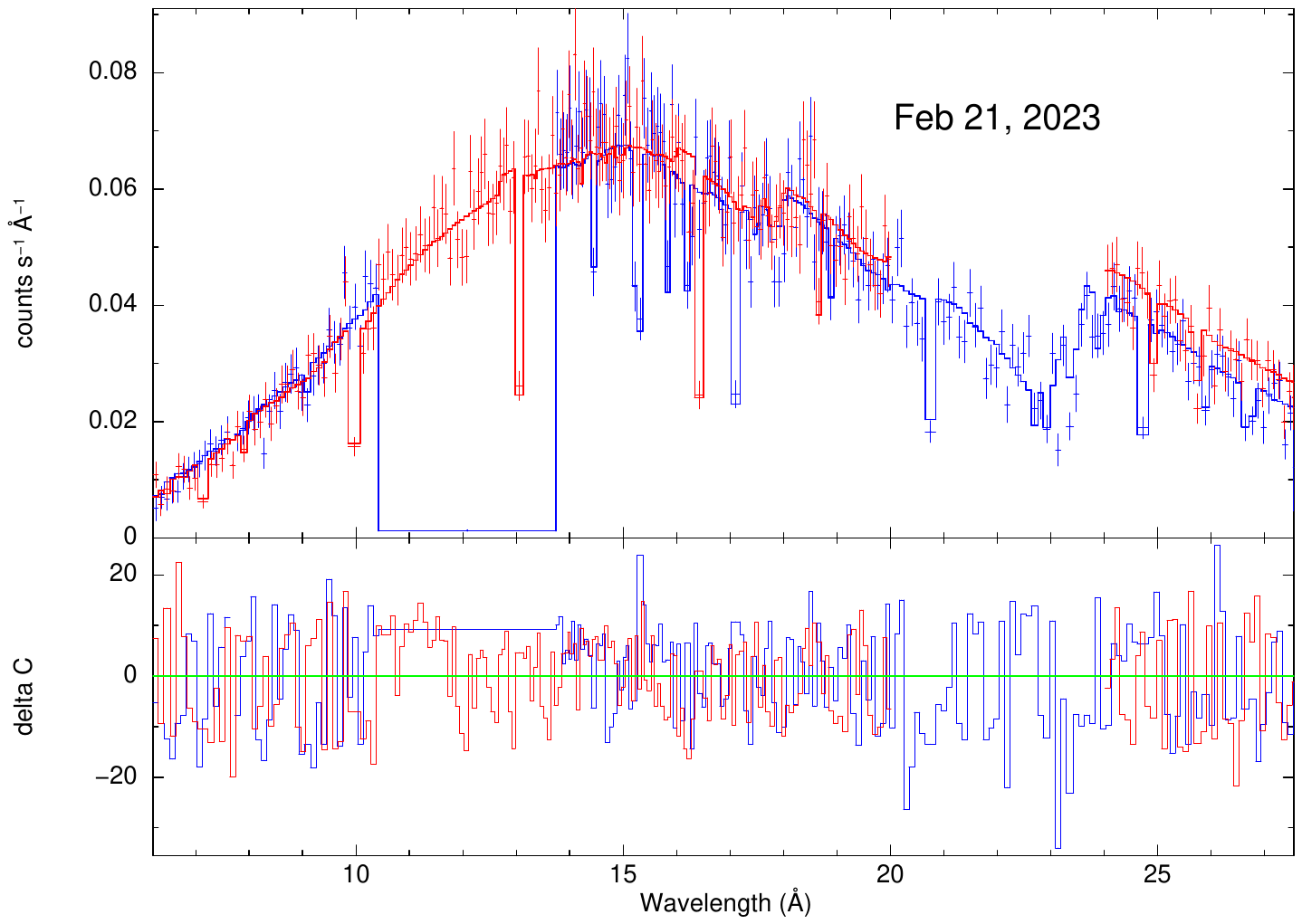}
    }
    \caption{Same as Fig \ref{fig:rgs_set1} for the observations taken on August 1, 2022, and Feb. 21, 2023.} 
    \label{fig:rgs_set3}
\end{figure*}

\begin{figure*}
    \hbox{
    \includegraphics[width=0.5\textwidth]{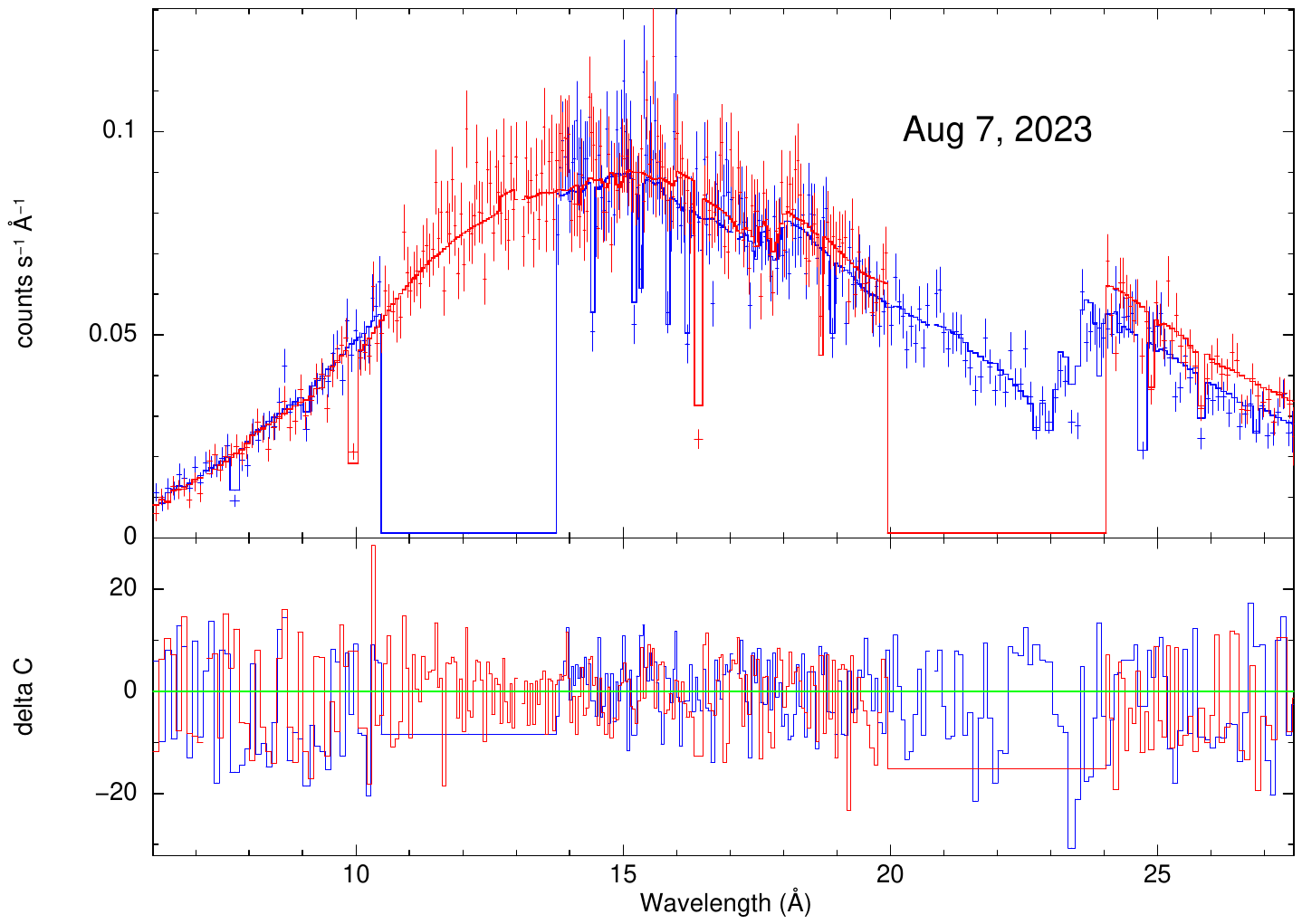}
    \includegraphics[width=0.5\textwidth]{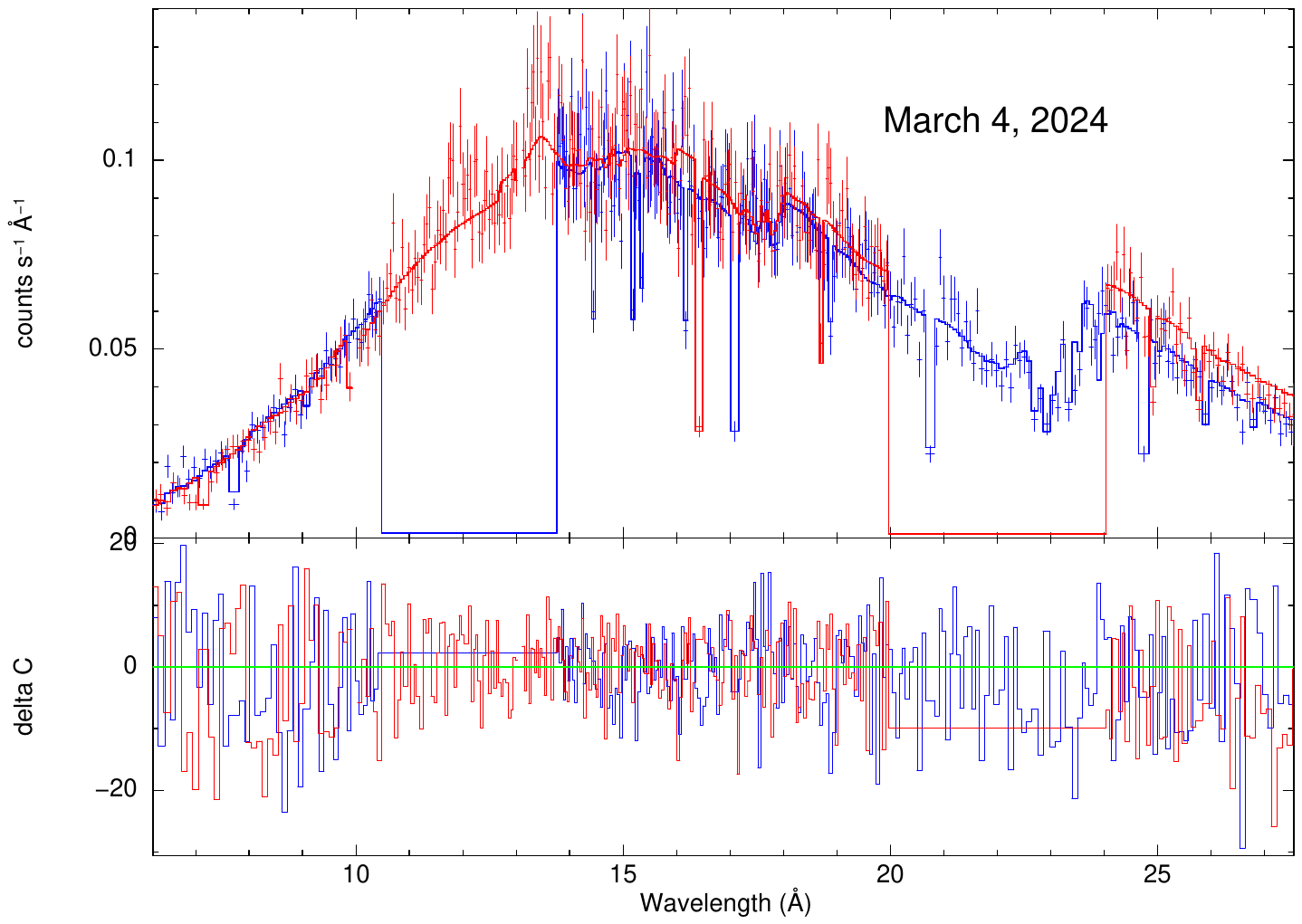}
    }
    \caption{Same as Fig. \ref{fig:rgs_set1} for the observations taken on August 7, 2023, and March 4, 2024. } 
    \label{fig:rgs_set4}
\end{figure*}

\begin{figure*}
    \hbox{
    \includegraphics[width=0.5\textwidth]{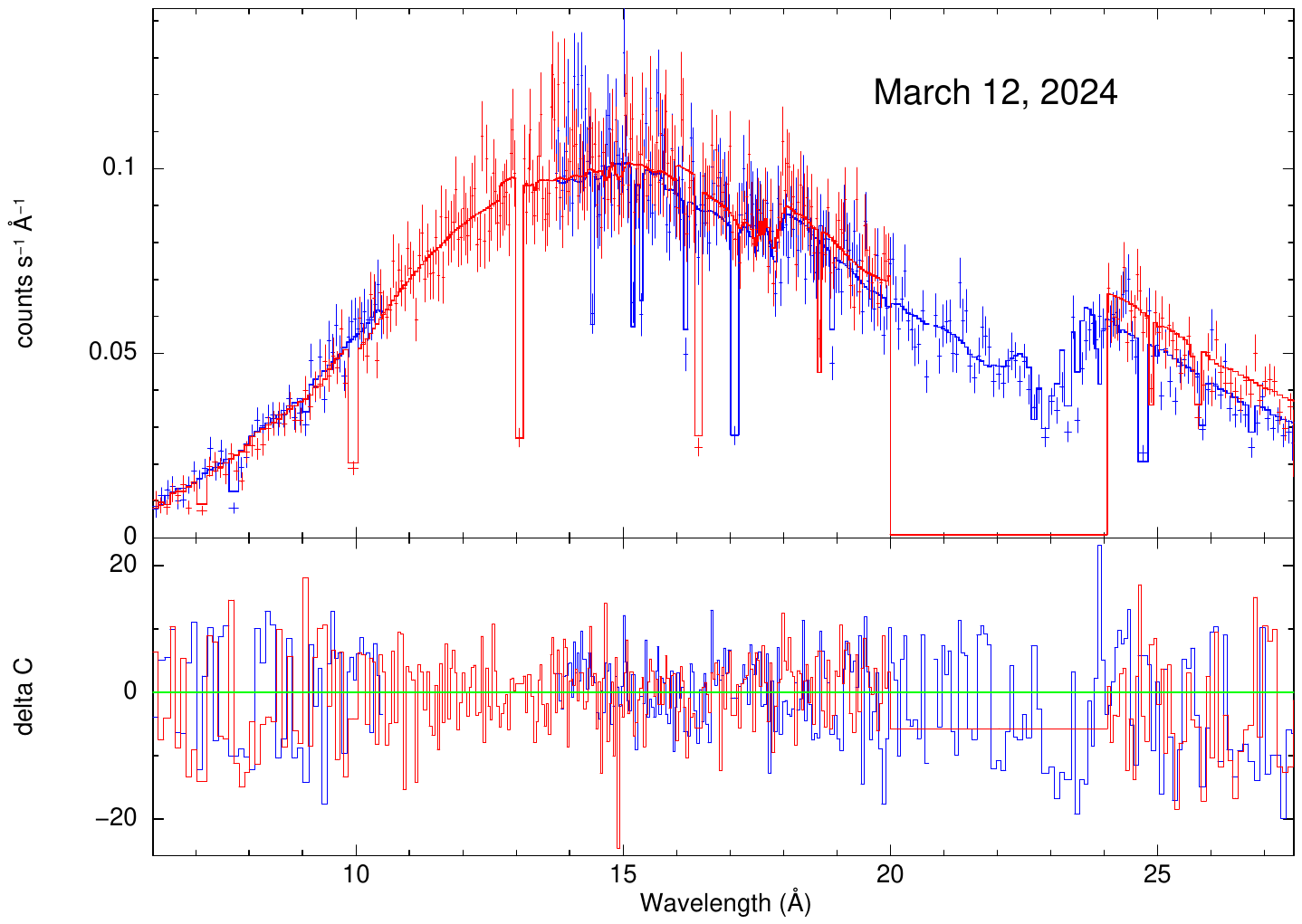}
    \includegraphics[width=0.5\textwidth]{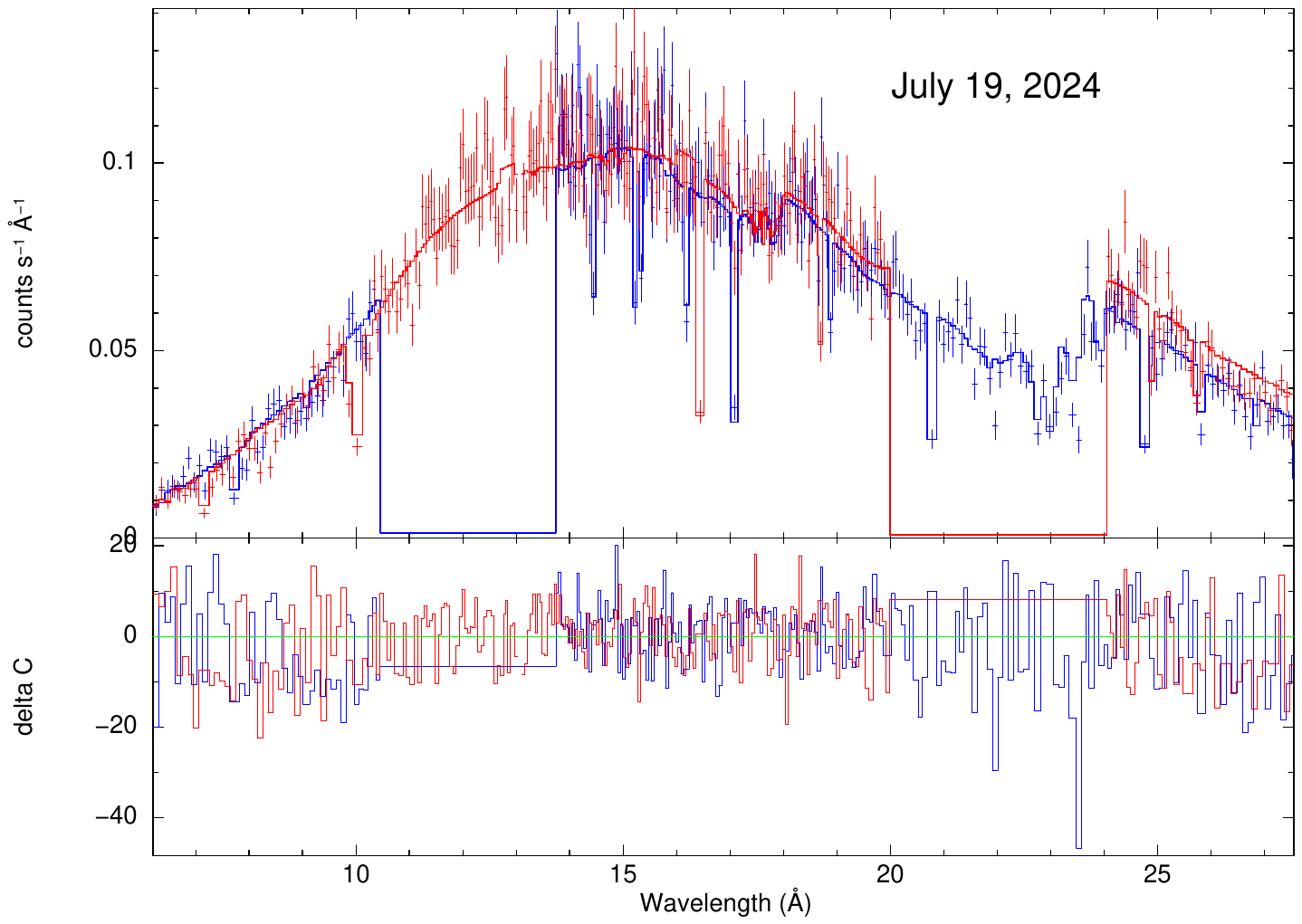}
    }
    \caption{Same as Fig. \ref{fig:rgs_set1} for the observations taken on March 12, 2024, and July 19, 2024.} 
    \label{fig:rgs_set5}
\end{figure*}

\begin{figure*}
    \hbox{
    \includegraphics[width=0.5\textwidth]{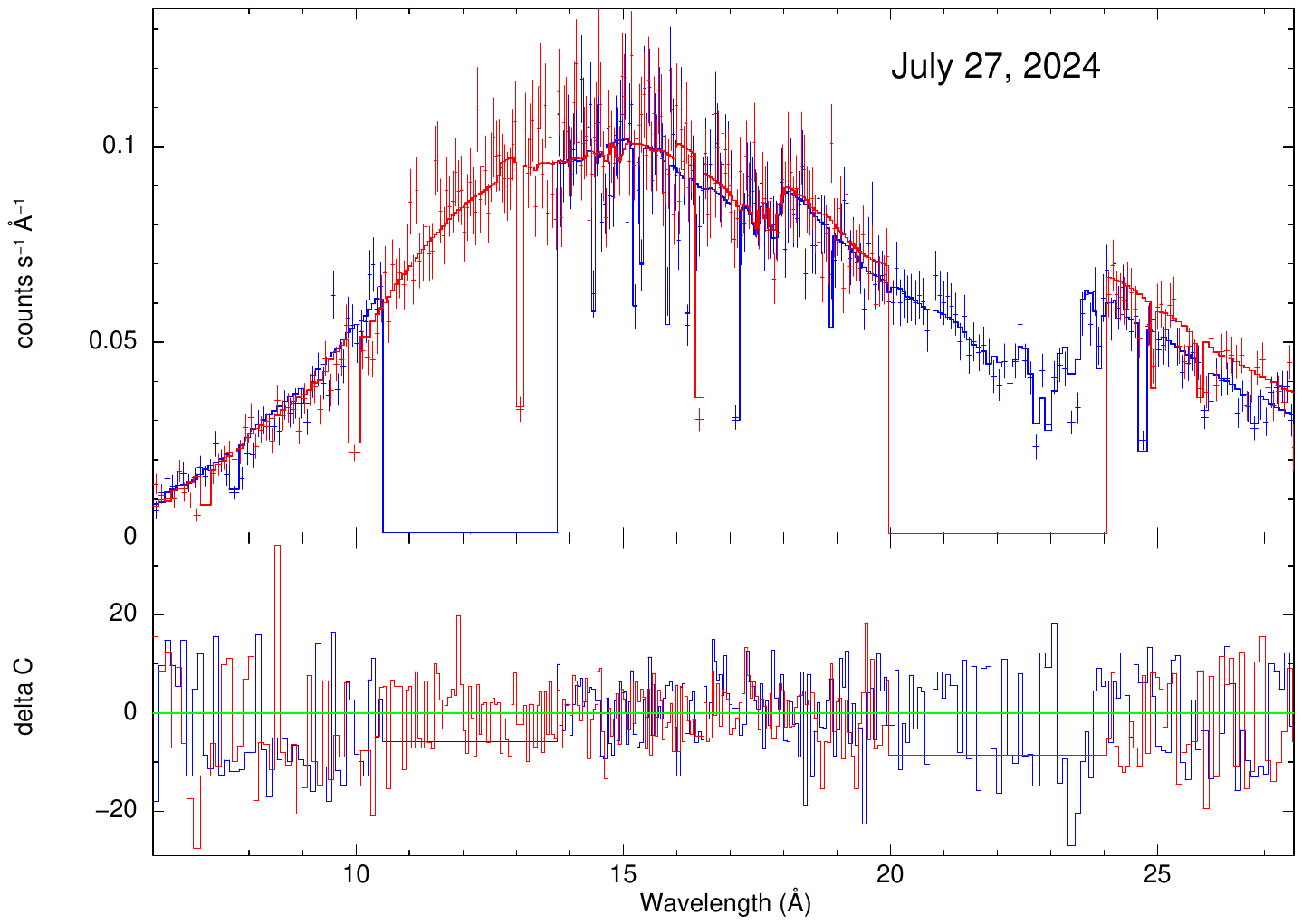}
    \includegraphics[width=0.5\textwidth]{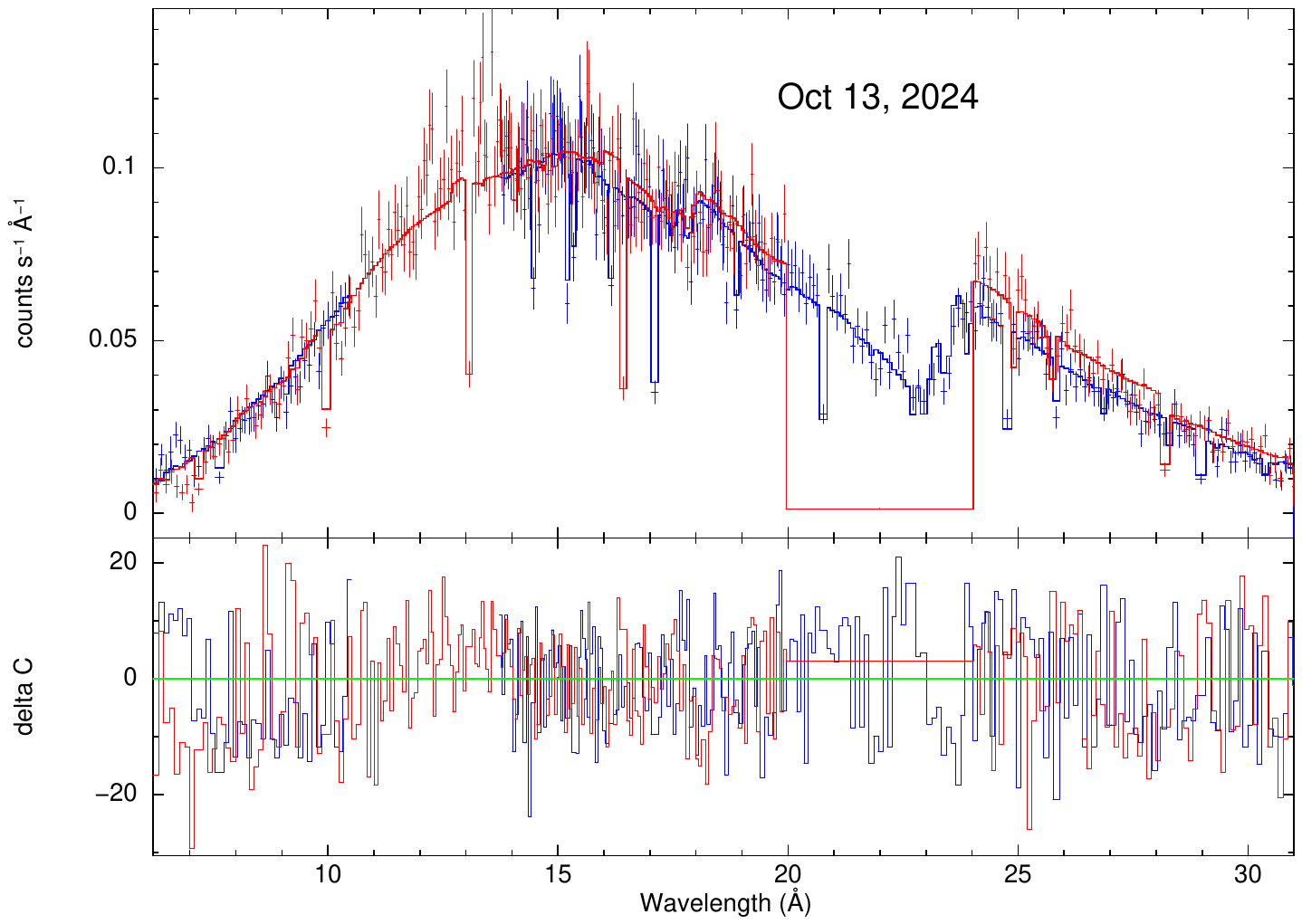}
    }
    \caption{Same as Fig \ref{fig:rgs_set1} for the observations taken on July 27, 2024, and Oct. 13, 2024. } 
    \label{fig:rgs_set6}
\end{figure*}

\begin{figure*}
    \hbox{
    \includegraphics[width=0.5\textwidth]{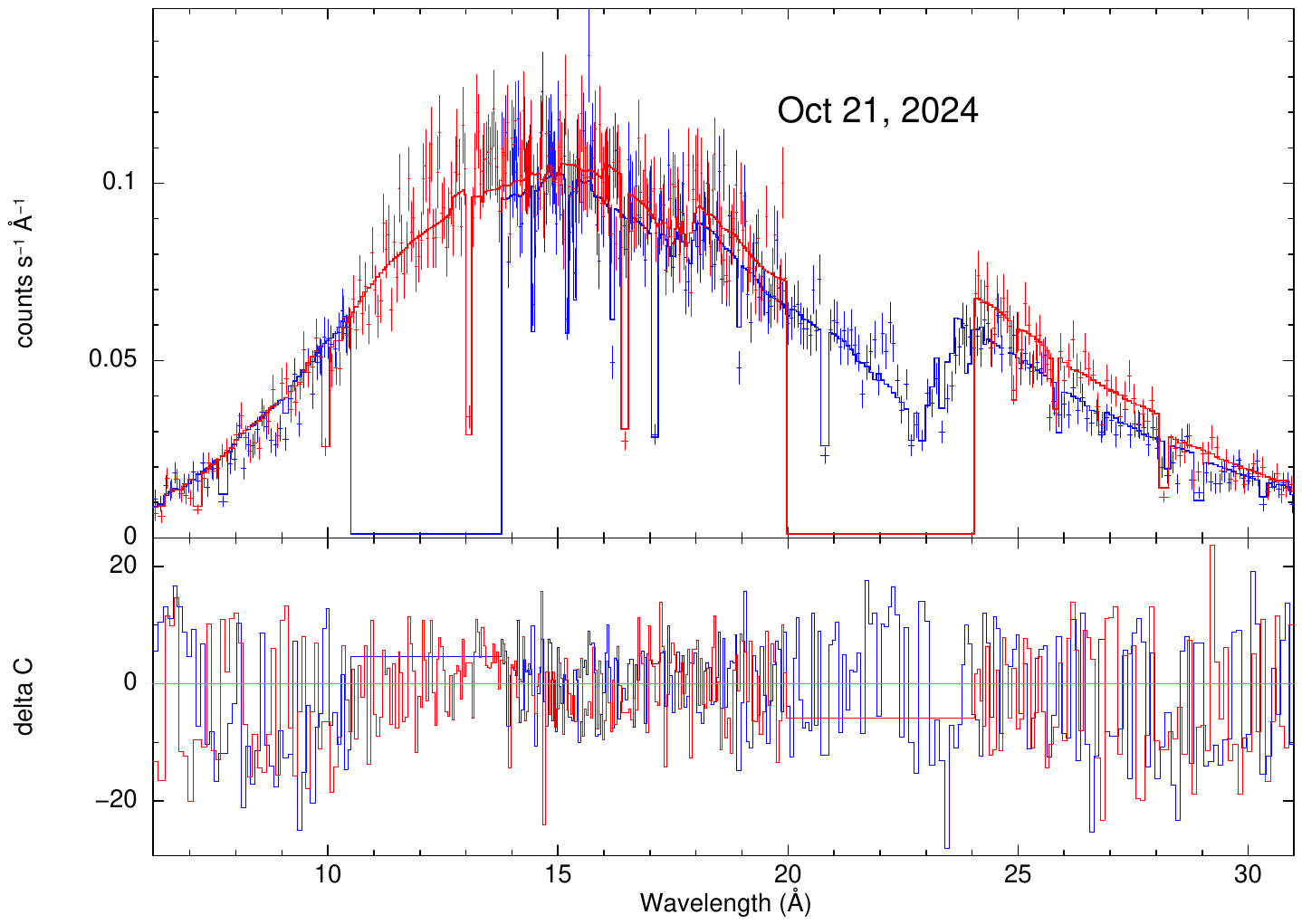}
    \includegraphics[width=0.5\textwidth]{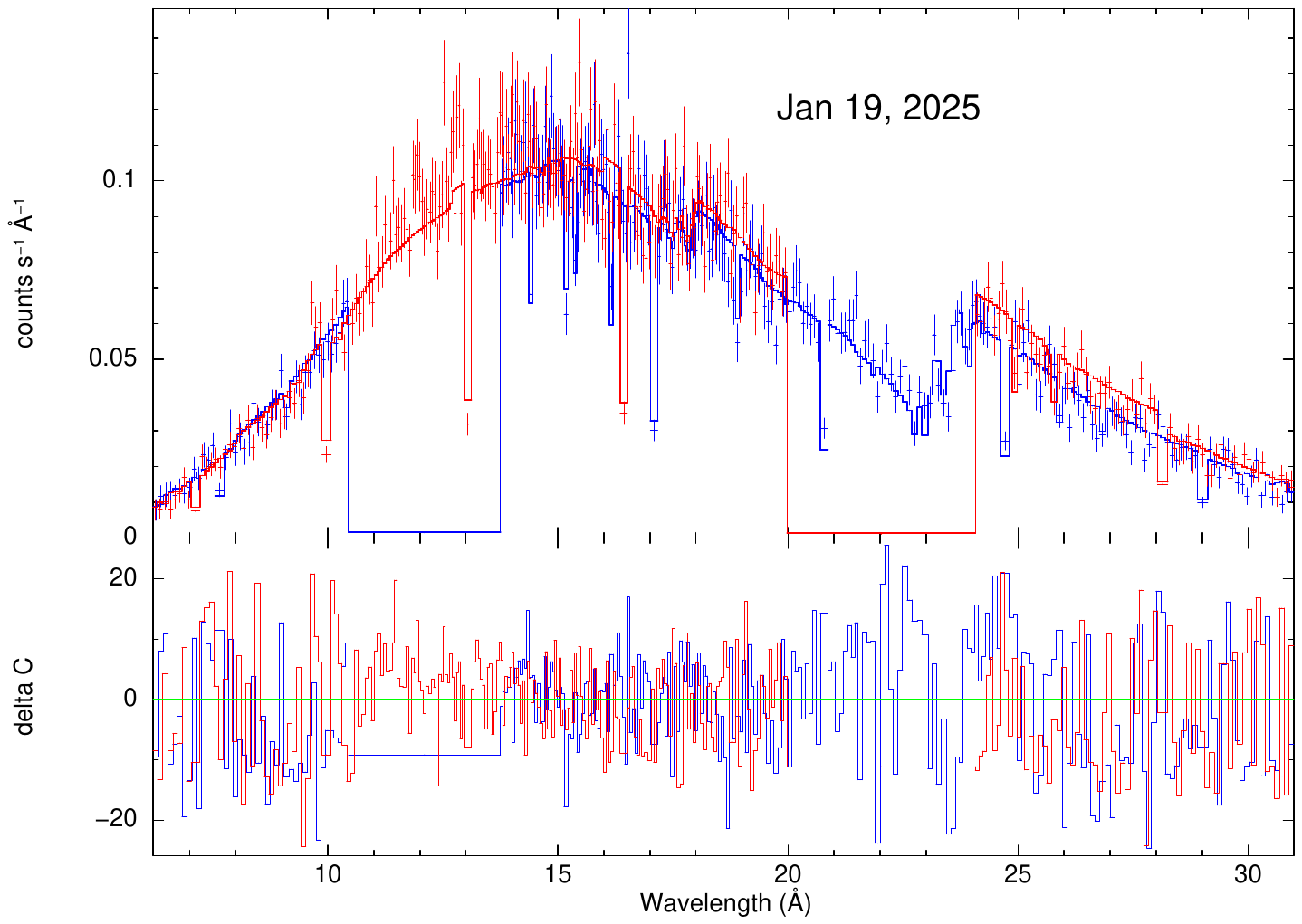}
    }
    \caption{Same as Fig \ref{fig:rgs_set1} for the observations taken on Oct. 21, 2024, and Jan. 19, 2025. } 
    \label{fig:rgs_set7}
\end{figure*}

\begin{figure*}
    \hbox{
    \includegraphics[width=0.5\textwidth]{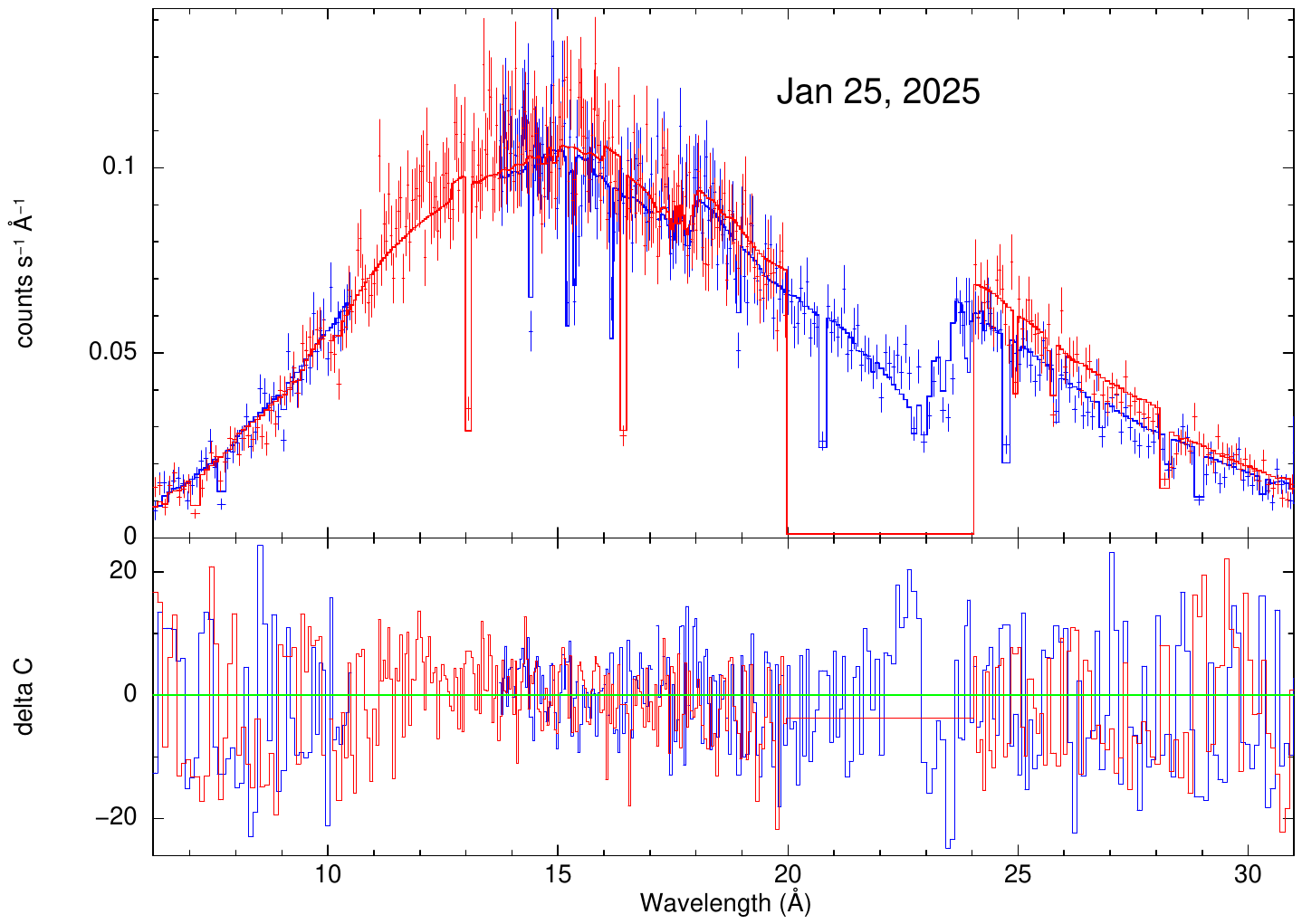}
    \includegraphics[width=0.5\textwidth]{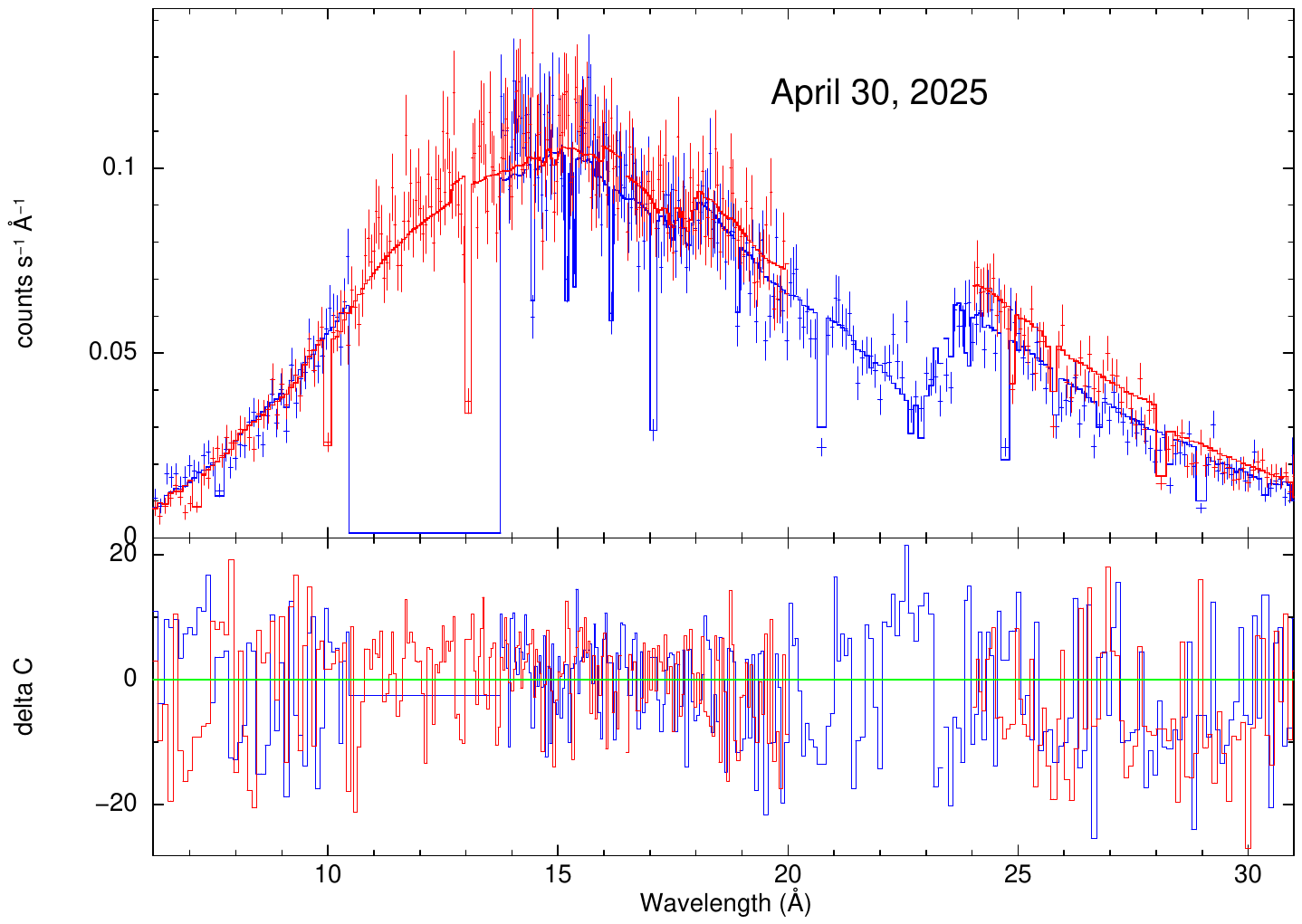}
    }
    \caption{Same as Fig \ref{fig:rgs_set1} for the observations taken on Jan. 25, 2025, and Apr. 30, 2025. } 
    \label{fig:rgs_set8}
\end{figure*}

\begin{figure*}
    \hbox{
    \includegraphics[width=0.5\textwidth]{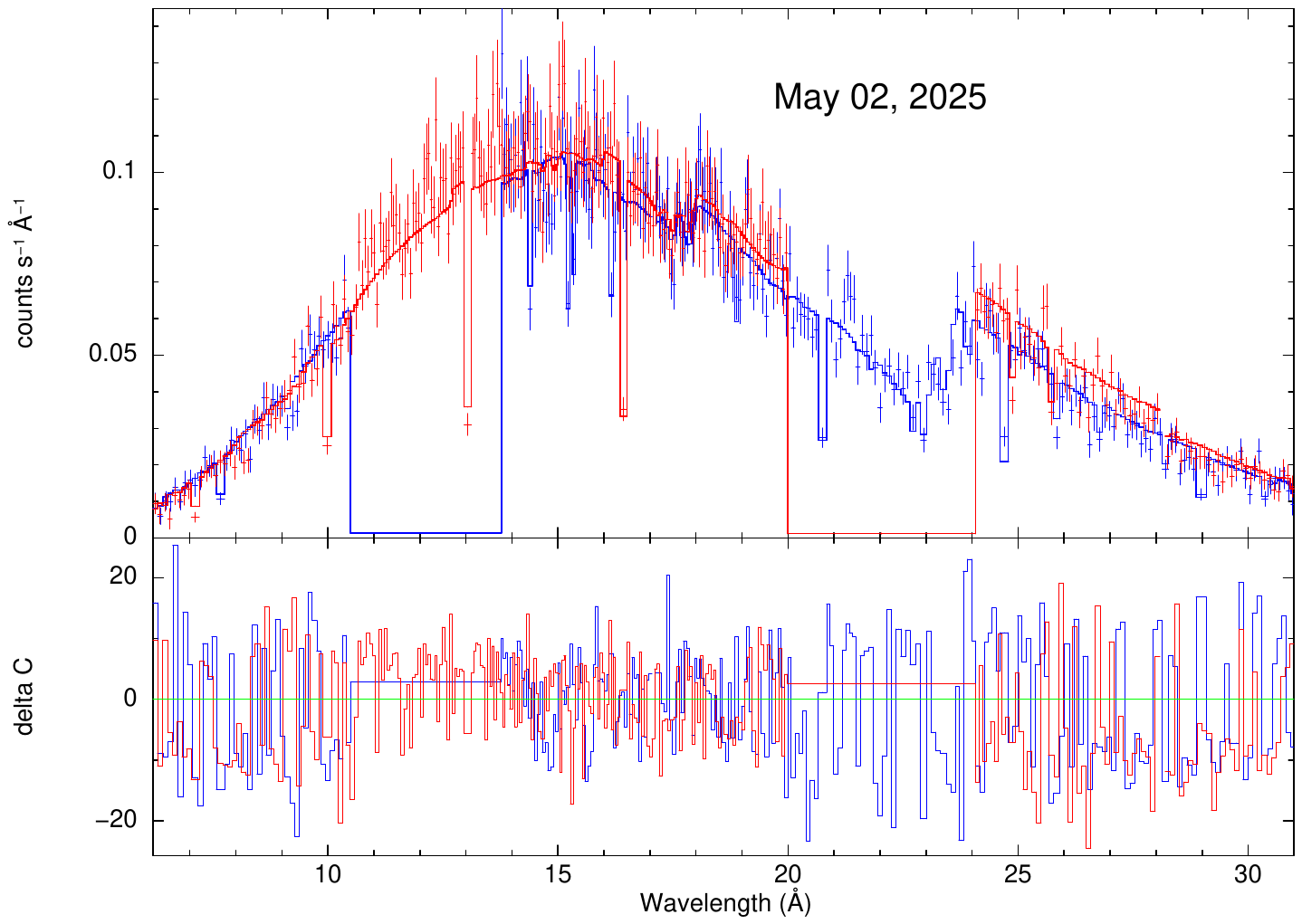}
    \includegraphics[width=0.5\textwidth]{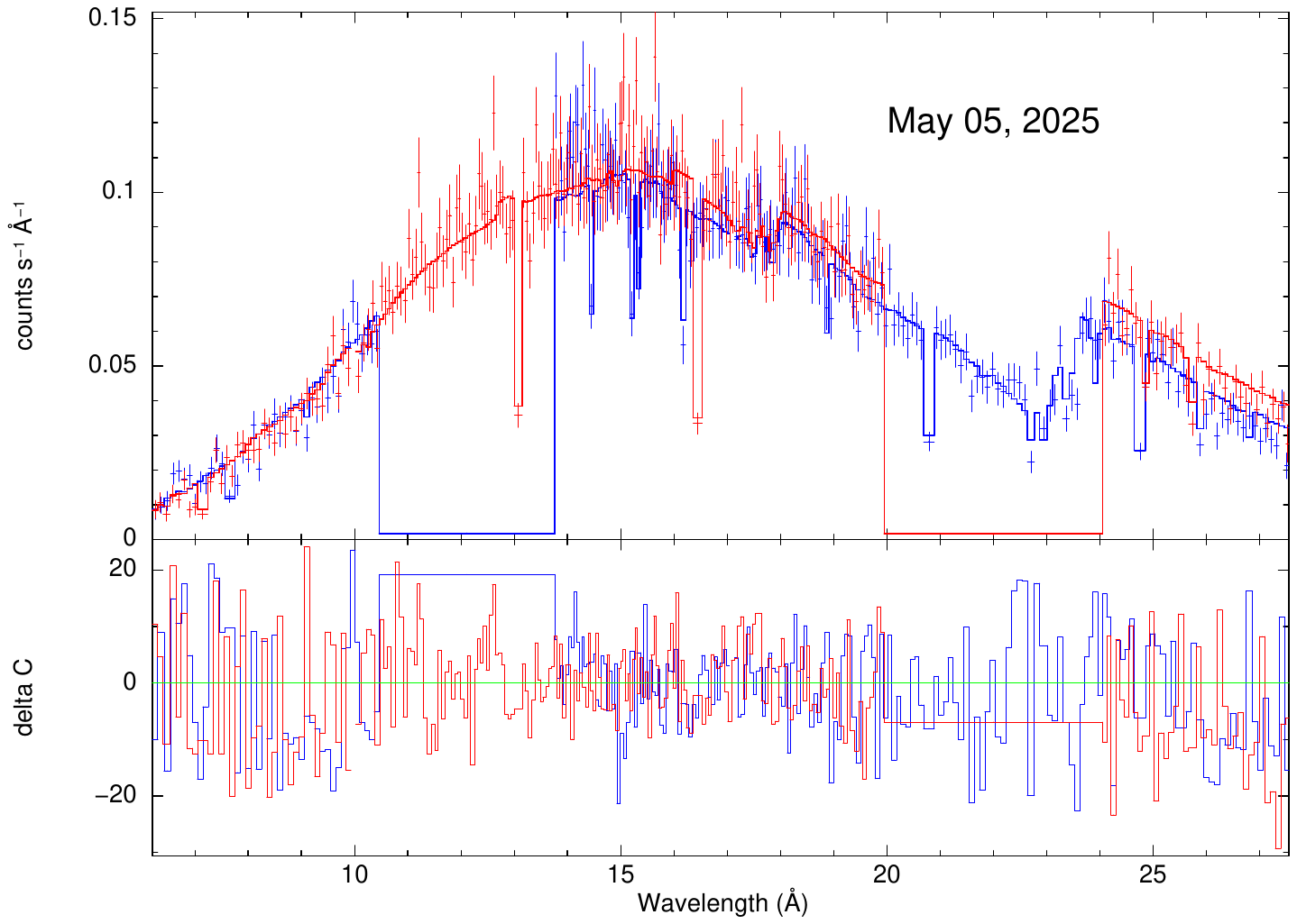}
    }
    \caption{Same as Fig \ref{fig:rgs_set1} for the observations taken on May 02, 2025, and May 05, 2025.} 
    \label{fig:rgs_set9}
\end{figure*}

\newpage

\section{PN individual spectrum}
\label{sec:app_pn_ind_specturm}

Here we have shown all the individual PN spectra for all the observations included in this work. The data, best fit model, background, and residual (ratio of data to model) with the continuum-only model, \texttt{tbabs*ztbabs(bb+pow)} and the ratio with the best fit model, \texttt{tbabs*ztbabs(zgauss*zgauss*zgauss*bb+pow)} are also shown. The observation taken on May 20, 2011, did not show any emission features, and the fit with the inclusion of the Gaussian lines did not improve. Other observations made in 2022, 2023, 2024, and 2025 needed emission lines $\sim 0.56 \kev$ and  $\sim 1 \kev$. In some of the observations, we added  $\sim (6-7) \kev$ line but the statistics did not improve significantly. It was less than 5.

\begin{figure*}
    \hbox{
    \includegraphics[width=0.5\textwidth]{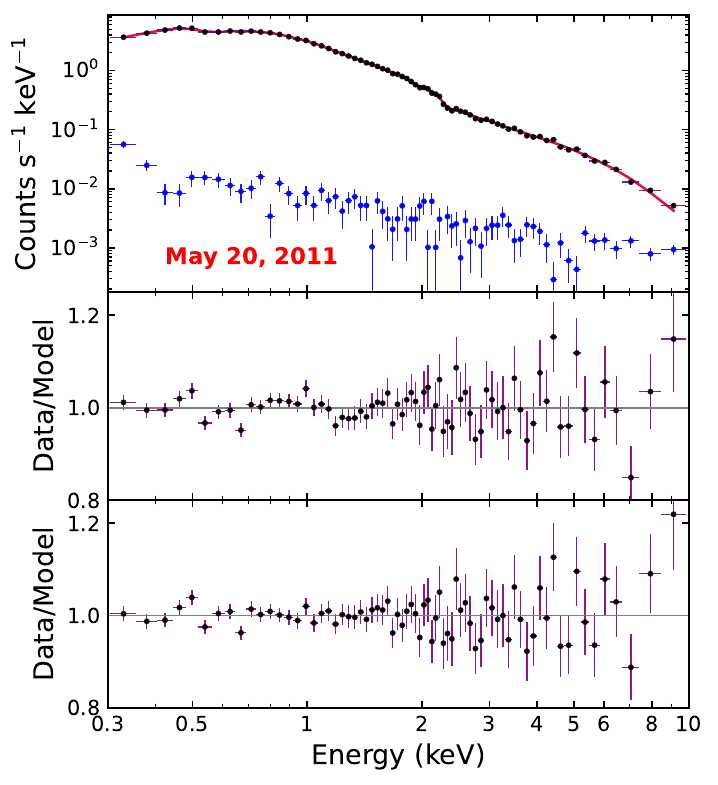}
    \includegraphics[width=0.5\textwidth]{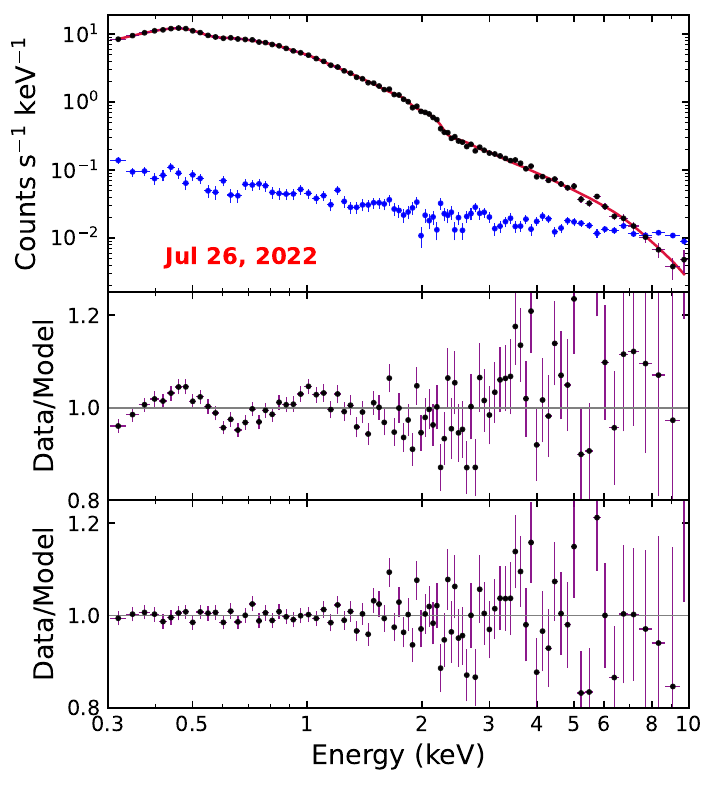}
    }
    \caption{{\tt Left:} The best fit EPIC PN spectrum and the model and the X-ray background (in black, red, and blue, respectively). This was the observation on May 20, 2011, the only pre-flare \xmm{} observation. The second and bottom-most panel is the ratio of data to model when fitted with the continuum model only, \texttt{tbabs*ztbabs(bb+pow)}. This observation did not require any Gaussian emission line component. {\tt Right:} Same as described for the left panel but for the observation taken on July 26, 2022. Here, we can clearly see positive residuals at $\sim 0.56 \kev$ and $\sim 1 \kev$. The bottom-most panel is the ratio when fitted with the model, \texttt{tbabs*ztbabs(zgauss+zgauss+zgauss+bb+pow)}. Three Gaussian lines are added in the continuum model at energies, $\sim 0.45 \kev$ and $\sim 1 \kev$. }
    \label{fig:pn_set1}
\end{figure*}

\begin{figure*}
    \hbox{
    \includegraphics[width=0.5\textwidth]{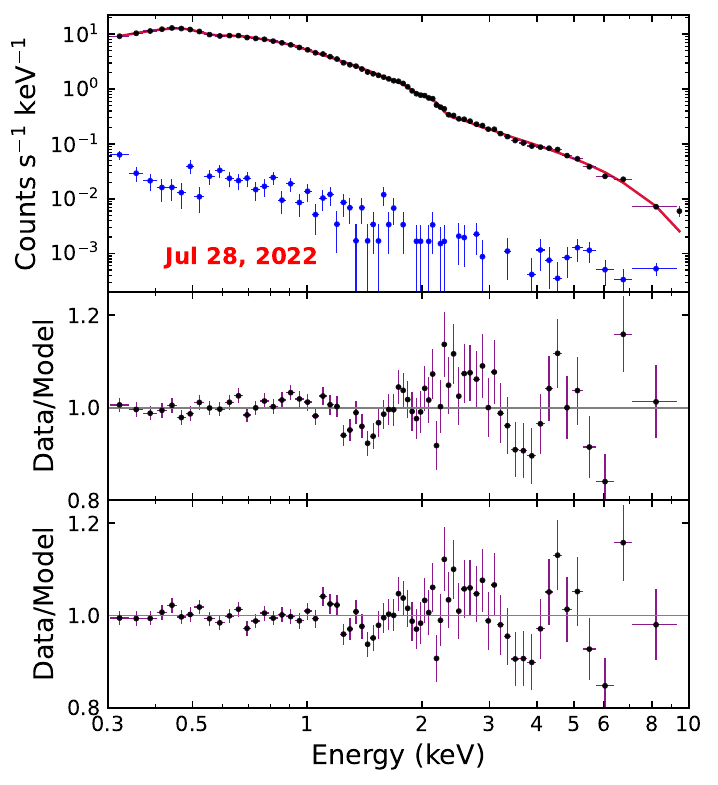}
    \includegraphics[width=0.5\textwidth]{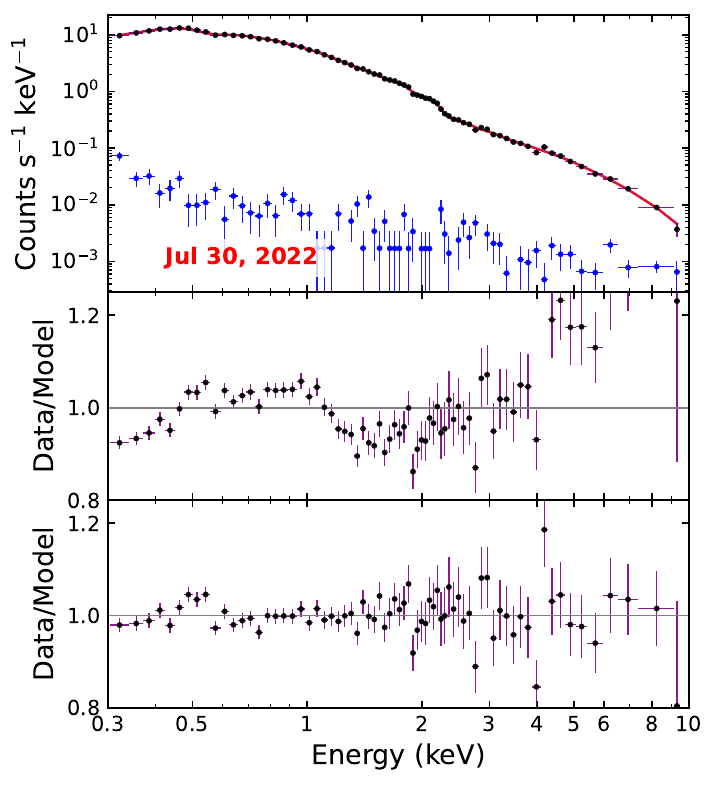}
    }
    \caption{Same as Fig. \ref{fig:pn_set1} right-panel, except for the observations taken on July 28, 2022, and July 30, 2022.}
    \label{fig:pn_set2}
\end{figure*}

\begin{figure*}
    \hbox{
    \includegraphics[width=0.5\textwidth]{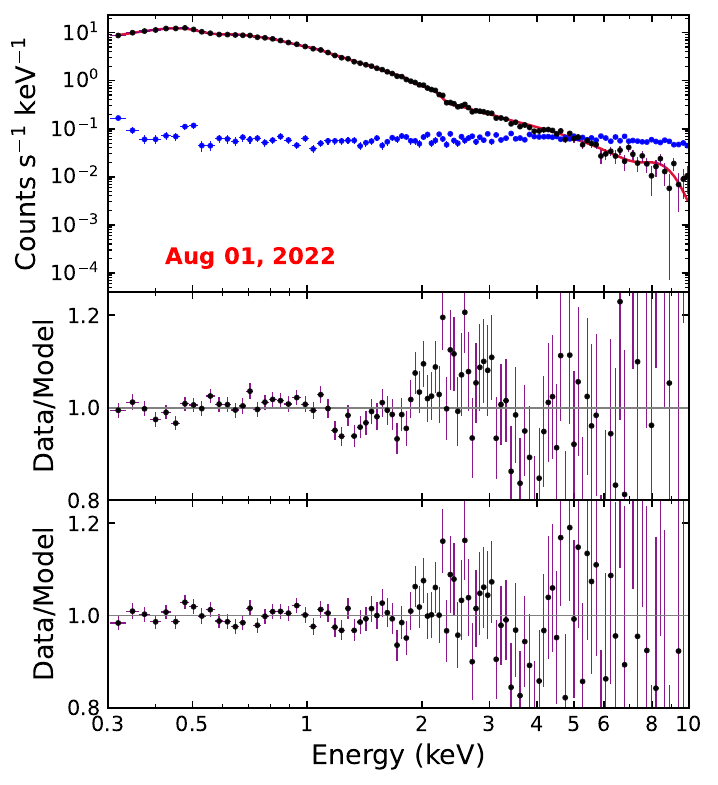}
    \includegraphics[width=0.5\textwidth]{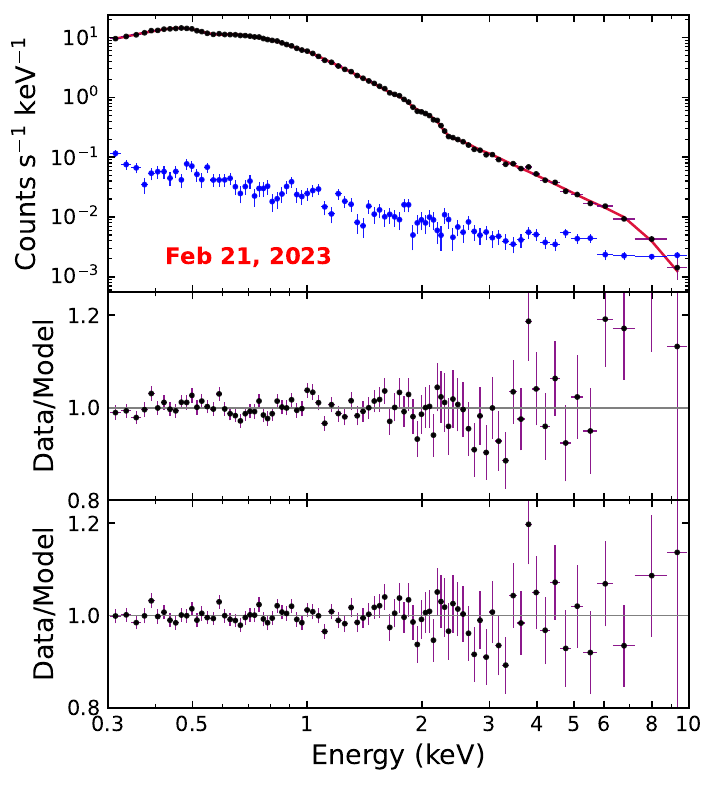}
    }
    \caption{Same as Fig. \ref{fig:pn_set2} for the observations taken on Aug 01, 2022, and Feb 21, 2023.}
    \label{fig:pn_set3}
\end{figure*}

\begin{figure*}
    \hbox{
    \includegraphics[width=0.5\textwidth]{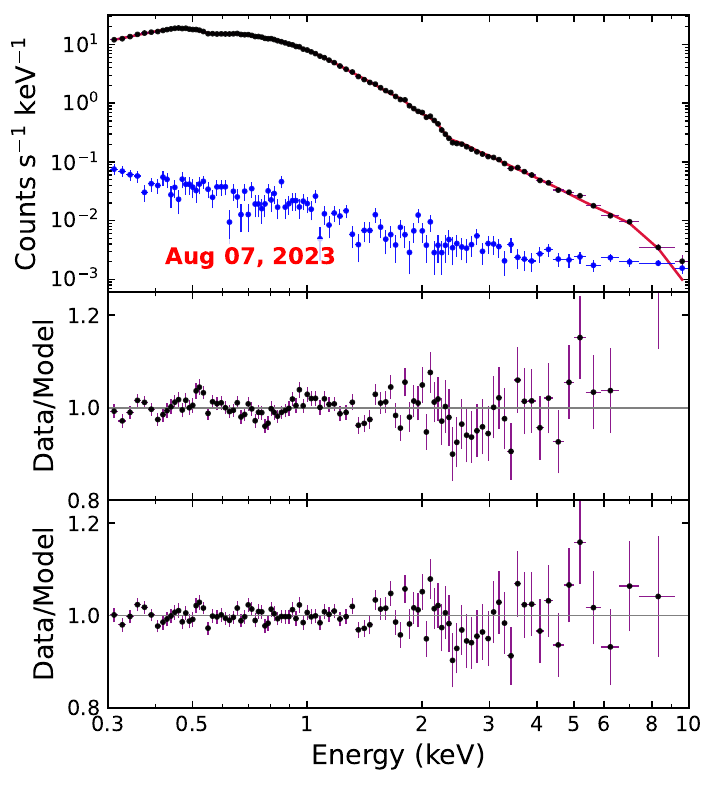}
    \includegraphics[width=0.5\textwidth]{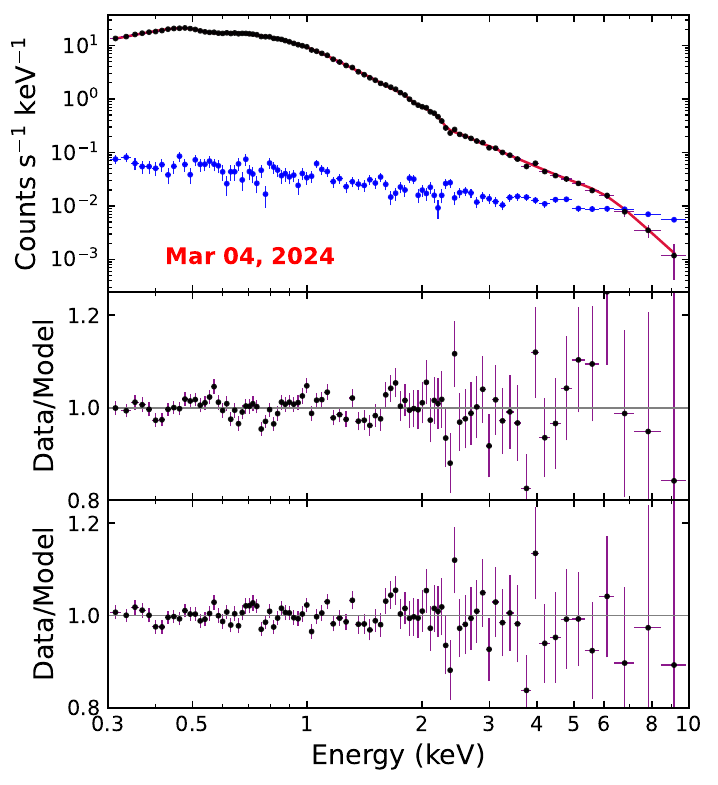}
    }
    \caption{Same as Fig. \ref{fig:pn_set2} for the observations taken on Aug 07, 2023, and Mar 04, 2024.}
    \label{fig:pn_set4}
\end{figure*}

\begin{figure*}
    \hbox{
    \includegraphics[width=0.5\textwidth]{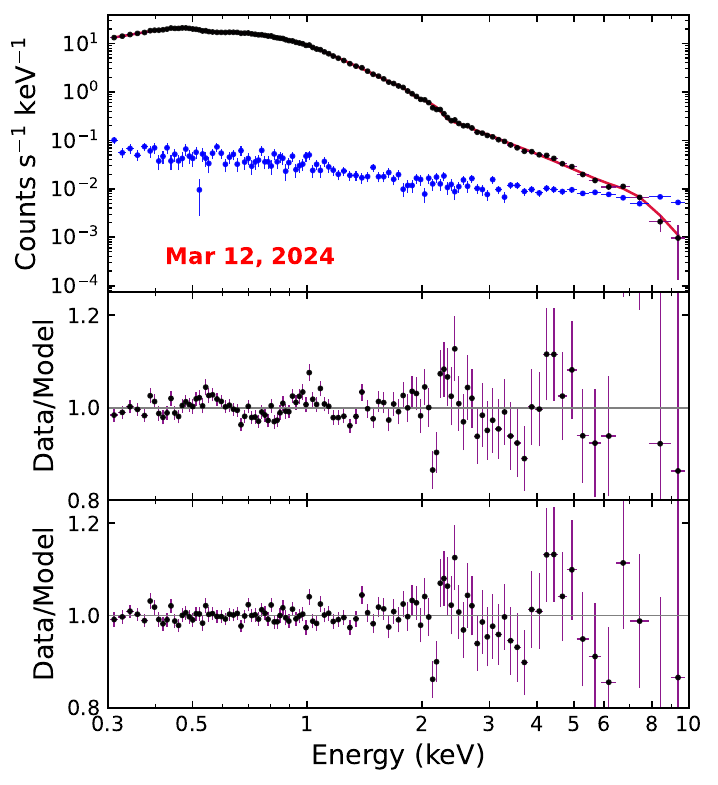}
    \includegraphics[width=0.5\textwidth]{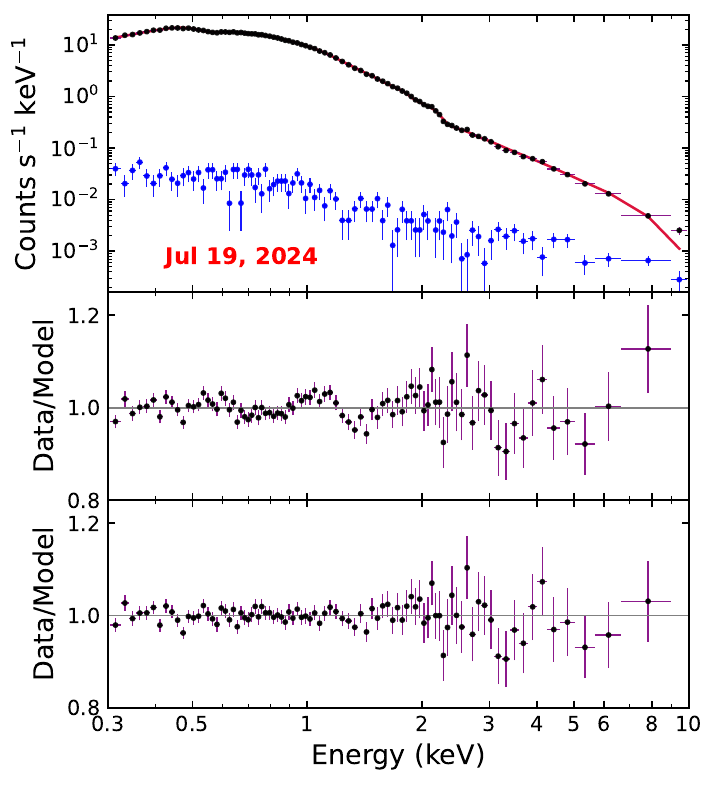}
    }
    \caption{Same as Fig. \ref{fig:pn_set2} for the observations taken on March 12, 2024, and July 19, 2024.}
    \label{fig:pn_set5}
\end{figure*}

\begin{figure*}
    \hbox{
    \includegraphics[width=0.5\textwidth]{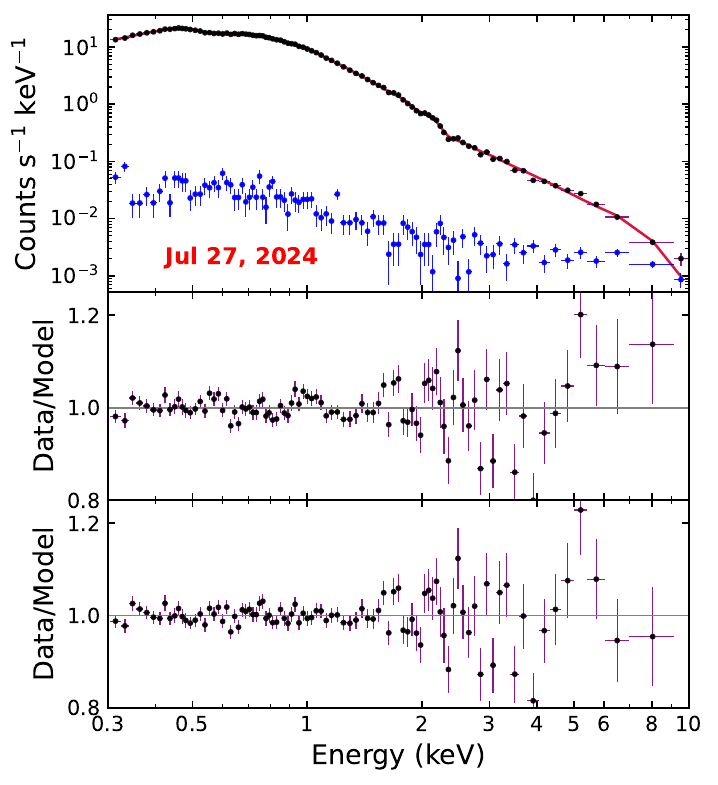}
    \includegraphics[width=0.5\textwidth]{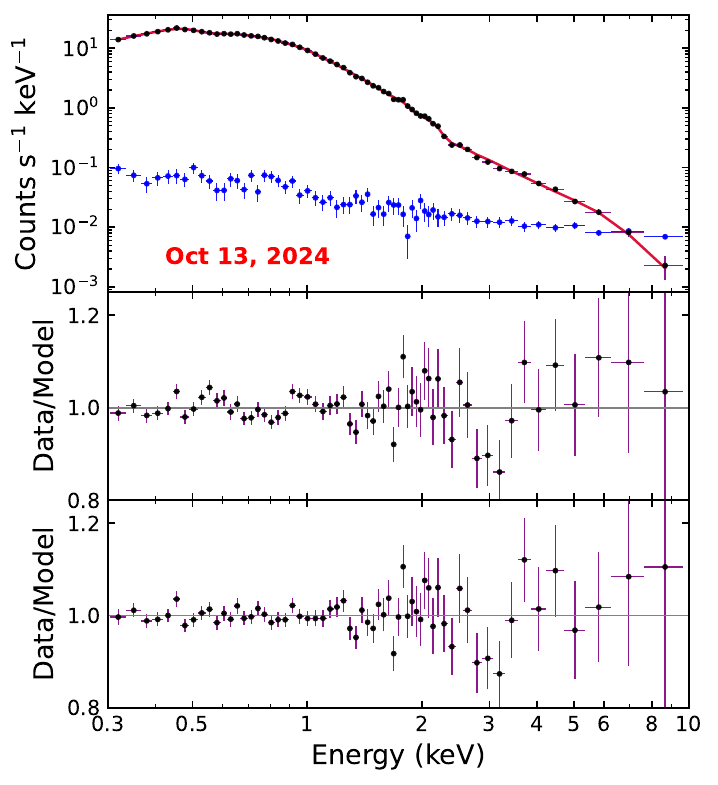}
    }
    \caption{Same as Fig. \ref{fig:pn_set2} for the observations taken on July 27, 2024, and Oct. 13, 2024.}
    \label{fig:pn_set6}
\end{figure*}

\begin{figure*}
    \hbox{
    \includegraphics[width=0.5\textwidth]{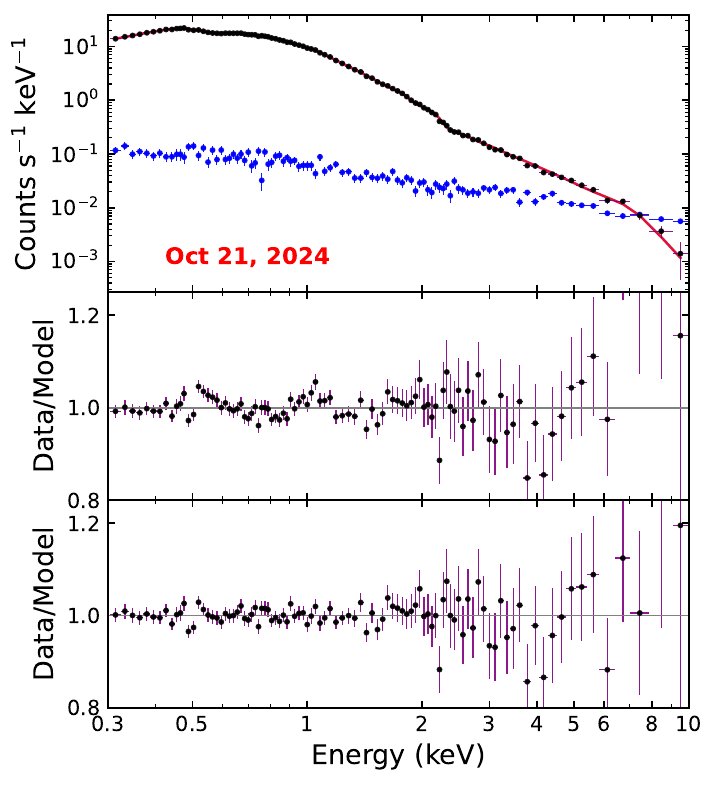}
    \includegraphics[width=0.5\textwidth]{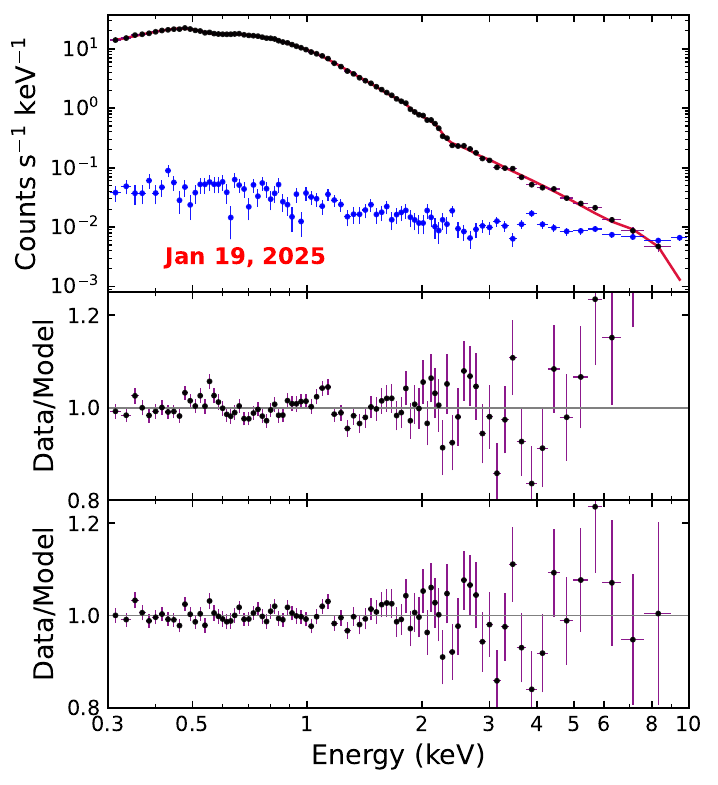}
    }
    \caption{Same as Fig. \ref{fig:pn_set2} for the observations taken on Oct 21, 2024, and Jan 19, 2025.}
    \label{fig:pn_set7}
\end{figure*}

\begin{figure*}
    \hbox{
    \includegraphics[width=0.5\textwidth]{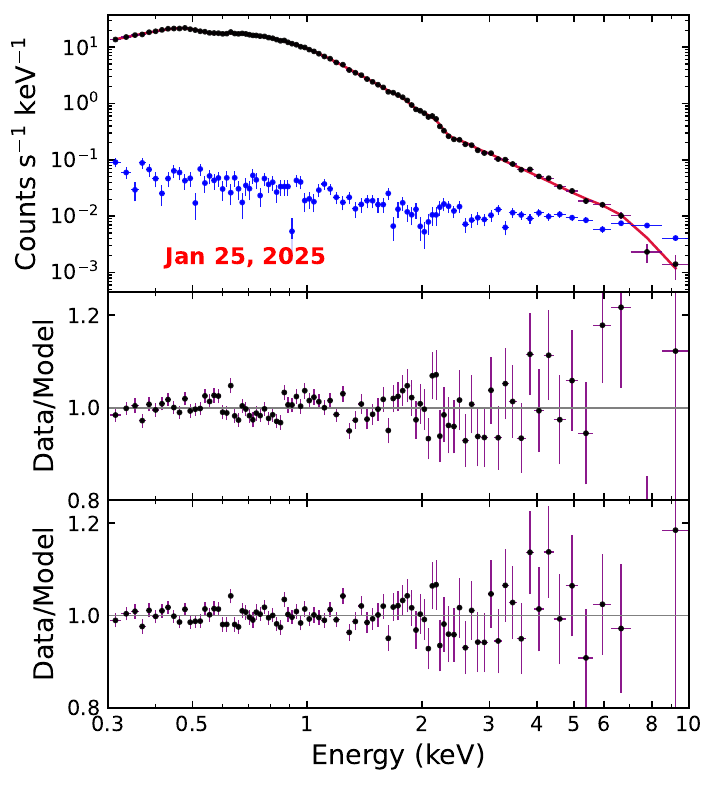}
    \includegraphics[width=0.5\textwidth]{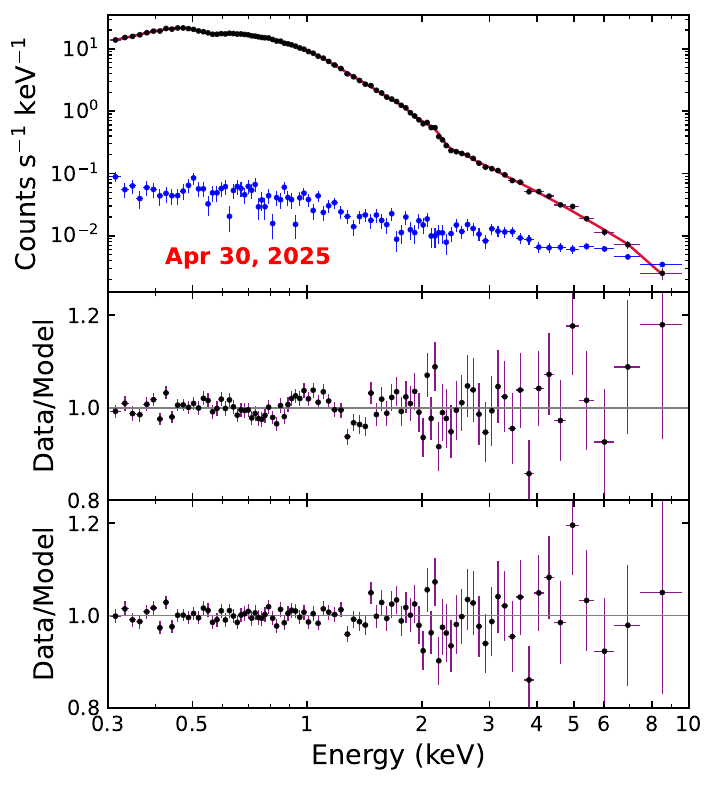}
    }
    \caption{Same as Fig. \ref{fig:pn_set2} for the observations taken on Jan 25, 2025, and April 30, 2025.}
    \label{fig:pn_set8}
\end{figure*}

\begin{figure*}
    \hbox{
    \includegraphics[width=0.5\textwidth]{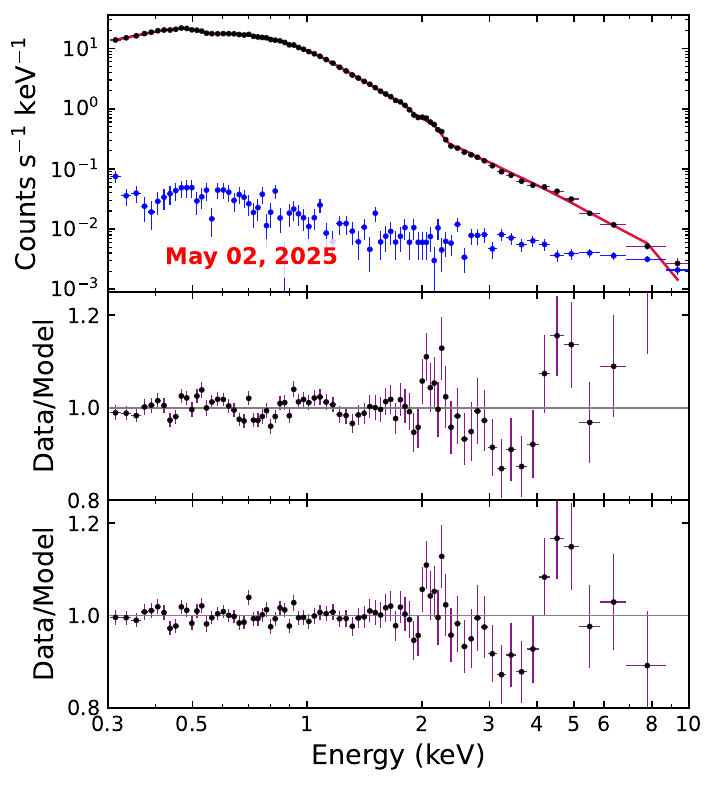}
    \includegraphics[width=0.5\textwidth]{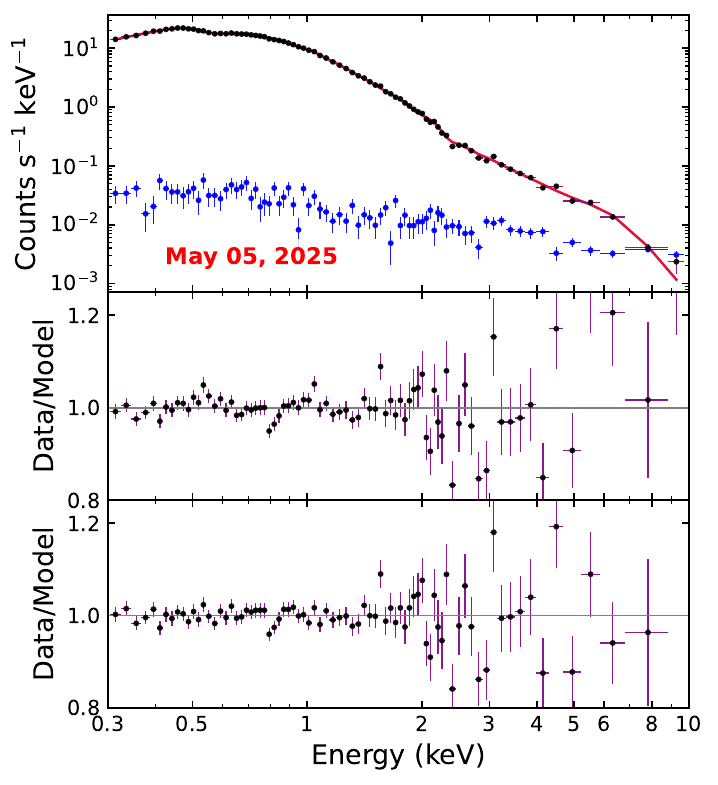}
    }
    \caption{Same as Fig. \ref{fig:pn_set2} for the observations taken on May 02, 2025, and May 05, 2025.}
    \label{fig:pn_set9}
\end{figure*}

\clearpage
\section{Additional optical spectroscopic data}
\label{sec:optical spectra evolution}

\begin{table}[!h]
\centering
\caption{Emission line fluxes measured from LBT observations. The line fluxes were estimated through spectral fitting. All fluxes are in units of $10^{-17}$ erg cm$^{-2}$ s$^{-1}$.}
\label{tab:line_fluxes_appendix}
\begin{tabular}{lrrrrr}
\hline
\textbf{Line ID} & \textbf{11-09-2023} & \textbf{09-11-2024} & \textbf{21-11-2024} & \textbf{13-10-2025} & \textbf{11-11-2025} \\
\hline
OIII4959 & $295 \pm \phantom{0}9$   & $316 \pm \phantom{0}9$   & $335 \pm 11$ & $357 \pm \phantom{0}7$   & $402 \pm 15$ \\
OIII5007 & $893 \pm 25$  & $957 \pm 26$  & $1016 \pm 32$ & $1055 \pm 22$ & $1184 \pm 45$ \\
H$\alpha$ & $723 \pm 24$  & $730 \pm 31$  & $439 \pm 16$ & $415 \pm 14$  & $482 \pm 30$ \\
NII6549  & $107 \pm \phantom{0}4$   & $121 \pm \phantom{0}5$   & $76 \pm \phantom{0}3$   & $48 \pm \phantom{0}2$    & $62 \pm \phantom{0}4$ \\
NII6585  & $49 \pm 12$   & $366 \pm 16$  & $238 \pm \phantom{0}9$  & $174 \pm \phantom{0}6$   & $224 \pm 14$ \\
\hline
\end{tabular}
\end{table}

\begin{figure*}
    \centering
    \includegraphics[width=0.9\linewidth]{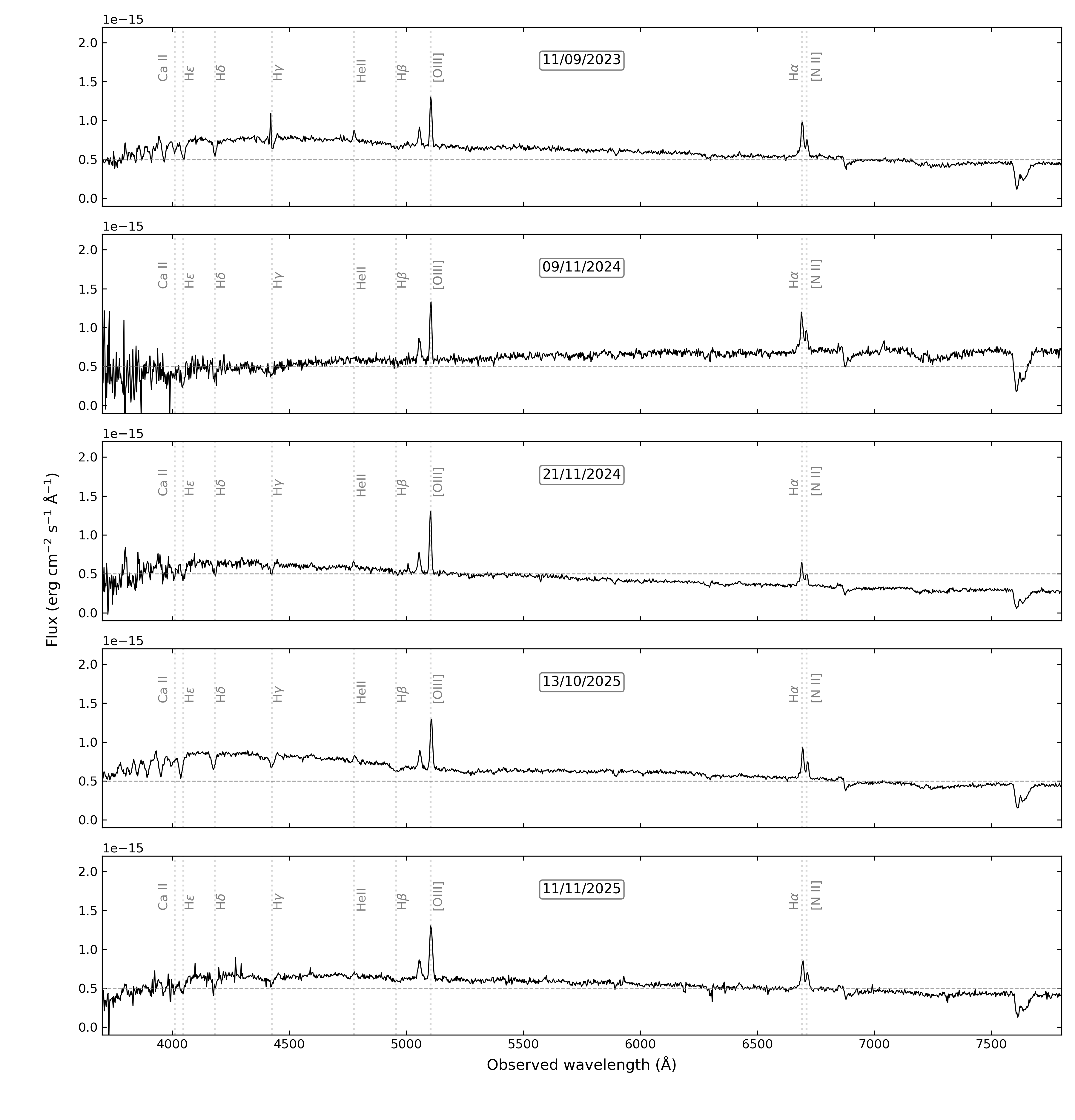}
    \caption{Optical spectra of 1ES 1927+654 obtained with DOLORES at the Telescopio Nazionale Galileo (TNG) at multiple epochs between 11 September 2023 and 11 November 2025. No broad emission-line component is detected in any of the epochs.}
\end{figure*}

\end{document}